\documentclass{ws-ijmpe}
\usepackage{amsmath,amssymb}
\usepackage{tensor}
\usepackage[makeroom]{cancel}
\usepackage{verbatim}
\let\trimmarks\relax
\def\onehalf{{\textstyle\frac{1}{2}}}
\def\quarter{{\textstyle\frac{1}{4}}}
\def\HC{\mathcal{H}}

\def\LC{\mathcal{L}}

\def\CO{\mathcal{O}}
\def\RB{\mathbb{R}}

\def\bx{x}

\def\bp{\pmb{p}}

\def\bj{\pmb{j}}

\def\ba{\pmb{a}}
\def\bA{\pmb{A}}

\def\bF{\pmb{F}}

\def\bS{\pmb{S}}

\def\bPhi{\pmb{\Phi}}
\def\bphi{\pmb{\phi}}

\def\bPi{\pmb{\Pi}}
\def\bpi{\pmb{\pi}}

\def\b0{\pmb{0}}
\DeclareMathOperator{\tr}{Tr}

\DeclareMathOperator{\diag}{diag}

\newcommand{\Eins}{\mathbf{1}}
\def\bpartial{\pmb{\partial}}

\def\d{\mathrm{d}}

\def\rmi{\mathrm{i}}\let\im\rmi
\newcommand{\pfrac}[2]{\frac{\partial{#1}}{\partial{#2}}}
\newcommand{\ppfrac}[3]{\frac{\partial^{2}{#1}}{\partial{#2}\partial{#3}}}
\newcommand{\pppfrac}[4]{\frac{\partial^{3}{#1}}{\partial{#2}\partial{#3}\partial{#4}}}

\begin{document}
\markboth{J.~Struckmeier, A.~Redelbach}{Covariant Hamiltonian Field Theory}
%
%
\title{COVARIANT HAMILTONIAN FIELD THEORY}
\author{J\"URGEN STRUCKMEIER and ANDREAS REDELBACH}
\address{GSI Helmholtzzentrum f\"ur Schwerionenforschung GmbH\\
Planckstr.~1, 64291~Darmstadt, Germany\\ and\\
Johann Wolfgang Goethe-Universit\"at Frankfurt am Main\\
Max-von-Laue-Str.~1, 60438~Frankfurt am Main, Germany\\
struckmeier@fias.uni-frankfurt.de}
\maketitle
\begin{history}
\received{18 July 2007}
\revised{27 October 2026}
\end{history}
\begin{abstract}
A local coordinate formulation of covariant Hamiltonian field theory is presented.
For regular first-order Lagrangians, the covariant canonical field equations are equivalent to the Euler-Lagrange equations.
The transformation theory is formulated for local changes of the field variables at fixed space-time coordinates.
The admissibility conditions on vector-valued generating functions are made explicit.
Canonical two-forms are invariant under these transformations, whereas component Poisson brackets and phase-space volume statements require the qualifications specified below.
The transformation rules are illustrated by scalar, vector, spinor, and gauge-field examples, with singular Legendre maps treated separately.
Infinitesimal canonical symmetries yield conserved Noether currents.
The affine extension of the unitary gauge transformation is developed for shift-invariant scalar matter, with an independent mass scale and a compensator interpretation of the massive vector fields.
The additional boundary terms required for a massless Dirac shift symmetry are distinguished from this scalar construction.
Space-time translations and the conservation of the symplectic current are discussed without identifying general field evolution with a finite-dimensional canonical map.
\end{abstract}
\ccode{PACS numbers: 11.10.Ef, 11.15Kc}
\section{Introduction}
Relativistic field theories and gauge theories are commonly
formulated on the basis of a Lagrangian density
$\LC$\cite{greiner85,weinberg96,ryder06,griffiths08}.
The space-time evolution of the fields is obtained by integrating
the Euler-Lagrange field equations that follow from the
four-dimensional representation of Hamilton's action principle.
A characteristic feature of this approach is that the four
independent variables of space and time are treated on equal
footing.
Relativistic covariance additionally requires the appropriate Lorentz transformation laws of the fields and a Lorentz-scalar Lagrangian density.
This is reflected by the fact that the Lagrangian density $\LC$
depends --- apart from a possible explicit dependence on the four
space-time coordinates $x^{\mu}$ --- on the set of fields $\phi_{I}$
and evenly on all four derivatives $\bpartial\phi_{I}$
of those fields with respect to the space-time coordinates,
i.e.\ $\LC=\LC(\phi_{I},\bpartial\phi_{I},x^{\mu})$.
Herein, the index ``$I$'' enumerates the individual fields
that are involved in the given physical system.

When the transition to a Hamiltonian description is made in
textbooks, the equal footing of the space-time coordinates
is abandoned\cite{saletan98,greiner85,weinberg96}.
In these presentations, the Hamiltonian density $\HC$ is
defined to depend on the set of scalar fields $\phi_{I}$
and on \emph{one} set of conjugate scalar fields $\pi_{I}$
that counterpart the \emph{time} derivatives of the $\phi_{I}$.
Keeping the dependencies on the three \emph{spatial} derivatives
$\partial_{\nu}\phi_{I}$ of the fields $\phi_{I}$, the functional
dependence of the Hamiltonian is then defined as
$\HC=\HC(\phi_{I},\bpi_{I},\partial_{\nu}\phi_{I},x^{\mu})$.
The canonical field equations then emerge as \emph{time
derivatives} of the scalar fields $\phi_{I}$ and $\pi_{I}$.
In other words, the \emph{time} variable is singled
out of the set of independent space-time variables.
This formulation supports a canonical transformation theory on the space of Cauchy data.
Its space-time split, however, is not manifestly covariant.
The present paper instead develops a local covariant formulation using polymomenta conjugate to all first derivatives of the fields.

On the other hand, numerous papers were published that formulate
a \emph{covariant} Hamiltonian description of field theories where
--- similar to the Lagrangian formalism --- the four
independent variables of space-time are treated equally.
These papers are generally based on the pioneering works
of De~Donder\cite{dedonder30} and Weyl\cite{weyl35}.
The key point of this approach is that the Hamiltonian
density $\HC$ is now defined to depend on a set of conjugate
$4$-vector fields $\pi_{I}^{\mu}$ that counterbalance the four
derivatives $\partial_{\mu}\phi_{I}$ of the Lagrangian density
$\LC$, so that $\HC=\HC(\phi_{I},\pi_{I}^{\mu},x^{\mu})$.
Corresponding to the Euler-Lagrange equations of field theory,
the canonical field equations then take on a symmetric form
with respect to the four independent variables of space-time.
This approach is commonly referred to as ``multisymplectic'' or
``polysymplectic field theory'', thereby labeling the covariant
extension of the symplectic geometry of the conventional Hamiltonian
theory\cite{kanatchikov98,gotay04,kastrup83,paufler01,forger03,eche96,sarda95,kisil04}.
Mathematically, the phase space of multisymplectic Hamiltonian
field theory is defined within modern differential geometry in
the language of ``jet bundles''\cite{saunders89,guenther87}.

The geometric formulations distinguish several phase spaces and several notions of covariant bracket.
Their respective domains and invariance properties must therefore be specified\cite{kanatchikov98,paufler01,sarda95,forger03}.
A local coordinate presentation is useful for identifying these restrictions explicitly.

With the present paper, we do \emph{not} pursue the differential
geometry path but provide a \emph{local coordinate} treatise
of De~Donder and Weyl's covariant Hamiltonian field theory.
The \emph{local} description enables us to keep the
mathematics on the level of tensor calculus.
The formulas below use Cartesian coordinates on Minkowski space with metric $(g_{\mu\nu})=(\eta_{\mu\nu})=\operatorname{diag}(1,-1,-1,-1)$.
They are Lorentz covariant when the fields transform in their specified representations.
An extension to arbitrary space-time coordinates requires the corresponding density weights and, for fields with space-time indices, the appropriate connection terms.
Similar to textbooks on Lagrangian gauge theories,
we maintain a close tie to physics throughout the paper.

Our paper is organized as follows.
In Sec.~\ref{sec:d-w-theory}, we give a brief review of
De~Donder and Weyl's approach to Hamiltonian field theory in order
to render the paper self-contained and to clarify notation.
After reviewing the covariant canonical field equations, we
identify an eigenvalue relation for the energy-momentum tensor of a single scalar field and
discuss equivalent momentum representations obtained by adding a total divergence to the Lagrangian.

The main benefit of the covariant Hamiltonian approach
is that it enables us to formulate a consistent theory
of covariant canonical transformations.
This is demonstrated in Sec.~\ref{sec:d-w-cantra}.
Strictly imitating the point mechanics' approach\cite{greiner03,goldstein02},
we set up the transformation rules on the basis of a
generating function by requiring the variational
principle to be maintained\cite{musi78}.
In contrast to point mechanics, the generating function
$F_{1}^{\mu}$ now emerges in our approach as a $4$-vector
function.
We recover a characteristic feature of canonical
transformations by deriving the symmetry relations of the
mutual partial derivatives of original and transformed fields.
We derive the transformation rules from generating functions $F_{2}^{\mu}$ and $F_{3}^{\mu}$.
The formal relations involving $F_{1}^{\mu}$ and $F_{4}^{\mu}$ are discussed together with the restrictions on their use as independent generating coordinates.
Mixed partial derivatives yield compatibility conditions between original and transformed fields.

The symmetry relations determine the invariance of the canonical two-forms and the conditions under which component Poisson and Lagrange brackets are preserved.
The volume element of each canonical field-momentum sector must be distinguished from the volume element of the full polymomentum space.
We conclude this section with the Jacobi identity, the limitations of a direct field-theoretic Poisson theorem, and the Hamilton-Jacobi equation.
The latter requires an additional embedding condition to reconstruct solutions of the field equations.

In Sec.~\ref{sec:examples-ham}, examples of Hamiltonian densities
are reviewed and their pertaining field equations are derived.
As the relativistic invariance of the resulting fields equations
is ensured if the Hamiltonian density $\HC$ is a Lorentz scalar,
various equations of relativistic quantum field theory are
demonstrated to embody, in fact, canonical field equations.
In particular, the Hamiltonian density engendering the Klein-Gordon
equation manifests itself as the covariant field theory analog
of the harmonic oscillator Hamiltonian of point mechanics.

Section~\ref{examples-ct} starts with simple examples of canonical transformations of Hamiltonian systems.
For the derivative-free generators at fixed space-time coordinates considered here, admissibility restricts the transformed fields to point transformations.
A canonical transformation relates Hamiltonian descriptions, whereas a canonical symmetry preserves the specified functional form of the Hamiltonian.
The associated infinitesimal generator then defines a conserved Noether current.
Space-time translation invariance yields the corresponding energy-momentum current.
The preservation of symplectic structure by field evolution is expressed through a conserved symplectic current and, with suitable boundary conditions, the induced symplectic form on Cauchy data.

A canonical change of field variables can also be used to expand a scalar potential about a chosen minimum.
Section~\ref{sec:higgs} distinguishes this reparametrization from the dynamical choice of a vacuum and from the Higgs mechanism itself.
Its scalar example gives a compensator formulation of massive complex vector fields, subject to the stated restrictions on the bare Hamiltonian and the mass matrix.
The proposed extension to Dirac fields is tested separately because their global shift symmetry involves a boundary term and additional momentum shifts.
We close our treatise with the scope of momentum transformations for the regularized Dirac field and its local gauge coupling.
\section{\label{sec:d-w-theory}Covariant Hamiltonian density}
\subsection{\label{sec:kfgln}Covariant canonical field equations}
The transition from particle dynamics to the dynamics of a
\emph{continuous} system is based on the assumption that a
\emph{continuum limit} exists for the given physical problem.
This limit is defined by letting the number of particles
involved in the system increase over all bounds
while letting their masses and distances go to zero.
In this limit, the information on the location of individual
particles is replaced by the \emph{value} of a smooth
function $\phi(\bx)$ that is \emph{uniquely} given at a spatial
location $x^{1},x^{2},x^{3}$ at time $t=x^{0}/c$.
The differentiable function $\phi(\bx)$ is called a \emph{primary field}.
In this notation, the index $\mu$ runs from $0$ to $3$, hence
distinguishes the four independent variables of space-time
$x^{\mu}\equiv(x^{0},x^{1},x^{2},x^{3})\equiv(ct,x,y,z)$, and
$x_{\mu}\equiv(x_{0},x_{1},x_{2},x_{3})\equiv(ct,-x,-y,-z)$.
We furthermore assume that the given physical problem can
be described in terms of $I=1,\ldots,N$ --- possibly
interacting --- scalar fields $\phi_{I}(\bx)$,
with the index ``$I$'' enumerating the individual fields.
In order to clearly distinguish scalar quantities from
vector quantities, we denote the latter with boldface letters.
Throughout our paper, the summation convention is used.
This means that whenever a pair of the same upper and
lower indices appears on one side of an equation,
this index is to be summed over.
If no confusion can arise, we omit the indices in the
argument list of functions in order to avoid the number
of indices to proliferate.

The Lagrangian description of the dynamics of a continuous
system is based on the Lagrangian density function $\LC$
that is supposed to carry the complete information
on the given physical system.
In a first-order field theory, the Lagrangian density $\LC$
is defined to depend on the $\phi_{I}$, possibly on the vector
of independent variables $\bx$, and on the four first
derivatives of the fields $\phi_{I}$ with respect to the
independent variables, i.e., on the $1$-forms
$$
\bpartial\phi_{I}\equiv(\partial_{ct}\phi_{I},\partial_{x}
\phi_{I},\partial_{y}\phi_{I},\partial_{z}\phi_{I}).
$$
The Euler-Lagrange field equations are then obtained
as the zero of the variation $\delta S$ of the action integral
\begin{equation}\label{action-int}
S=\int\LC(\phi_{I},\bpartial\phi_{I},\bx)\,d\bx
\end{equation}
as
\begin{equation}\label{elgl}
\pfrac{}{x^{\alpha}}\pfrac{\LC}{(\partial_{\alpha}\phi_{I})}-
\pfrac{\LC}{\phi_{I}}=0.
\end{equation}
To derive the equivalent \emph{covariant} Hamiltonian
description of continuum dynamics, we first define for
each primary field $\phi_{I}(\bx)$ a $4$-vector of conjugate
momentum fields $\pi_{I}^{\mu}(\bx)$.
Its components are given by
\begin{equation}\label{p-def}
\pi_{I}^{\mu}=\pfrac{\LC}{(\partial_{\mu}\phi_{I})}
\equiv\pfrac{\LC}{\left(\pfrac{\phi_{I}}{x^{\mu}}\right)}.
\end{equation}
The $4$-vector $\bpi_{I}$ is thus induced by the
Lagrangian $\LC$ as the \emph{dual counterpart} of
the $1$-form $\bpartial\phi_{I}$.
For the entire set of $N$ scalar fields $\phi_{I}(\bx)$,
this establishes a set of $N$ conjugate $4$-vector fields.
With this definition of the $4$-vectors of canonical momenta
$\bpi_{I}(\bx)$, we can now define the Hamiltonian
density $\HC(\phi_{I},\bpi_{I},\bx)$ as the
covariant Legendre transform of the Lagrangian density
$\LC(\phi_{I},\bpartial\phi_{I},\bx)$
\begin{equation}\label{H-def}
\HC(\phi_{I},\bpi_{I},\bx)=\pi_{J}^{\alpha}
\pfrac{\phi_{J}}{x^{\alpha}}-\LC(\phi_{I},\bpartial\phi_{I},\bx).
\end{equation}
At this point we suppose that $\LC$ is \emph{regular},
hence that the full $4N\times4N$ Hessian
$$
W_{I\mu,J\nu}=\frac{\partial^{2}\LC}{\partial(\partial_{\mu}\phi_{I})\,\partial(\partial_{\nu}\phi_{J})}
$$
is non-singular.
This condition includes the mixed derivatives between different fields.
It ensures local invertibility of the complete Legendre map.
Singular theories require constraints or a separately justified equivalent regular Lagrangian.
This ensures that $\HC$ takes over the complete
information on the given dynamical system from $\LC$
by means of the Legendre transformation.
The definition of $\HC$ by Eq.~(\ref{H-def}) is referred to
in literature as the ``De~Donder-Weyl'' Hamiltonian density.

Obviously, the dependencies of $\HC$ and $\LC$ on the
$\phi_{I}$ and the $x^{\mu}$ only differ by a sign,
$$
\pfrac{\HC}{\phi_{I}}=-\pfrac{\LC}{\phi_{I}},\qquad
\left.\pfrac{\HC}{x^{\mu}}\right\vert_{\text{expl}}=
-\left.\pfrac{\LC}{x^{\mu}}\right\vert_{\text{expl}}.
$$
These variables do not take part in the Legendre
transformation of Eqs.~(\ref{p-def}), (\ref{H-def}).
With regard to this transformation, the Hamiltonian
density $\HC$ is, therefore, to be considered as a function of
the $\pi_{I}^{\mu}$ only, and, correspondingly, the Lagrangian
density $\LC$ as a function of the $\partial_{\mu}\phi_{I}$ only.
In order to derive the canonical field equations, we calculate
from Eq.~(\ref{H-def}) the partial derivative of $\HC$ with
respect to $\pi_{I}^{\mu}$,
$$
\pfrac{\HC}{\pi_{I}^{\mu}}=\delta_{IJ}\,\delta_{\mu}^{\alpha}\,
\pfrac{\phi_{J}}{x^{\alpha}}=\pfrac{\phi_{I}}{x^{\mu}}.
$$
In differentiating Eq.~(\ref{H-def}), the terms containing derivatives of $\partial_{\alpha}\phi_{J}$ with respect to $\pi_{I}^{\mu}$ cancel by Eq.~(\ref{p-def}).
In conjunction with the Euler-Lagrange equation, we obtain
the set of covariant canonical field equations finally as
\begin{equation}\label{fgln}
\pfrac{\HC}{\pi_{I}^{\mu}}=\pfrac{\phi_{I}}{x^{\mu}},\qquad
\pfrac{\HC}{\phi_{I}}=-\pfrac{\pi_{I}^{\alpha}}{x^{\alpha}}.
\end{equation}
This pair of first-order partial differential equations
is equivalent to the set of second-order differential
equations of Eq.~(\ref{elgl}).
We observe that in this formulation of the canonical
field equations all coordinates of space-time
appear symmetrically --- similar to the Lagrangian
formulation of Eq.~(\ref{elgl}).
Provided that the Lagrangian density $\LC$ is a Lorentz
scalar, the dynamics of the fields is invariant with
respect to Lorentz transformations.
The covariant Legendre transformation~(\ref{H-def})
passes this property to the Hamiltonian density $\HC$.
It thus ensures \emph{a priori} the relativistic
invariance of the fields that emerge as integrals of
the canonical field equations if $\LC$
--- and hence $\HC$ --- represents a Lorentz scalar.
\subsection{\label{sec:emtensor}Energy-Momentum Tensor}
In the Lagrangian description, the \emph{canonical energy-momentum
tensor} $\theta\indices{_{\mu}^{\nu}}$ is defined as the following
mixed second rank tensor
\begin{equation}\label{e-i-def}
\theta\indices{_{\mu}^\nu}=\pfrac{\LC}{(\partial_{\nu}\phi_{I})}
\pfrac{\phi_{I}}{x^{\mu}}-\delta_{\mu}^{\nu}\,\LC.
\end{equation}
With the definition~(\ref{p-def}) of the conjugate
momentum fields $\pi_{I}^{\mu}$, and the Hamiltonian density
of Eq.~(\ref{H-def}), the energy-momentum
tensor~(\ref{e-i-def}) is equivalently expressed as
\begin{equation}\label{e-i-def-h}
\theta\indices{_{\mu}^{\nu}}=\pi_{I}^{\nu}\,\pfrac{\phi_{I}}{x^{\mu}}+
\delta_{\mu}^{\nu}\left(\HC-\pi_{I}^{\alpha}\pfrac{\phi_{I}}{x^{\alpha}}\right).
\end{equation}
If the Hamiltonian describes a \emph{single} scalar field, direct contraction of Eq.~(\ref{e-i-def-h}) gives
$$
\theta\indices{_{\mu}^{\nu}}\pfrac{\phi}{x^{\nu}}=\HC\pfrac{\phi}{x^{\mu}},
\qquad
\pi^{\mu}\theta\indices{_{\mu}^{\nu}}=\HC\pi^{\nu}.
$$
Whenever the relevant vector is nonzero, $\partial_{\mu}\phi$ is a right eigenvector and $\pi^{\mu}$ is a left eigenvector of the mixed matrix $\theta\indices{_{\mu}^{\nu}}$, with eigenvalue $\HC$.
For a collection of interacting fields, the sum over field labels generally prevents this eigenvalue relation.
The scalar contraction with $\partial_{\nu}\phi\,\pi^{\mu}$ alone would not establish an eigenvalue equation, and a rank-two eigentensor is not required.

An important property of the energy-momentum tensor
is revealed by calculating the divergence
$\partial\theta\indices{_{\mu}^{\alpha}}/\partial x^{\alpha}$.
From the definition~(\ref{e-i-def-h}), we find
\begin{align*}
\pfrac{\theta\indices{_{\mu}^{\alpha}}}{x^{\alpha}}&=
\delta_{\mu}^{\alpha}\left(\pfrac{\HC}{\phi_{I}}
\pfrac{\phi_{I}}{x^{\alpha}}+\pfrac{\HC}{\pi_{I}^{\beta}}
\pfrac{\pi_{I}^{\beta}}{x^{\alpha}}+\left.\pfrac{\HC}{x^{\alpha}}
\right\vert_{\text{expl}}\right)+
\pfrac{\pi_{I}^{\alpha}}{x^{\alpha}}\pfrac{\phi_{I}}{x^{\mu}}\\
&\quad+\pi_{I}^{\alpha}\ppfrac{\phi_{I}}{x^{\mu}}{x^{\alpha}}-
\delta_{\mu}^{\alpha}\left(\pfrac{\pi_{I}^{\beta}}{x^{\alpha}}
\pfrac{\phi_{I}}{x^{\beta}}+\pi_{I}^{\beta}\ppfrac{\phi_{I}}
{x^{\alpha}}{x^{\beta}}\right)
\end{align*}
Inserting the canonical field equations~(\ref{fgln}), this becomes
\begin{equation}\label{deltaH-expl}
\pfrac{\theta\indices{_{\mu}^{\alpha}}}{x^{\alpha}}=\left.\pfrac{\HC}{x^{\mu}}
\right\vert_{\text{expl}}.
\end{equation}
The second canonical field equation determines the divergence of each momentum field.
This fact alone does not leave its components arbitrary.
For a fixed regular Lagrangian, the complete Legendre map fixes the momenta from the field and its first derivatives, and the energy-momentum tensor is then fixed by its definition.

If the Hamiltonian density $\HC$ does not \emph{explicitly}
depend on the independent variable $x^{\mu}$, then $\HC$ is
obviously invariant with respect to a shift of the reference
system along the $x^{\mu}$ axis.
Then, the components of the $\mu$-th column of the
energy-momentum tensor satisfy the continuity equation
$$
\pfrac{\theta\indices{_{\mu}^{\alpha}}}{x^{\alpha}}=0\qquad\Longleftrightarrow
\qquad\left.\pfrac{\HC}{x^{\mu}}\right\vert_{\text{expl}}=0.
$$
Equation~(\ref{deltaH-expl}) is the local balance law associated with space-time translations.
It must be distinguished from a conservation law for the De~Donder-Weyl Hamiltonian density itself.
In particular, the ordinary-mechanical relation $dH/dt=\partial H/\partial t$ has no direct counterpart for all four derivatives of $\HC$.
As we easily convince ourselves, the derivative of $\HC$
with respect to $x^{\mu}$ is \emph{not} uniquely determined
by its explicit dependence on $x^{\mu}$
\begin{align}
\pfrac{\HC}{x^{\mu}}&=\left.\pfrac{\HC}{x^{\mu}}\right
\vert_{\text{expl}}+\pfrac{\HC}{\pi_{I}^{\alpha}}
\pfrac{\pi_{I}^{\alpha}}{x^{\mu}}+
\pfrac{\HC}{\phi_{I}}\pfrac{\phi_{I}}{x^{\mu}}\nonumber\\
&=\left.\pfrac{\HC}{x^{\mu}}\right\vert_{\text{expl}}+
\pfrac{\phi_{I}}{x^{\alpha}}\pfrac{\pi_{I}^{\alpha}}{x^{\mu}}-
\pfrac{\phi_{I}}{x^{\mu}}\pfrac{\pi_{I}^{\beta}}{x^{\beta}}\nonumber\\
&=\left.\pfrac{\HC}{x^{\mu}}\right\vert_{\text{expl}}+
k\indices{_{I\mu}^{\alpha}}\pfrac{\phi_{I}}{x^{\alpha}},\qquad
k\indices{_{I\mu}^{\nu}}=\pfrac{\pi_{I}^{\nu}}{x^{\mu}}-
\delta_{\mu}^{\nu}\pfrac{\pi_{I}^{\beta}}{x^{\beta}}.
\label{nicht-geeicht}
\end{align}
Owing to the fact that the number of independent variables
is greater than one, the two rightmost terms of
Eq.~(\ref{nicht-geeicht}) constitute a \emph{sum}.
In contrast to the case of point mechanics,
these terms generally do not cancel by virtue of the
canonical field equations.
\subsection{\label{dis-pi}Equivalent representations of the conjugate vector fields $\bpi_{I}$}
The second canonical field equation fixes the divergence of the conjugate momentum field, but it does not by itself imply a freedom to change the momenta while retaining a fixed Hamiltonian function.
An equivalent representation is obtained by changing the Lagrangian by a total divergence.
Let $\eta_{I}^{\mu}(x)$ be a prescribed smooth vector field satisfying
\begin{equation}\label{eichung}
\tilde{\pi}_{I}^{\mu}=\pi_{I}^{\mu}-\eta_{I}^{\mu},\qquad
\pfrac{\eta_{I}^{\alpha}}{x^{\alpha}}=0.
\end{equation}
The corresponding Lagrangian is
$$
\tilde{\LC}=\LC-\eta_{I}^{\alpha}\pfrac{\phi_{I}}{x^{\alpha}}
=\LC-\pfrac{}{x^{\alpha}}\big(\eta_{I}^{\alpha}\phi_{I}\big).
$$
For boundary variations that annihilate this total divergence, the Euler-Lagrange equations are unchanged.
The Legendre transformation yields
\begin{align*}
\tilde{\pi}_{I}^{\mu}&=\pfrac{\tilde{\LC}}{(\partial_{\mu}\phi_{I})}
=\pi_{I}^{\mu}-\eta_{I}^{\mu},\\
\tilde{\HC}(\bphi,\tilde{\bpi},x)&=
\HC(\bphi,\tilde{\bpi}+\boldsymbol{\eta}(x),x).
\end{align*}
Thus the numerical values of the two Hamiltonians coincide at corresponding phase-space points, although their functional dependence on their momentum arguments is generally different.
The canonical field equations transform consistently because
$$
\pfrac{\tilde{\HC}}{\tilde{\pi}_{I}^{\mu}}=
\pfrac{\HC}{\pi_{I}^{\mu}},\qquad
\pfrac{\tilde{\HC}}{\phi_{I}}=\pfrac{\HC}{\phi_{I}}
=-\pfrac{\tilde{\pi}_{I}^{\alpha}}{x^{\alpha}}.
$$
In contrast, their explicit coordinate derivatives obey
$$
\left.\pfrac{\tilde{\HC}}{x^{\mu}}\right|_{\mathrm{expl}}
=\left.\pfrac{\HC}{x^{\mu}}\right|_{\mathrm{expl}}
+\pfrac{\HC}{\pi_{I}^{\alpha}}\pfrac{\eta_{I}^{\alpha}}{x^{\mu}}.
$$
A divergence-free momentum shift is therefore a change of Hamiltonian representation, and is a symmetry of a specified Hamiltonian only when the additional symmetry condition is satisfied.
\section{\label{sec:d-w-cantra}Canonical
transformations in covariant Hamiltonian field theory}
\subsection{Generating functions of type $\bF_{1}(\bphi,\bPhi,\bx)$ and their scope}
Similar to the canonical formalism of point mechanics, a local invertible transformation of the fields
$(\bphi,\bpi)\mapsto(\bPhi,\bPi)$ is called \emph{canonical} when it preserves the Hamiltonian variational principle with an appropriately transformed Hamiltonian,
\begin{equation}\label{varprinzip}
\delta\int_R\left(\pi_I^\alpha\pfrac{\phi_I}{x^\alpha}-\HC\right)\d^4x
=\delta\int_R\left(\Pi_I^\alpha\pfrac{\Phi_I}{x^\alpha}-\HC'\right)\d^4x.
\end{equation}
We consider transformations at fixed space-time coordinates with no dependence on derivatives of the canonical variables.
A sufficient local condition is that the two integrands differ by a total divergence whose boundary integral has vanishing variation.
Compactly supported variations provide one consistent choice of boundary conditions.
The transformed equations then retain the form
\begin{equation}\label{fgln-p}
\pfrac{\HC'}{\Pi_I^\mu}=\pfrac{\Phi_I}{x^\mu},\qquad
\pfrac{\HC'}{\Phi_I}=-\pfrac{\Pi_I^\alpha}{x^\alpha}.
\end{equation}
Writing the boundary vector as $F_1^\mu$, the integrand identity is
\begin{align}
\LC&=\LC'+\pfrac{F_1^\alpha}{x^\alpha},\label{intbedlag}\\
\pi_I^\alpha\pfrac{\phi_I}{x^\alpha}-\HC
&=\Pi_I^\alpha\pfrac{\Phi_I}{x^\alpha}-\HC'
+\pfrac{F_1^\alpha}{x^\alpha}.\label{intbed}
\end{align}
If $F_1^\mu$ can be represented using independent coordinates $(\bphi,\bPhi,x)$, its divergence is formally
\begin{equation}\label{divF}
\pfrac{F_1^\alpha}{x^\alpha}
=\pfrac{F_1^\alpha}{\phi_I}\pfrac{\phi_I}{x^\alpha}
+\pfrac{F_1^\alpha}{\Phi_I}\pfrac{\Phi_I}{x^\alpha}
+\left.\pfrac{F_1^\alpha}{x^\alpha}\right|_{\mathrm{expl}}.
\end{equation}
Comparison of independent derivative coefficients would give
\begin{equation}\label{genF1}
\pi_I^\mu=\pfrac{F_1^\mu}{\phi_I},\qquad
\Pi_I^\mu=-\pfrac{F_1^\mu}{\Phi_I},\qquad
\HC'=\HC+\left.\pfrac{F_1^\alpha}{x^\alpha}\right|_{\mathrm{expl}},
\end{equation}
and the mixed-derivative relation
\begin{equation}\label{symmF1}
\pfrac{\pi_I^\mu}{\Phi_J}=-\pfrac{\Pi_J^\mu}{\phi_I}.
\end{equation}
The independence qualification is essential.
At fixed $x$, the full polymomentum space has dimension $5N$, while $(\bphi,\bPhi)$ supplies only $2N$ variables.
These variables therefore cannot parametrize a general graph of an invertible transformation of the full polymomentum space.
For a point transformation, they are moreover related by $\Phi_I=f_I(\bphi,x)$.
Consequently, Eqs.~(\ref{genF1}) and (\ref{symmF1}) are formal generating relations on a suitable restricted domain, and do not furnish a general generating chart equivalent to type $F_2$.
A boundary vector evaluated on the transformation graph must not be differentiated as if $\bphi$ and $\bPhi$ were independent there.
The type-$F_2$ and type-$F_3$ descriptions below avoid this problem.

\subsection{\label{sec:genf2}Generating functions of type $\bF_{2}(\bphi,\bPi,\bx)$}
The useful mixed variables are the original fields and the transformed polymomenta.
On the transformation graph one may write the formal Legendre relation
\begin{equation}\label{legendre1}
F_2^\mu=F_1^\mu+\Phi_J\Pi_J^\mu.
\end{equation}
The transformation rules can be derived directly, without assuming that a nonsingular type-$F_1$ representation exists.
Using the product rule in Eq.~(\ref{intbed}) gives
$$
\pi_I^\alpha\pfrac{\phi_I}{x^\alpha}-\HC
=-\Phi_I\pfrac{\Pi_I^\alpha}{x^\alpha}-\HC'
+\pfrac{F_2^\alpha}{x^\alpha}.
$$
Comparing coefficients in the independent mixed variables yields
\begin{equation}\label{genF2}
\pi_I^\mu=\pfrac{F_2^\mu}{\phi_I},\qquad
\Phi_I\delta_\nu^\mu=\pfrac{F_2^\mu}{\Pi_I^\nu},\qquad
\HC'=\HC+\left.\pfrac{F_2^\alpha}{x^\alpha}\right|_{\mathrm{expl}}.
\end{equation}
These equations impose a restriction on the admissible generating functions.
For $\nu\ne\mu$, the second equation implies that $F_2^\mu$ is independent of $\Pi_I^\nu$.
Its derivatives with respect to $\Pi_I^\mu$ must nevertheless give the same $\Phi_I$ for all four values of $\mu$.
Since the four momentum sectors are independent, $\Phi_I$ cannot depend on any of them.
The local general form is therefore
\begin{equation}\label{admissibleF2}
F_2^\mu(\bphi,\bPi,x)=f_I(\bphi,x)\Pi_I^\mu+b^\mu(\bphi,x),
\qquad
\det\left(\pfrac{f_I}{\phi_J}\right)\ne0.
\end{equation}
Thus the generating function is affine in the momentum component carrying its free space-time index.
The nonsingular field Jacobian ensures that the full transformation is locally invertible.
Writing $A_{JI}=\partial f_J/\partial\phi_I$, the rules become
\begin{align*}
\Phi_J&=f_J(\bphi,x),\qquad
\pi_I^\mu=A_{JI}\Pi_J^\mu+\pfrac{b^\mu}{\phi_I},\\
\HC'(\bPhi,\bPi,x)&=\HC(\bphi,\bpi,x)
+\Pi_J^\alpha\left.\pfrac{f_J}{x^\alpha}\right|_{\mathrm{expl}}
+\left.\pfrac{b^\alpha}{x^\alpha}\right|_{\mathrm{expl}}.
\end{align*}
The old variables on the right are expressed in terms of the new variables when $\HC'$ is regarded as a function on the new phase space.
A canonical transformation need not preserve the specified functional form of $\HC$.
That additional requirement defines a symmetry.

The symmetry of mixed partial derivatives gives
\begin{equation}\label{symmF2}
\left.\pfrac{\pi_I^\mu}{\Pi_J^\nu}\right|_{\bphi,x}
=\pfrac{\Phi_J}{\phi_I}\delta_\nu^\mu.
\end{equation}
The variables held fixed are part of this statement.
It must not be combined with derivatives belonging to a different generating chart without applying the chain rule.

For completeness, the canonical energy-momentum tensors transform according to
\begin{align*}
\Theta\indices{_\mu^\nu}
&=\theta\indices{_\mu^\nu}
+\Pi_I^\nu\left.\pfrac{f_I}{x^\mu}\right|_{\mathrm{expl}}
+\left.\pfrac{b^\nu}{x^\mu}\right|_{\mathrm{expl}}
-k\indices{_\mu^\nu},\\
k\indices{_\mu^\nu}
&=\pfrac{}{x^\alpha}\left(\delta_\mu^\alpha b^\nu-\delta_\mu^\nu b^\alpha\right),
\qquad \pfrac{k\indices{_\mu^\nu}}{x^\nu}=0.
\end{align*}
The last identity follows from commuting partial derivatives and antisymmetry in the two upper indices of the expression in parentheses.
The divergence-free term is an improvement of the tensor.
It may be omitted only when explicitly passing to an improved representative and when the corresponding surface contribution to an integrated quantity is controlled.

\subsection{\label{sec:genf3}Generating functions of type $\bF_{3}(\bPhi,\bpi,\bx)$}
The inverse point transformation can be written as $\phi_I=h_I(\bPhi,x)$.
A type-$F_3$ generator is
\begin{equation}\label{legendre2}
F_3^\mu(\bPhi,\bpi,x)
=b^\mu(\boldsymbol h(\bPhi,x),x)-h_I(\bPhi,x)\pi_I^\mu.
\end{equation}
On the graph this is the formal relation $F_3^\mu=F_1^\mu-\phi_I\pi_I^\mu$.
Differentiation, or a direct comparison of the action densities after applying the product rule, gives
\begin{equation}\label{genF3}
\Pi_I^\mu=-\pfrac{F_3^\mu}{\Phi_I},\qquad
\phi_I\delta_\nu^\mu=-\pfrac{F_3^\mu}{\pi_I^\nu},\qquad
\HC'=\HC+\left.\pfrac{F_3^\alpha}{x^\alpha}\right|_{\mathrm{expl}}.
\end{equation}
The explicit derivatives in the $F_2$ and $F_3$ formulas are taken at their respective independent mixed variables.
After expressing all quantities at corresponding phase-space points, they yield the same transformed Hamiltonian.
The associated mixed-derivative relation is
\begin{equation}\label{symmF3}
\left.\pfrac{\Pi_I^\mu}{\pi_J^\nu}\right|_{\bPhi,x}
=\pfrac{\phi_J}{\Phi_I}\delta_\nu^\mu.
\end{equation}
As in the type-$F_2$ representation, each component of $F_3$ is affine in the corresponding momentum component.

\subsection{Generating functions of type $\bF_{4}(\bPi,\bpi,\bx)$ and compatibility}
A second formal Legendre relation is
\begin{equation}\label{legendre3}
F_4^\mu=F_3^\mu+\Phi_J\Pi_J^\mu
=F_1^\mu-\phi_J\pi_J^\mu+\Phi_J\Pi_J^\mu.
\end{equation}
If independent momentum coordinates are available on an appropriate restricted domain, coefficient comparison would produce
\begin{equation}\label{genF4}
\Phi_I\delta_\nu^\mu=\pfrac{F_4^\mu}{\Pi_I^\nu},\qquad
\phi_I\delta_\nu^\mu=-\pfrac{F_4^\mu}{\pi_I^\nu},\qquad
\HC'=\HC+\left.\pfrac{F_4^\alpha}{x^\alpha}\right|_{\mathrm{expl}},
\end{equation}
and
\begin{equation}\label{symmF4}
\left.\pfrac{\Phi_I}{\pi_J^\alpha}\right|_{\bPi,x}\delta_\beta^\mu
=-\left.\pfrac{\phi_J}{\Pi_I^\beta}\right|_{\bpi,x}\delta_\alpha^\mu.
\end{equation}
These are again formal relations, not a general independent generating chart for the transformation class in Eq.~(\ref{admissibleF2}).
In particular, the $8N$ entries $(\bpi,\bPi)$ cannot all be independent coordinates on a $5N$-dimensional transformation graph.
Contracting Eq.~(\ref{symmF4}) and reinterpreting the resulting mixed-chart derivatives as ordinary transformation Jacobians is not justified.
The actual point-transformation restriction follows directly from Eq.~(\ref{genF2}) and gives
\begin{equation}\label{symmF4a}
\left.\pfrac{\Phi_I}{\pi_J^\mu}\right|_{\bphi,x}=0,\qquad
\left.\pfrac{\phi_I}{\Pi_J^\mu}\right|_{\bPhi,x}=0.
\end{equation}
This restriction applies to the derivative-free generators at fixed space-time coordinates studied here.
It is not a claim excluding canonical transformations on Cauchy data or more general transformations involving derivatives or space-time coordinates.

A further restriction that will be relevant to component brackets is
\begin{equation}\label{symmF1a}
\left.\pfrac{\Pi_J^\mu}{\phi_I}\right|_{\bpi,x}=0,\qquad
\left.\pfrac{\pi_I^\mu}{\Phi_J}\right|_{\bPi,x}=0.
\end{equation}
This property is not shared by general point transformations.
Its role in bracket invariance is specified in Sect.~\ref{sec:can-inv-pb}.

\subsection{Consistency check of the canonical transformation rules}
For the admissible generator~(\ref{admissibleF2}), the chain rule gives
$$
\pfrac{\Phi_I}{x^\alpha}
=A_{IJ}\pfrac{\phi_J}{x^\alpha}
+\left.\pfrac{f_I}{x^\alpha}\right|_{\mathrm{expl}}.
$$
Substitution of the transformed momenta and Hamiltonian into the action density yields
\begin{align*}
\Pi_I^\alpha\pfrac{\Phi_I}{x^\alpha}-\HC'
&=\left(\pi_J^\alpha-\pfrac{b^\alpha}{\phi_J}\right)
\pfrac{\phi_J}{x^\alpha}-\HC
-\left.\pfrac{b^\alpha}{x^\alpha}\right|_{\mathrm{expl}}\\
&=\pi_J^\alpha\pfrac{\phi_J}{x^\alpha}-\HC
-\pfrac{b^\alpha}{x^\alpha}.
\end{align*}
This verifies Eq.~(\ref{varprinzip}) directly with the total divergence $\partial_\alpha b^\alpha$ and the stated boundary conditions.
It also establishes equivalence of the $F_2$ and $F_3$ descriptions without assuming a nonsingular $F_1$ or $F_4$ chart.
The product-rule identity
\begin{equation}\label{equiv-act}
\pi_I^\alpha\pfrac{\phi_I}{x^\alpha}
=\pfrac{}{x^\alpha}(\pi_I^\alpha\phi_I)
-\phi_I\pfrac{\pi_I^\alpha}{x^\alpha}
\end{equation}
explains the alternative expressions of the action used in these derivations.
When exchanging them, the variation of the corresponding total divergence must vanish on the boundary.
Fixing only the fields at the boundary does not automatically fix products involving independently varied momenta.
Compactly supported variations avoid that ambiguity.
\subsection{\label{sec:p-k}Poisson brackets, Lagrange brackets}
For a system with given Hamiltonian density $\HC(\phi_{I},\bpi_{I},\bx)$,
and for two differentiable functions $f(\phi_{I},\bpi_{I},\bx)$,
$g(\phi_{I},\bpi_{I},\bx)$ of the fields $\phi_{I}$, $\pi_{I}^{\mu}$ and
the independent variables $x^{\mu}$, we define the $\mu$-th
component of the \emph{Poisson bracket} of $f$ and $g$ as follows
\begin{equation}\label{pbdef}
\big[f,g\big]_{\bphi,\bpi^{\mu}}=
\pfrac{f}{\phi_{I}}\,\pfrac{g}{\pi_{I}^{\mu}}-
\pfrac{f}{\pi_{I}^{\mu}}\,\pfrac{g}{\phi_{I}}.
\end{equation}
With this definition, the four Poisson brackets
$[f,g]_{\bphi,\bpi^{\mu}}$ constitute the components
of a dual $4$-vector, i.e., a $1$-form.
Obviously, the Poisson bracket~(\ref{pbdef}) satisfies
the following algebraic rules
\begin{align*}
\big[f,g\big]_{\bphi,\bpi^{\mu}}&=-\big[g,f\big]_{\bphi,\bpi^{\mu}}\\
\big[cf,g\big]_{\bphi,\bpi^{\mu}}&=c\big[f,g\big]_{\bphi,\bpi^{\mu}},
\quad c\in\mathbb{R}\\
\big[f,g\big]_{\bphi,\bpi^{\mu}}+\big[h,g\big]_{\bphi,\bpi^{\mu}}&=
\big[f+h,g\big]_{\bphi,\bpi^{\mu}}
\end{align*}
The Leibniz rule is obtained from~(\ref{pbdef}) via
\begin{align*}
\big[f,gh\big]_{\bphi,\bpi^{\mu}}&=
\pfrac{f}{\phi_{I}}\pfrac{}{\pi_{I}^{\mu}}(gh)-
\pfrac{f}{\pi_{I}^{\mu}}\pfrac{}{\phi_{I}}(gh)\\
&=\pfrac{f}{\phi_{I}}\left(\pfrac{g}{\pi_{I}^{\mu}}h+
g\pfrac{h}{\pi_{I}^{\mu}}\right)-\pfrac{f}{\pi_{I}^{\mu}}
\left(g\pfrac{h}{\phi_{I}}+\pfrac{g}{\phi_{I}}h\right)\\
&=\left(\pfrac{f}{\phi_{I}}\pfrac{g}{\pi_{I}^{\mu}}-
\pfrac{f}{\pi_{I}^{\mu}}\pfrac{g}{\phi_{I}}\right)h+
g\left(\pfrac{f}{\phi_{I}}\pfrac{h}{\pi_{I}^{\mu}}-
\pfrac{f}{\pi_{I}^{\mu}}\pfrac{h}{\phi_{I}}\right)\\
&=\big[f,g\big]_{\bphi,\bpi^{\mu}}h+g\big[f,h\big]_{\bphi,\bpi^{\mu}}
\end{align*}
For an arbitrary differentiable function $f(\phi_{I},\bpi_{I},\bx)$
of the field variables, we can, in particular, set up the Poisson
brackets with the canonical fields $\phi_{I}$, and $\bpi_{I}$.
As the individual field variables $\phi_{I}$ and $\pi_{I}^{\mu}$ are
independent by assumption, we immediately get
\begin{align*}
\big[\phi_{I},f\big]_{\bphi,\bpi^{\mu}}&=
\pfrac{\phi_{I}}{\phi_{J}}
\pfrac{f}{\pi_{J}^{\mu}}-
\pfrac{\phi_{I}}{\pi_{J}^{\mu}}
\pfrac{f}{\phi_{J}}=\hphantom{-}\delta_{IJ}
\pfrac{f}{\pi_{J}^{\mu}}=\pfrac{f}{\pi_{I}^{\mu}}\\
\big[\pi_{I}^{\nu},f\big]_{\bphi,\bpi^{\mu}}&=
\pfrac{\pi_{I}^{\nu}}{\phi_{J}}\pfrac{f}{\pi_{J}^{\mu}}-
\pfrac{\pi_{I}^{\nu}}{\pi_{J}^{\mu}}
\pfrac{f}{\phi_{J}}=-\delta_{IJ}\delta_{\mu}^{\nu}\,
\pfrac{f}{\phi_{J}}=-\delta_{\mu}^{\nu}\,
\pfrac{f}{\phi_{I}}.
\end{align*}
The Poisson bracket of a function $f$ of the field variables
with a particular field variable thus corresponds to the derivative
of that function $f$ with respect to the conjugate field variable.
For the particular case $f\equiv\HC$, this means
\begin{align}
\big[\phi_{I},\HC\big]_{\bphi,\bpi^{\mu}}&=\hphantom{-\delta_{\mu}^{\nu}\,}
\pfrac{\HC}{\pi_{I}^{\mu}}=\hphantom{\delta_{\mu}^{\nu}}\pfrac{\phi_{I}}{x^{\mu}}\nonumber\\
\big[\pi_{I}^{\nu},\HC\big]_{\bphi,\bpi^{\mu}}&=-\delta_{\mu}^{\nu}\,
\pfrac{\HC}{\phi_{I}}=\delta_{\mu}^{\nu}\,\pfrac{\pi_{I}^{\alpha}}{x^{\alpha}}.
\label{Heisenberg-eqs}
\end{align}
The last equation determines the divergence of the momentum vector, rather than its individual space-time derivatives.
It does not imply a freedom to change the momenta while keeping a prescribed Hamiltonian function fixed, as discussed in Sect.~\ref{dis-pi}.
In particular, the equality
$[\pi_{I}^{\nu},\HC]_{\bphi,\bpi^{\mu}}=\partial_{\mu}\pi_{I}^{\nu}$
would require
$\partial_{\mu}\pi_{I}^{\nu}=\delta_{\mu}^{\nu}\partial_{\alpha}\pi_{I}^{\alpha}$.
Contraction of $\mu$ and $\nu$ would then give
$\partial_{\alpha}\pi_{I}^{\alpha}=4\partial_{\alpha}\pi_{I}^{\alpha}$,
which is incompatible with a general nonzero momentum divergence.
Consequently, the component brackets with $\HC$ do not by themselves generate the complete space-time derivatives of arbitrary phase-space functions.
The correct chain rule along a solution is
\begin{equation}\label{equiv-PB}
\pfrac{f}{x^{\mu}}=\left.\pfrac{f}{x^{\mu}}\right|_{\mathrm{expl}}+[f,\HC]_{\bphi,\bpi^{\mu}}
+\pfrac{f}{\pi_{I}^{\nu}} \left(\pfrac{\pi_{I}^{\nu}}{x^{\mu}}
-\delta_{\mu}^{\nu}\pfrac{\pi_{I}^{\alpha}}{x^{\alpha}}\right).
\end{equation}
The last term must be retained unless additional conditions make it vanish.

The \emph{fundamental} Poisson brackets are constituted
by pairing field variables $\phi_{I}$ and $\pi_{I}^{\mu}$,
\begin{align}
\big[\phi_{I},\phi_{J}\big]_{\bphi,\bpi^{\mu}}&=
\pfrac{\phi_{I}}{\phi_{K}}\pfrac{\phi_{J}}{\pi_{K}^{\mu}}-
\pfrac{\phi_{I}}{\pi_{K}^{\mu}}\pfrac{\phi_{J}}{\phi_{K}}=0,\nonumber\\
\big[\phi_{I},\pi_{J}^{\nu}\big]_{\bphi,\bpi^{\mu}}&=
\pfrac{\phi_{I}}{\phi_{K}}\pfrac{\pi_{J}^{\nu}}{\pi_{K}^{\mu}}-
\pfrac{\phi_{I}}{\pi_{K}^{\mu}}\pfrac{\pi_{J}^{\nu}}{\phi_{K}}=
\pfrac{\pi_{J}^{\nu}}{\pi_{I}^{\mu}}=\delta_{\mu}^{\nu}\,\delta_{IJ},
\label{fundpk1}\\
\big[\pi_{I}^{\alpha},\pi_{J}^{\beta}\big]_{\bphi,\bpi^{\mu}}&=
\pfrac{\pi_{I}^{\alpha}}{\phi_{K}}\pfrac{\pi_{J}^{\beta}}{\pi_{K}^{\mu}}-
\pfrac{\pi_{I}^{\alpha}}{\pi_{K}^{\mu}}\pfrac{\pi_{J}^{\beta}}{\phi_{K}}=0.
\nonumber
\end{align}
The Lagrange brackets are the coordinate components of the vertical canonical $2$-forms.
Choose independent parameters $y^{a}$ that form a complete local coordinate chart of the phase-space fiber at fixed $x$.
For $f=y^{a}$ and $g=y^{b}$, with the other parameters held fixed, define
\begin{equation}\label{lag-def}
\{f,g\}^{\bphi,\bpi^{\mu}}=\pfrac{\phi_{I}}{f}\pfrac{\pi_{I}^{\mu}}{g}-
\pfrac{\pi_{I}^{\mu}}{f}\pfrac{\phi_{I}}{g}.
\end{equation}
The fundamental Lagrange brackets then emerge as
\begin{align}\label{fundlk1}
\big\{\phi_{I},\phi_{J}\big\}^{\bphi,\bpi^{\mu}}&=
\pfrac{\phi_{K}}{\phi_{I}}\pfrac{\pi_{K}^{\mu}}{\phi_{J}}-
\pfrac{\pi_{K}^{\mu}}{\phi_{I}}\pfrac{\phi_{K}}{\phi_{J}}=0,\nonumber\\
\big\{\phi_{I},\pi_{J}^{\nu}\big\}^{\bphi,\bpi^{\mu}}&=
\pfrac{\phi_{K}}{\phi_{I}}\pfrac{\pi_{K}^{\mu}}{\pi_{J}^{\nu}}-
\pfrac{\pi_{K}^{\mu}}{\phi_{I}}\pfrac{\phi_{K}}{\pi_{J}^{\nu}}=
\pfrac{\pi_{I}^{\mu}}{\pi_{J}^{\nu}}=\delta_{\nu}^{\mu}\,\delta_{IJ},\\
\big\{\pi_{I}^{\alpha},\pi_{J}^{\beta}\big\}^{\bphi,\bpi^{\mu}}&=
\pfrac{\phi_{K}}{\pi_{I}^{\alpha}}\pfrac{\pi_{K}^{\mu}}{\pi_{J}^{\beta}}-
\pfrac{\pi_{K}^{\mu}}{\pi_{I}^{\alpha}}\pfrac{\phi_{K}}{\pi_{J}^{\beta}}=0.
\nonumber
\end{align}
The partial derivatives in Eq.~(\ref{lag-def}) are defined relative to this complete parameter chart.
Two arbitrary functions without a specified set of complementary parameters do not determine a Lagrange bracket uniquely.
The next section establishes the canonical invariance of these $2$-form components and states the additional conditions needed for invariance of the component Poisson brackets.
\subsection{Canonical invariance of Poisson and Lagrange brackets\label{sec:can-inv-pb}}
Consider an admissible, locally invertible transformation generated by
\[
F_{2}^{\mu}=f_{J}(\bphi,x)\Pi_{J}^{\mu}+b^{\mu}(\bphi,x),
\qquad
A_{JI}=\pfrac{f_{J}}{\phi_{I}},\qquad \det A\ne0.
\]
Its transformation rules are
\[
\Phi_{J}=f_{J}(\bphi,x),\qquad
\pi_{I}^{\mu}=A_{JI}\Pi_{J}^{\mu}+\pfrac{b^{\mu}}{\phi_{I}}.
\]
All differentials and partial derivatives in the following argument are taken at fixed $x$.
Writing $\d_{v}$ for the exterior derivative with respect to the fields and momenta gives
\[
\pi_{I}^{\mu}\d_{v}\phi_{I}
=\Pi_{J}^{\mu}\d_{v}\Phi_{J}+\d_{v}b^{\mu}.
\]
Applying $-\d_{v}$ proves the invariance of the four canonical $2$-forms
\begin{equation}\label{canonical-two-forms}
\omega^{\mu}=\d_{v}\phi_{I}\wedge\d_{v}\pi_{I}^{\mu}
=\d_{v}\Phi_{I}\wedge\d_{v}\Pi_{I}^{\mu}.
\end{equation}
These forms are degenerate on the full $5N$-dimensional phase-space fiber.
Their invariance does not imply that the four bivectors defining Eq.~(\ref{pbdef}) are inverse invariant tensors on that entire fiber.

The first two fundamental Poisson brackets are maintained,
\begin{align}
[\Phi_{I},\Phi_{J}]_{\bphi,\bpi^{\mu}}&=0,\label{pk1}\\
[\Phi_{I},\Pi_{J}^{\nu}]_{\bphi,\bpi^{\mu}}&=
\delta_{IJ}\delta_{\mu}^{\nu}.\label{pk2}
\end{align}
To evaluate the momentum brackets, put
$B_{KI}=\partial\phi_{K}/\partial\Phi_{I}$, so that $B=A^{-1}$.
Since
$\partial\Pi_{I}^{\alpha}/\partial\pi_{K}^{\mu}
=B_{KI}\delta_{\mu}^{\alpha}$,
one obtains
\begin{align}\label{pk3}
[\Pi_{I}^{\alpha},\Pi_{J}^{\beta}]_{\bphi,\bpi^{\mu}}
&=\delta_{\mu}^{\beta}B_{KJ}\pfrac{\Pi_{I}^{\alpha}}{\phi_{K}}
-\delta_{\mu}^{\alpha}B_{KI}\pfrac{\Pi_{J}^{\beta}}{\phi_{K}}.
\end{align}
For $\alpha=\beta=\mu$, this expression vanishes by the canonical $2$-form identity.
For different momentum sectors, it need not vanish.
For example, the single-field transformation
\[
\Phi=e^{\phi},\qquad \Pi^{\alpha}=e^{-\phi}\pi^{\alpha}
\]
is generated by $F_{2}^{\mu}=e^{\phi}\Pi^{\mu}$, but for $\alpha\ne\mu$ it yields
\[
[\Pi^{\alpha},\Pi^{\mu}]_{\phi,\pi^{\mu}}
=-e^{-2\phi}\pi^{\alpha}\ne0
\]
in general.

For arbitrary differentiable functions of all momentum components, the chain rule therefore gives
\begin{equation}\label{kpfg}
[f,g]_{\bphi,\bpi^{\mu}}=[f,g]_{\bPhi,\bPi^{\mu}}
+\pfrac{f}{\Pi_{I}^{\alpha}}\pfrac{g}{\Pi_{J}^{\beta}}
[\Pi_{I}^{\alpha},\Pi_{J}^{\beta}]_{\bphi,\bpi^{\mu}}.
\end{equation}
The last term must not be discarded for a general canonical transformation.
All four component brackets are preserved on the full function algebra if
$\partial\Pi_{I}^{\alpha}/\partial\phi_{K}=0$ for every index.
Within the admissible invertible class above, this condition is also necessary.
Indeed, choose $\beta=\mu\ne\alpha$ in Eq.~(\ref{pk3}) and use the invertibility of $B$.
On an open phase-space set, these transformations have the affine form
\[
\bPhi=A(x)\bphi+\pmb{c}(x),\qquad
\bPi^{\mu}=A(x)^{-T}\big(\bpi^{\mu}-\pmb{d}^{\mu}(x)\big),
\]
where $A(x)$ is invertible and $\pmb{c}(x),\pmb{d}^{\mu}(x)$ are prescribed vectors of field-space coefficients.
For a general admissible point transformation, the bracket in a fixed sector $\mu$ is still invariant for functions depending only on $\bphi$ and $\bpi^{\mu}$.
Such functions transform into functions of $\bPhi$ and $\bPi^{\mu}$ only, and the correction in Eq.~(\ref{kpfg}) then vanishes.
The abbreviation $[f,g]_{\mu}$ used below refers to the specified canonical chart unless one of these invariance conditions applies.

The Lagrange brackets have a different status.
By Eq.~(\ref{canonical-two-forms}), their fundamental values are
\begin{align}
\{\Phi_{I},\Phi_{J}\}^{\bphi,\bpi^{\mu}}&=0,\label{lk1}\\
\{\Phi_{I},\Pi_{J}^{\nu}\}^{\bphi,\bpi^{\mu}}&=
\delta_{IJ}\delta_{\nu}^{\mu},\label{lk2}\\
\{\Pi_{I}^{\alpha},\Pi_{J}^{\beta}\}^{\bphi,\bpi^{\mu}}&=0.\label{lk3}
\end{align}
More generally, for any fixed complete parameter chart $y^{a}$,
\[
\omega^{\mu}=\frac12\{y^{a},y^{b}\}^{\bphi,\bpi^{\mu}}
\d y^{a}\wedge\d y^{b}
=\frac12\{y^{a},y^{b}\}^{\bPhi,\bPi^{\mu}}
\d y^{a}\wedge\d y^{b}.
\]
Equality of the coefficients proves the canonical invariance of the Lagrange brackets with that parameter chart held fixed.
Thus the invariant $2$-forms and Lagrange brackets coexist with the more restricted invariance of the component Poisson brackets.
\subsection{Canonical volume forms}
The invariant forms~(\ref{canonical-two-forms}) yield a version of Liouville's theorem for each fixed momentum sector.
For fixed $\mu$, with no summation over that index, define
\[
\d V_{\mu}=\d\phi_{1}\wedge\cdots\wedge\d\phi_{N}
\wedge\d\pi_{1}^{\mu}\wedge\cdots\wedge\d\pi_{N}^{\mu}.
\]
Up to a constant sign, this is $\big(\omega^{\mu}\big)^{N}/N!$.
It is a $2N$-form on the full phase-space fiber and becomes a volume form on a sector with coordinates $(\bphi,\bpi^{\mu})$.
The transformation considered in Sect.~\ref{sec:can-inv-pb} acts separately on each such sector.
Its Jacobian matrix has the block form
\begin{equation}\label{jacdet1}
J_{\mu}=\pfrac{(\bPhi,\bPi^{\mu})}{(\bphi,\bpi^{\mu})}
=\begin{pmatrix}A&0\\ C_{\mu}&A^{-T}\end{pmatrix},
\qquad
(C_{\mu})_{IJ}=\pfrac{\Pi_{I}^{\mu}}{\phi_{J}}.
\end{equation}
Consequently,
\begin{equation}\label{gen-liou}
\det J_{\mu}=\det A\det A^{-T}=1,
\qquad
\d V_{\mu}=\d\Phi_{1}\wedge\cdots\wedge\d\Phi_{N}
\wedge\d\Pi_{1}^{\mu}\wedge\cdots\wedge\d\Pi_{N}^{\mu}.
\end{equation}
This result concerns each sector separately.
The volume element of the complete $5N$-dimensional fiber contains one copy of the fields and four independent copies of the momenta.
Its Jacobian is instead
\begin{equation}\label{full-fiber-jacobian}
\det\pfrac{(\bPhi,\bPi^{0},\bPi^{1},\bPi^{2},\bPi^{3})}
{(\bphi,\bpi^{0},\bpi^{1},\bpi^{2},\bpi^{3})}
=\det A\big(\det A^{-T}\big)^{4}=(\det A)^{-3}.
\end{equation}
For example, $\Phi=2\phi$ and $\Pi^{\mu}=\pi^{\mu}/2$ preserve every $\d V_{\mu}$, whereas the full fiber volume changes by a factor $1/8$.
A general invariance of that full coordinate volume does not follow from canonicality.
Nor does the sector result by itself establish a volume-preserving evolution for a complete set of field data on a space-time hypersurface.
\subsection{Jacobi's identity and Poisson's theorem
in canonical field theory}
In order to derive the canonical field theory analog
of Jacobi's identity of point mechanics, we let
$f(\bphi,\bpi,\bx)$, $g(\bphi,\bpi,\bx)$, and
$h(\bphi,\bpi,\bx)$ denote arbitrary differentiable
functions of the canonical fields.
Within a fixed canonical chart, the sum of the three cyclically permuted nested
Poisson brackets be denoted by $a_{\mu\nu}$,
\begin{equation}\label{jac1}
a_{\mu\nu}=
{\Big[f,{\big[g,h\big]}_{\mu}\Big]}_{\nu}+
{\Big[h,{\big[f,g\big]}_{\mu}\Big]}_{\nu}+
{\Big[g,{\big[h,f\big]}_{\mu}\Big]}_{\nu}.
\end{equation}
We will now show that the $a_{\mu\nu}$ are the components
of a skew-symmetric $(0,2)$ tensor, hence that
\begin{equation}\label{jac2}
a_{\mu\nu}+a_{\nu\mu}=0.
\end{equation}
Writing Eq.~(\ref{jac1}) explicitly, we get a sum of
$24$ terms, each of them consisting of a triple product
of \emph{two} first-order derivatives and \emph{one}
second-order derivative of the functions $f$, $g$, and $h$
\begin{align*}
a_{\mu\nu}&=
\pfrac{f}{\phi_{J}}\pfrac{}{\pi_{J}^{\nu}}\left(
\pfrac{g}{\phi_{I}}\pfrac{h}{\pi_{I}^{\mu}}-
\pfrac{g}{\pi_{I}^{\mu}}\pfrac{h}{\phi_{I}}\right)\\
&\mbox{}-%
\pfrac{f}{\pi_{J}^{\nu}}\pfrac{}{\phi_{J}}\left(
\pfrac{g}{\phi_{I}}\pfrac{h}{\pi_{I}^{\mu}}-
\pfrac{g}{\pi_{I}^{\mu}}\pfrac{h}{\phi_{I}}\right)\\
&\mbox{}+%
\pfrac{h}{\phi_{J}}\pfrac{}{\pi_{J}^{\nu}}\left(
\pfrac{f}{\phi_{I}}\pfrac{g}{\pi_{I}^{\mu}}-
\pfrac{f}{\pi_{I}^{\mu}}\pfrac{g}{\phi_{I}}\right)\\
&\mbox{}-%
\pfrac{h}{\pi_{J}^{\nu}}\pfrac{}{\phi_{J}}\left(
\pfrac{f}{\phi_{I}}\pfrac{g}{\pi_{I}^{\mu}}-
\pfrac{f}{\pi_{I}^{\mu}}\pfrac{g}{\phi_{I}}\right)\\
&\mbox{}+%
\pfrac{g}{\phi_{J}}\pfrac{}{\pi_{J}^{\nu}}\left(
\pfrac{h}{\phi_{I}}\pfrac{f}{\pi_{I}^{\mu}}-
\pfrac{h}{\pi_{I}^{\mu}}\pfrac{f}{\phi_{I}}\right)\\
&\mbox{}-%
\pfrac{g}{\pi_{J}^{\nu}}\pfrac{}{\phi_{J}}\left(
\pfrac{h}{\phi_{I}}\pfrac{f}{\pi_{I}^{\mu}}-
\pfrac{h}{\pi_{I}^{\mu}}\pfrac{f}{\phi_{I}}\right).
\end{align*}
The proof can be simplified making use of the fact that
the terms of $a_{\mu\nu}$ from Eq.~(\ref{jac1}) emerge as
\emph{cyclic} permutations of the functions $f$, $g$, and $h$.
With regard to the explicit form of Eq.~(\ref{jac1}) from above
it suffices to show that Eq.~(\ref{jac2}) is fulfilled for all
terms containing second derivatives of, for instance, $f$,
\begin{align}
a_{\mu\nu}&=
\pfrac{h}{\phi_{J}}\pfrac{g}{\pi_{I}^{\mu}}
\ppfrac{f}{\pi_{J}^{\nu}}{\phi_{I}}-
\pfrac{h}{\phi_{J}}\pfrac{g}{\phi_{I}}
\ppfrac{f}{\pi_{J}^{\nu}}{\pi_{I}^{\mu}}\nonumber\\
&\mbox{}-%
\pfrac{h}{\pi_{J}^{\nu}}\pfrac{g}{\pi_{I}^{\mu}}
\ppfrac{f}{\phi_{J}}{\phi_{I}}+
\pfrac{h}{\pi_{J}^{\nu}}\pfrac{g}{\phi_{I}}
\ppfrac{f}{\phi_{J}}{\pi_{I}^{\mu}}\nonumber
\\
&\mbox{}+%
\pfrac{g}{\phi_{J}}\pfrac{h}{\phi_{I}}
\ppfrac{f}{\pi_{J}^{\nu}}{\pi_{I}^{\mu}}-
\pfrac{g}{\phi_{J}}\pfrac{h}{\pi_{I}^{\mu}}
\ppfrac{f}{\pi_{J}^{\nu}}{\phi_{I}}\nonumber\\
&\mbox{}-%
\pfrac{g}{\pi_{J}^{\nu}}\pfrac{h}{\phi_{I}}
\ppfrac{f}{\phi_{J}}{\pi_{I}^{\mu}}+
\pfrac{g}{\pi_{J}^{\nu}}\pfrac{h}{\pi_{I}^{\mu}}
\ppfrac{f}{\phi_{J}}{\phi_{I}}+\ldots\label{jac3}
\end{align}
Resorting and interchanging the sequence of
differentiations yields
\begin{align}
a_{\mu\nu}&=-%
\pfrac{h}{\phi_{I}}\pfrac{g}{\pi_{J}^{\nu}}
\ppfrac{f}{\pi_{I}^{\mu}}{\phi_{J}}+
\pfrac{h}{\phi_{I}}\pfrac{g}{\phi_{J}}
\ppfrac{f}{\pi_{I}^{\mu}}{\pi_{J}^{\nu}}\nonumber\\
&\mbox{}+%
\pfrac{h}{\pi_{I}^{\mu}}\pfrac{g}{\pi_{J}^{\nu}}
\ppfrac{f}{\phi_{I}}{\phi_{J}}-
\pfrac{h}{\pi_{I}^{\mu}}\pfrac{g}{\phi_{J}}
\ppfrac{f}{\phi_{I}}{\pi_{J}^{\nu}}\nonumber\\
&\mbox{}-%
\pfrac{g}{\phi_{I}}\pfrac{h}{\phi_{J}}
\ppfrac{f}{\pi_{I}^{\mu}}{\pi_{J}^{\nu}}+
\pfrac{g}{\phi_{I}}\pfrac{h}{\pi_{J}^{\nu}}
\ppfrac{f}{\pi_{I}^{\mu}}{\phi_{J}}\nonumber\\
&\mbox{}+%
\pfrac{g}{\pi_{I}^{\mu}}\pfrac{h}{\phi_{J}}
\ppfrac{f}{\phi_{I}}{\pi_{J}^{\nu}}-
\pfrac{g}{\pi_{I}^{\mu}}\pfrac{h}{\pi_{J}^{\nu}}
\ppfrac{f}{\phi_{I}}{\phi_{J}}+\ldots\label{jac4}
\end{align}
Mutually renaming the formal summation indices $I$ and $J$,
the right hand sides of Eqs.~(\ref{jac3}) and (\ref{jac4})
differ only by the sign and the interchange of the
indices $\mu$ and $\nu$.
Thereby, Eq.~(\ref{jac2}) is proved.

The explicit space-time derivative, taken with the canonical coordinates fixed, is a derivation of the component brackets.
Thus
\begin{equation}\label{th-poisson}
\left.\pfrac{}{x^{\nu}}[f,g]_{\mu}\right|_{\mathrm{expl}}
=\left[\left.\pfrac{f}{x^{\nu}}\right|_{\mathrm{expl}},g\right]_{\mu}
+\left[f,\left.\pfrac{g}{x^{\nu}}\right|_{\mathrm{expl}}\right]_{\mu}.
\end{equation}
This identity follows directly by differentiating Eq.~(\ref{pbdef}) and commuting the partial derivatives.
It implies that the bracket of two explicitly $x^{\nu}$-independent functions is also explicitly $x^{\nu}$-independent.
It does not establish a conservation law along the solutions of the field equations.

If a chosen family of field configurations supplies a local vector field representing space-time differentiation, write
\[
D_{\nu}=\left.\pfrac{}{x^{\nu}}\right|_{\mathrm{expl}}
+V_{\nu}^{A}(z,x)\pfrac{}{z^{A}},
\qquad z=(\bphi,\bpi^{0},\bpi^{1},\bpi^{2},\bpi^{3}).
\]
Such an operator is a derivation of $[\ ,\ ]_{\mu}$ only if its flow preserves the corresponding Poisson bivector $P_{\mu}$, defined by
$[f,g]_{\mu}=P_{\mu}^{AB}(\partial_{A}f)(\partial_{B}g)$.
Since its coefficients are constant in the chosen chart, the exact defect is
\begin{align*}
D_{\nu}[f,g]_{\mu}-[D_{\nu}f,g]_{\mu}-[f,D_{\nu}g]_{\mu}
&=-\left(P_{\mu}^{CB}\partial_{C}V_{\nu}^{A}
+P_{\mu}^{AC}\partial_{C}V_{\nu}^{B}\right)
(\partial_{A}f)(\partial_{B}g).
\end{align*}
When this expression vanishes for all $f,g$, two first integrals satisfying
$D_{\nu}f=D_{\nu}g=0$ as local phase-space identities have a bracket that is also a first integral.
This is the applicable form of Poisson's theorem.
For a Hamiltonian flow generated with one fixed bracket, the condition follows from its ordinary Jacobi identity, obtained by putting $\nu=\mu$ in Eq.~(\ref{jac2}).
The De~Donder--Weyl equations alone do not supply four such complete Hamiltonian flows.
In particular, the undetermined individual momentum derivatives in Eq.~(\ref{equiv-PB}) cannot be treated as phase-space constants when interchanging a total space-time derivative with a phase-space derivative.
\subsection{\label{sec:hj-eqn}Hamilton-Jacobi equation}
The De~Donder--Weyl Hamilton-Jacobi equation can be formulated directly in terms of a vector function $S^{\mu}(\bphi,x)$\cite{vonRieth84}.
The following argument assumes a regular Hamiltonian theory with independent polymomenta.
Additional parameters may label a family of such functions and are held fixed in the following derivatives.
Define a momentum graph by
\[
\pi_{I}^{\mu}=\pfrac{S^{\mu}}{\phi_{I}}.
\]
The Hamilton-Jacobi equation is
\begin{equation}\label{hj-eqn}
\HC\left(\bphi,\pfrac{\bS}{\bphi},x\right)
+\left.\pfrac{S^{\alpha}}{x^{\alpha}}\right|_{\mathrm{expl}}=0.
\end{equation}
The formal condition $\HC^{\prime}=0$ in a generating-function description motivates this equation.
However, its solutions need not define invertible canonical transformations of the full phase-space fiber.
Even if $\HC^{\prime}=0$, the canonical equations would imply only
\[
\partial_{\mu}\Phi_{I}=0,\qquad
\partial_{\mu}\Pi_{I}^{\mu}=0,
\]
not the vanishing of every derivative of $\Pi_{I}^{\mu}$.

To reconstruct a solution of the field equations, one must find fields $\phi_{I}(x)$ satisfying the embedding equations
\begin{equation}\label{hj-embedding}
\pfrac{\phi_{I}}{x^{\mu}}=V_{I\mu}(\bphi,x),
\qquad
V_{I\mu}(\bphi,x)=
\left.\pfrac{\HC}{\pi_{I}^{\mu}}\right|_{\bpi=\partial\bS/\partial\bphi}.
\end{equation}
These first-order equations must be mutually compatible.
For an integrable local family of solutions, the corresponding condition is
\begin{align*}
\left.\pfrac{V_{I\nu}}{x^{\mu}}\right|_{\mathrm{expl}}
+V_{J\mu}\pfrac{V_{I\nu}}{\phi_{J}}
=\left.\pfrac{V_{I\mu}}{x^{\nu}}\right|_{\mathrm{expl}}
+V_{J\nu}\pfrac{V_{I\mu}}{\phi_{J}}.
\end{align*}
It is an additional integrability requirement and is not guaranteed by the single scalar equation~(\ref{hj-eqn}).

Once an embedded solution exists, differentiation of Eq.~(\ref{hj-eqn}) with respect to $\phi_{I}$ gives
\[
\left.\ppfrac{S^{\mu}}{x^{\mu}}{\phi_{I}}\right|_{\mathrm{expl}}
+\pfrac{\HC}{\phi_{I}}
+\pfrac{\HC}{\pi_{J}^{\mu}}\ppfrac{S^{\mu}}{\phi_{J}}{\phi_{I}}=0,
\]
with the Hamiltonian derivatives evaluated on the momentum graph.
Together with Eq.~(\ref{hj-embedding}), this yields
\[
\pfrac{\pi_{I}^{\mu}}{x^{\mu}}
=\left.\ppfrac{S^{\mu}}{x^{\mu}}{\phi_{I}}\right|_{\mathrm{expl}}
+\ppfrac{S^{\mu}}{\phi_{I}}{\phi_{J}}\pfrac{\phi_{J}}{x^{\mu}}
=-\pfrac{\HC}{\phi_{I}}.
\]
Thus both canonical equations hold for the embedded fields and their momenta.
Along that solution,
\begin{align*}
\pfrac{S^{\alpha}}{x^{\alpha}}
&=\pfrac{S^{\alpha}}{\phi_{I}}\pfrac{\phi_{I}}{x^{\alpha}}
+\left.\pfrac{S^{\alpha}}{x^{\alpha}}\right|_{\mathrm{expl}}\\
&=\pi_{I}^{\alpha}\pfrac{\phi_{I}}{x^{\alpha}}-\HC=\LC.
\end{align*}
This is the field-theoretic counterpart of the relation $\d S/\d t=L$ for Hamilton's principal function in mechanics.
It does not replace the hypersurface data and compatibility conditions required to determine a field solution.
For a singular gauge theory, the momentum graph must also satisfy the primary constraints and the embedding equations must be projected onto the independent momentum variations.
For example, the Maxwell constraint $p^{\mu\nu}+p^{\nu\mu}=0$ requires
\[
\pfrac{S^{\nu}}{a_{\mu}}+\pfrac{S^{\mu}}{a_{\nu}}=0.
\]
The corresponding embedding relation determines the curl of $a_{\mu}$, rather than every individual derivative.
A constrained Hamilton-Jacobi construction therefore requires additional reduction beyond the regular argument above\cite{pietrzyk22}.
\section{\label{sec:examples-ham}Examples
for Hamiltonian densities in covariant field theory}
\subsection{Ginzburg-Landau Hamiltonian density}
We consider the scalar field $\phi(x,t)$ whose
\emph{Lagrangian density} $\LC$ is given by
\begin{equation}\label{ex1-ld}
\LC(\phi,\partial_{t}\phi,\partial_{x}\phi)=
\onehalf\left[{(\partial_{t}\phi)}^{2}-v^{2}
{(\partial_{x}\phi)}^{2}\right]+
\lambda{\left(\phi^{2}-1\right)}^{2}.
\end{equation}
Herein, $v\ne0$ and $\lambda$ are constant quantities.
With the sign convention of Eq.~(\ref{ex1-ld}), the potential is $V(\phi)=-\lambda(\phi^{2}-1)^{2}$.
The usual stable double-well potential therefore corresponds to $\lambda<0$.
The particular Euler-Lagrange equation for this Lagrangian
density simplifies the general form of Eq.~(\ref{elgl}) to
$$
\pfrac{}{t}\pfrac{\LC}{(\partial_{t}\phi)}+
\pfrac{}{x}\pfrac{\LC}{(\partial_{x}\phi)}-
\pfrac{\LC}{\phi}=0.
$$
The resulting field equation is
\begin{equation}\label{ex1-fg1}
\pfrac{^{2}\phi}{t^{2}}-v^{2}\pfrac{^{2}\phi}{x^{2}}-
4\lambda\,\phi\left(\phi^{2}-1\right)=0.
\end{equation}
In order to derive the equivalent Hamiltonian representation,
we first define the conjugate momentum fields from $\LC$
$$
\pi_{t}(x,t)=\pfrac{\LC}{(\partial_{t}\phi)}=\pfrac{\phi}{t},\qquad
\pi_{x}(x,t)=\pfrac{\LC}{(\partial_{x}\phi)}=-v^{2}\pfrac{\phi}{x}.
$$
The Hamiltonian density $\HC$ now follows
as the Legendre transform of the Lagrangian density $\LC$
$$
\HC(\phi,\pi_{t},\pi_{x})=\pi_{t}\pfrac{\phi}{t}+
\pi_{x}\pfrac{\phi}{x}-\LC(\phi,\partial_{t}\phi,
\partial_{x}\phi).
$$
The Ginzburg-Landau \emph{Hamiltonian density} $\HC$
is thus given by
\begin{equation}\label{ex1-hd}
\HC(\phi,\pi_{t},\pi_{x})=\onehalf\left[\pi_{t}^{2}-
\frac{1}{v^{2}}\pi_{x}^{2}\right]-\lambda{\left(\phi^{2}-1\right)}^{2}.
\end{equation}
The canonical field equations for the density $\HC$
of Eq.~(\ref{ex1-hd}) are
$$
\pfrac{\HC}{\pi_{t}}=\pfrac{\phi}{t},\qquad
\pfrac{\HC}{\pi_{x}}=\pfrac{\phi}{x},\qquad
\pfrac{\HC}{\phi}=-\pfrac{\pi_{t}}{t}-\pfrac{\pi_{x}}{x},
$$
from which we derive the following set of coupled first order
equations
$$
\pi_{t}=\pfrac{\phi}{t},\qquad
\pi_{x}=-v^{2}\pfrac{\phi}{x},\qquad
\pfrac{\pi_{t}}{t}+\pfrac{\pi_{x}}{x}-
4\lambda\,\phi\left(\phi^{2}-1\right)=0.
$$
As usual, the canonical field equations for the scalar field
$\phi(x,t)$ just reproduce the definition of the momentum
fields $\pi_{t}$ and $\pi_{x}$ from the Lagrangian density $\LC$.

By inserting $\pi_{t}$ and $\pi_{x}$ into the second field equation
the coupled set of first order field equations is converted into
a single second order equation for $\phi(x,t)$
$$
\pfrac{^{2}\phi}{t^{2}}-v^{2}\pfrac{^{2}\phi}{x^{2}}-
4\lambda\,\phi\left(\phi^{2}-1\right)=0,
$$
which coincides with Eq.~(\ref{ex1-fg1}), as expected.
\subsection{\label{sec:k-g-real}Klein-Gordon Hamiltonian density for a real scalar field}
We first consider the Klein-Gordon \emph{Lagrangian density}
$\LC_{\text{KG}}$ for a \emph{real} scalar field $\phi(\bx)$
\begin{equation}\label{KG-lag-real}
\LC_{\text{KG}}\left(\phi,\partial_{\mu}\phi\right)=
\onehalf\pfrac{\phi}{x^{\alpha}}\,\pfrac{\phi}{x_{\alpha}}-
\onehalf\omega^{2}\,\phi^{2}(\bx).
\end{equation}
The Euler-Lagrange equation~(\ref{elgl}) for
$\phi(\bx)$ follows from this Lagrangian density as
\begin{equation}\label{KG-lag-fg1}
\ppfrac{\phi}{x_{\alpha}}{x^{\alpha}}=-\omega^{2}\,\phi.
\end{equation}
We now derive the corresponding covariant Hamiltonian density
$\HC_{\text{KG}}(\phi,\pi_{\mu})$ that contains the identical
information on the dynamical system and thus yields with the
covariant canonical equations~(\ref{fgln}) the equivalent
description of the system's dynamics.
To this end, we must first define from $\LC_{\text{KG}}$
the conjugate momentum fields,
$$
\pi^{\mu}=\pfrac{\LC_{\text{KG}}}
{\left(\pfrac{\phi}{x^{\mu}}\right)}=\pfrac{\phi}{x_{\mu}},\qquad
\pi_{\mu}=\pfrac{\LC_{\text{KG}}}
{\left(\pfrac{\phi}{x_{\mu}}\right)}=\pfrac{\phi}{x^{\mu}}.
$$
The Hamiltonian density $\HC$ then follows as
the Legendre transform of the Lagrangian density
$$
\HC(\phi,\pi^{\mu})=\pfrac{\phi}{x^{\alpha}}\pi^{\alpha}-\LC(\phi,\partial_{\mu}\phi),
$$
provided that the Hessian matrix associated with the Legendre transformation
is non-singular.
For the actual case, the components of the Hessian matrix are
$$
\ppfrac{\LC_{\text{KG}}}{\left(\pfrac{\phi}{x^{\mu}}\right)}
{\left(\pfrac{\phi}{x^{\nu}}\right)}=g^{\mu\nu},
$$
which is nonsingular for the Minkowski metric.
The covariant Klein-Gordon \emph{Hamiltonian density}
$\HC_{\text{KG}}$ thus exists and is given by
\begin{equation}\label{hd-kg}
\HC_{\text{KG}}(\phi,\pi^{\mu})=\onehalf\pi_{\alpha}\pi^{\alpha}+
\onehalf\omega^{2}\,\phi^{2}(\bx).
\end{equation}
For the Hamiltonian density~(\ref{hd-kg}), the canonical field
equations~(\ref{fgln}) provide the following set of coupled
first order partial differential equations
\begin{equation}\label{kg-eqm1}
\pfrac{\phi}{x_{\mu}}=\pfrac{\HC_{\text{KG}}}{\pi_{\mu}}=\pi^{\mu},\qquad
-\pfrac{\pi^{\alpha}}{x^{\alpha}}=\pfrac{\HC_{\text{KG}}}{\phi}=\omega^{2}\,\phi.
\end{equation}
Obviously, the canonical field equation for the scalar field $\phi(\bx)$
coincides with the definition of the momentum field $\pi_{\mu}(\bx)$
from the Lagrangian density $\LC_{\text{KG}}$.
Inserting $\pi_{\mu}$ from the first canonical field equation into the
second then yields the Euler-Lagrange equation of Eq.~(\ref{KG-lag-fg1}).

We note that the Hamiltonian density $\HC_{\text{KG}}$ resembles
the Hamiltonian function $H$ of a conservative Hamiltonian
system of classical particle mechanics, which is given by
$H=T+V$ as the sum of kinetic energy $T$ and potential energy $V$.
The analogy is algebraic, since $\pi_{\alpha}\pi^{\alpha}$ is not positive definite.
The covariant Hamiltonian density is distinct from the energy density $\theta\indices{_{0}^{0}}$.
Spatial Fourier modes of the free Klein-Gordon field obey harmonic-oscillator equations.

The Heisenberg equations~(\ref{Heisenberg-eqs}) compute to
\begin{align*}
\pfrac{\phi}{x^{\mu}}&={\left[\phi,\HC_{\text{KG}}\right]}_{\mu}
={\left[\phi,\onehalf\pi^{\alpha}\pi_{\alpha}+\onehalf\omega^{2}\,\phi^{2}\right]}_{\mu}\\
&=\underbrace{{\left[\phi,\pi^{\alpha}\right]}_{\mu}}_{=\delta_{\mu}^{\alpha}}\pi_{\alpha}+
\omega^{2}\,\underbrace{{\left[\phi,\phi\right]}_{\mu}}_{=0}\,\phi\\
&=\pi_{\mu}
\end{align*}
and
\begin{align*}
\delta_{\mu}^{\nu}\pfrac{\pi^{\alpha}}{x^{\alpha}}&={\left[\pi^{\nu},\HC_{\text{KG}}\right]}_{\mu}
={\left[\pi^{\nu},\onehalf\pi^{\alpha}\pi_{\alpha}+\onehalf\omega^{2}\,\phi^{2}\right]}_{\mu}\\
&=\underbrace{{\left[\pi^{\nu},\pi^{\alpha}\right]}_{\mu}}_{=0}\pi_{\alpha}+
\omega^{2}\,\underbrace{{\left[\pi^{\nu},\phi\right]}_{\mu}}_{=-\delta_{\mu}^{\nu}}\,\phi\\
&=-\delta_{\mu}^{\nu}\omega^{2}\,\phi\\
\Rightarrow\quad\pfrac{\pi^{\alpha}}{x^{\alpha}}&=-\omega^{2}\,\phi,\qquad
\ppfrac{\phi}{x_{\alpha}}{x^{\alpha}}+\omega^{2}\phi=0.
\end{align*}
As expected, the Klein-Gordon field equation emerges.
\subsection{\label{sec:k-g-complex}
Klein-Gordon Hamiltonian density for a set of $N$ complex scalar fields}
We first consider the Klein-Gordon \emph{Lagrangian density}
$\LC_{\text{KG}}$ for a set of \emph{complex} scalar fields $\phi_{I}$
(see, for instance, Ref.~\cite{greiner85})
\begin{equation}\label{ex3-ld}
\LC_{\text{KG}}\left(\bphi,\overline{\bphi},\partial^{\mu}\bphi,
\partial_{\mu}\overline{\bphi}\,\right)=
\pfrac{\overline{\phi}_{I}}{x^{\alpha}}\,\pfrac{\phi_{I}}{x_{\alpha}}-
\overline{\phi}_{I}\,\omega_{IJ}^{2}\,\phi_{J}.
\end{equation}
Herein $\overline{\phi}_{I}$ denotes the complex conjugate field of $\phi_{I}$
and $\omega_{IJ}^{2}$ the (diagonal) mass matrix.
The quantities $\phi_{I}$ and $\overline{\phi}_{I}$ are varied independently and the complex-conjugacy condition is imposed on physical configurations.
The Euler-Lagrange equations~(\ref{elgl}) for
$\phi_{I}$ and $\overline{\phi}_{I}$ follow from this Lagrangian density as
\begin{equation}\label{ex3-fg1}
\ppfrac{\overline{\phi}_{I}}{x_{\alpha}}{x^{\alpha}}=
-\overline{\phi}_{J}\,\omega_{JI}^{2},\qquad
\ppfrac{\phi_{I}}{x_{\alpha}}{x^{\alpha}}=-\omega_{IJ}^{2}\,\phi_{J}.
\end{equation}
As a prerequisite for deriving the corresponding
Hamiltonian density $\HC_{\text{KG}}$ we must first
define from $\LC_{\text{KG}}$ the conjugate momentum fields,
$$
\pi_{I}^{\mu}=\pfrac{\LC_{\text{KG}}}
{\left(\pfrac{\overline{\phi}_{I}}{x^{\mu}}\,\right)}=
\pfrac{\phi_{I}}{x_{\mu}},\qquad
\overline{\pi}_{I\mu}=\pfrac{\LC_{\text{KG}}}
{\left(\pfrac{\phi_{I}}{x_{\mu}}\right)}=
\pfrac{\overline{\phi}_{I}}{x^{\mu}}.
$$
Thereby, the canonical momenta $\pi_{I}^{\mu}$ are defined as the
\emph{dual quantities} of the partial derivatives of the fields $\overline{\phi}_{I}$.

The Hamiltonian density $\HC$ then follows as
the Legendre transform of the Lagrangian density
$$
\HC(\bpi^{\mu},\overline{\bpi}^{\mu},\bphi,\overline{\bphi})=
\pfrac{\overline{\phi}_{I}}{x^{\alpha}}\pi_{I}^{\alpha}+
\overline{\pi}_{I}^{\alpha}\pfrac{\phi_{I}}{x^{\alpha}}-\LC.
$$
The Klein-Gordon \emph{Hamiltonian density}
$\HC_{\text{KG}}$ is thus given by
\begin{equation}\label{hd-kg2}
\HC_{\text{KG}}(\bpi_{\mu},\overline{\bpi}^{\mu},\bphi,\overline{\bphi})=
\overline{\pi}_{I\alpha}\pi_{I}^{\alpha}+\overline{\phi}_{I}\,\omega_{IJ}^{2}\,\phi_{J}.
\end{equation}
For the Hamiltonian density~(\ref{hd-kg2}), the canonical field
equations~(\ref{fgln}) provide the following set of coupled
first order partial differential equations
\begin{align}
\pfrac{\HC_{\text{KG}}}{\overline{\pi}_{I}^{\mu}}&=\pfrac{\phi_{I}}{x^{\mu}}=
\pi_{I\mu},&
\pfrac{\HC_{\text{KG}}}{\pi_{I\mu}}&=
\pfrac{\overline{\phi}_{I}}{x_{\mu}}=\overline{\pi}_{I}^{\mu}\nonumber\\
\pfrac{\HC_{\text{KG}}}{\phi_{I}}&=-\pfrac{\overline{\pi}_{I}^{\alpha}}{x^{\alpha}}=
\overline{\phi}_{J}\omega_{JI}^{2},&
\pfrac{\HC_{\text{KG}}}{\overline{\phi}_{I}}&=-\pfrac{\pi_{I}^{\alpha}}{x^{\alpha}}=
\omega_{IJ}^{2}\phi_{J}.
\label{caneq-kg-compl}
\end{align}
As a result of the Legendre transformation, the canonical field equations
for the scalar fields $\phi_{I}$ and $\overline{\phi}_{I}$ coincide with
the definitions of the momentum fields $\pi_{I\mu}$ and $\overline{\pi}_{I}^{\mu}$
from the Lagrangian density $\LC_{\text{KG}}$.
Eliminating the $\pi_{I\mu}$, $\overline{\pi}_{I}^{\mu}$ from the
canonical field equations then yields the Euler-Lagrange
equations of Eq.~(\ref{ex3-fg1}).

For \emph{complex} fields, the energy-momentum tensor in the
Lagrangian formalism is defined analogously to the \emph{real}
case of Eq.~(\ref{e-i-def})
$$
\theta\indices{_{\mu}^{\nu}}=
\pfrac{\LC}{\left(\pfrac{\phi_{I}}{x^{\nu}}\,\right)}
\pfrac{\phi_{I}}{x^{\mu}}+\pfrac{\overline{\phi}_{I}}{x^{\mu}}
\pfrac{\LC}{\left(\pfrac{\overline{\phi}_{I}}{x^{\nu}}\right)}-\delta_{\mu}^{\nu}\,\LC.
$$
Expressed by means of the complex Hamiltonian density $\HC$,
this means
$$
\theta\indices{_{\mu}^{\nu}}=\overline{\pi}_{I}^{\nu}\pfrac{\phi_{I}}{x^{\mu}}+
\pfrac{\overline{\phi}_{I}}{x^{\mu}}\,\pi_{I}^{\nu}-
\delta_{\mu}^{\nu}\,\overline{\pi}_{I}^{\alpha}\,\pfrac{\phi_{I}}{x^{\alpha}}-
\delta_{\mu}^{\nu}\,\pfrac{\overline{\phi}_{I}}{x^{\alpha}}\,\pi_{I}^{\alpha}+
\delta_{\mu}^{\nu}\,\HC.
$$
For the Klein-Gordon Hamiltonian density $\HC_{\text{KG}}$
from Eq.~(\ref{hd-kg2}), we thus get the particular canonical
energy-momentum tensor $\theta\indices{_{\mu}^{\nu}}$
\begin{equation}\label{e-i-def-komplex}
\theta\indices{_{\mu}^{\nu}}=
\overline{\pi}_{I}^{\nu}\pi_{I\mu}+\overline{\pi}_{I\mu}\pi_{I}^{\nu}-
\delta_{\mu}^{\nu}\,\overline{\pi}_{I}^{\alpha}\pi_{I\alpha}+
\delta_{\mu}^{\nu}\,\overline{\phi}_{I}\,\omega^{2}_{IJ}\,\phi_{J}.
\end{equation}
For complex fields, the component bracket includes both canonical pairs,
$$
[f,g]_{\mu}=\pfrac{f}{\phi_I}\pfrac{g}{\overline{\pi}_I^{\mu}}
-\pfrac{f}{\overline{\pi}_I^{\mu}}\pfrac{g}{\phi_I}
+\pfrac{f}{\overline{\phi}_I}\pfrac{g}{\pi_I^{\mu}}
-\pfrac{f}{\pi_I^{\mu}}\pfrac{g}{\overline{\phi}_I}.
$$
The equations of motion for the sets of fields $\phi_{I},\bpi_{I}$ can be
derived as well from the covariant Heisenberg equations~(\ref{Heisenberg-eqs})
$$
\pfrac{\phi_{I}}{x^{\mu}}=\big[\phi_{I},\HC\big]_{\mu},\quad
\delta_{\mu}^{\nu}\,\pfrac{\pi_{I}^{\alpha}}{x^{\alpha}}=
\big[\pi_{I}^{\nu},\HC\big]_{\mu},\quad
\pfrac{\overline{\phi}_{I}}{x^{\mu}}=\big[\,\overline{\phi}_{I},\HC\big]_{\mu},\quad
\delta_{\mu}^{\nu}\,\pfrac{\overline{\pi}_{I}^{\alpha}}{x^{\alpha}}=
\big[\overline{\pi}_{I}^{\nu},\HC\big]_{\mu},
$$
in conjunction with the \emph{fundamental Poisson bracket rules}
from Eqs.~(\ref{fundpk1}).
Here, the fields $(\phi_{I},\overline{\bpi}_{I})$ and,
correspondingly, $(\,\overline{\phi}_{I},\bpi_{I})$
represent the pairs of canonical conjugate quantities, while $\phi_{I}$
and $\overline{\phi}_{I}$ are to be treated as \emph{independent fields}.
This applies as well for the conjugate fields, $\bpi_{I}$ and
$\overline{\bpi}_{I}$.
The equations for $\phi_{I}$ and $\overline{\phi}_{I}$ follow as
\begin{align}
\pfrac{\phi_{I}}{x^{\mu}}&=\left[\phi_{I},\HC_{\text{KG}}\right]_{\mu}=
\left[\phi_{I},\overline{\pi}_{J\alpha}\pi_{J}^{\alpha}+
\overline{\phi}_{J}\,\omega_{JK}^{2}\,\phi_{K}\right]_{\mu}\nonumber\\
&=\overline{\pi}_{J\alpha}\underbrace{\left[\phi_{I},\pi_{J}^{\alpha}
\right]_{\mu}}_{=0}+
\underbrace{\left[\phi_{I},\overline{\pi}_{J}^{\alpha}
\right]_{\mu}}_{=\delta_{\mu}^{\alpha}\delta_{IJ}}\pi_{J\alpha}+
\overline{\phi}_{J}\,\omega_{JK}^{2}
\underbrace{\left[\phi_{I},\phi_{K}\right]_{\mu}}_{=0}+
\underbrace{\left[\phi_{I},
\overline{\phi}_{J}\right]_{\mu}}_{=0}\omega_{JK}^{2}\,\phi_{K}\nonumber\\
&=\pi_{I\mu}.
\label{kg-eqm3-qu}
\end{align}
\begin{align}
\pfrac{{\overline{\phi}}_{I}}{x^{\mu}}&=
\left[\,{\overline{\phi}}_{I},{\HC}_{\text{KG}}\right]_{\mu}=
\left[{\overline{\phi}}_{I},{\overline{\pi}}_{J\alpha}{\pi}_{J}^{\alpha}+
{\overline{\phi}}_{J}\,\omega_{JK}^{2}\,{\phi}_{K}\right]_{\mu}\nonumber\\
&={\overline{\pi}}_{J\alpha}
\underbrace{\left[{\overline{\phi}}_{I},{\pi}_{J}^{\alpha}
\right]_{\mu}}_{=\delta_{\mu}^{\alpha}\delta_{IJ}}+
\underbrace{\left[{\overline{\phi}}_{I},{\overline{\pi}}_{J}^{\alpha}
\right]_{\mu}}_{=0}{\pi}_{J\alpha}+
{\overline{\phi}}_{J}\,\omega_{JK}^{2}\,
\underbrace{\left[{\overline{\phi}}_{I},{\phi}_{K}\right]_{\mu}}_{=0}+
\underbrace{\left[{\overline{\phi}}_{I},
{\overline{\phi}}_{J}\right]_{\mu}}_{=0}\omega_{JK}^{2}\,{\phi}_{K}\nonumber\\
&=\overline{\pi}_{I\mu}.
\label{kg-eqm4-qu}
\end{align}
Similarly, the equations for the divergences of
$\bpi_{I}$ and $\overline{\bpi}_{I}$ are
\begin{align}
\delta_{\mu}^{\nu}\pfrac{{\pi}_{I}^{\alpha}}{x^{\alpha}}&=
\left[{\pi}_{I}^{\nu},{\HC}_{\text{KG}}\right]_{\mu}=
\left[{\pi}_{I}^{\nu},{\overline{\pi}}_{J}^{\alpha}{\pi}_{J\alpha}+
{\overline{\phi}}_{J}\,\omega_{JK}^{2}\,{\phi}_{K}\right]_{\mu}\nonumber\\
&={\overline{\pi}}_{J\alpha}\underbrace{\left[{\pi}_{I}^{\nu},
{\pi}_{J}^{\alpha}\right]_{\mu}}_{=0}+
\underbrace{\left[{\pi}_{I}^{\nu},{\overline{\pi}}_{J}^{\alpha}
\right]_{\mu}}_{=0}{\pi}_{J\alpha}+{\overline{\phi}}_{J}\,\omega_{JK}^{2}\,
\underbrace{\left[{\pi}_{I}^{\nu},{\phi}_{K}\right]_{\mu}}_{=0}+
\underbrace{\left[{\pi}_{I}^{\nu},
{\overline{\phi}}_{J}\right]_{\mu}}_{=-\delta_{\mu}^{\nu}\delta_{IJ}}
\omega^{2}_{JK}\,{\phi}_{K}\nonumber\\
&=-\delta_{\mu}^{\nu}\,\omega_{IK}^{2}\,{\phi}_{K}
\label{kg-eqm5-qu}
\end{align}
\begin{align}
\delta_{\mu}^{\nu}\pfrac{{\overline{\pi}}_{I}^{\alpha}}{x^{\alpha}}&=
\left[{\overline{\pi}}_{I}^{\nu},{\HC}_{\text{KG}}\right]_{\mu}=
\left[{\overline{\pi}}_{I}^{\nu},{\overline{\pi}}_{J}^{\alpha}
{\pi}_{J\alpha}+
{\overline{\phi}}_{J}\,\omega_{JK}^{2}{\phi}_{K}\right]_{\mu}\nonumber\\
&={\overline{\pi}}_{J\alpha}\underbrace{\left[{\overline{\pi}}_{I}^{\nu},
{\pi}_{J}^{\alpha}\right]_{\mu}}_{=0}+
\underbrace{\left[{\overline{\pi}}_{I}^{\nu},{\overline{\pi}}_{J}^{\alpha}
\right]_{\mu}}_{=0}{\pi}_{J\alpha}+{\overline{\phi}}_{J}\,\omega_{JK}^{2}
\underbrace{\left[{\overline{\pi}}_{I}^{\nu},{\phi}_{K}
\right]_{\mu}}_{=-\delta_{\mu}^{\nu}\delta_{IK}}+
\underbrace{\left[{\overline{\pi}}_{I}^{\nu},
{\overline{\phi}}_{J}\right]_{\mu}}_{=0}
\omega_{JK}^{2}{\phi}_{K}\nonumber\\
&=-\delta_{\mu}^{\nu}\,\overline{\phi}_{J}\,\omega_{JI}^{2}.
\label{kg-eqm6-qu}
\end{align}
Obviously, the obtained equations coincide with the canonical equations
from Eqs.~(\ref{caneq-kg-compl}).
\subsection{\label{sec:maxwell}Maxwell's equations
as canonical field equations}
The Lagrangian density $\LC_{\text{M}}$ of the
electromagnetic field is given by
\begin{equation}\label{ld-maxwell}
\LC_{\text{M}}(\ba,\bpartial\ba,\bx)=-\quarter f_{\alpha\beta}f^{\alpha\beta}-
\frac{4\pi}{c}j^{\alpha}(\bx)\,a_{\alpha},\quad
f_{\mu\nu}=\pfrac{a_{\nu}}{x^{\mu}}-\pfrac{a_{\mu}}{x^{\nu}}.
\end{equation}
Herein, the four components $a_{\mu}$ of the $4$-vector potential
$\ba$ now take the place of the scalar fields
$\phi_{I}\equiv a_{\mu}$ in the notation used so far.
The Lagrangian density (\ref{ld-maxwell}) thus entails
a set of \emph{four} Euler-Lagrange equations, i.e.,
an equation for each component $a_{\mu}$.
The source vector $\bj=(c\rho,j_{x},j_{y},j_{z})$ denotes
the $4$-vector of electric currents combining the usual
current density vector $(j_{x},j_{y},j_{z})$ of
configuration space with the charge density $\rho$.
In Cartesian coordinates on Minkowski space, the
Euler-Lagrange equations~(\ref{elgl}) take on the form
\begin{equation}\label{elgl1}
\pfrac{}{x^{\alpha}}\pfrac{\LC_{\text{M}}}{(\partial_{\alpha}a_{\mu})}-
\pfrac{\LC_{\text{M}}}{a_{\mu}}=0,\qquad\mu=0,\ldots,3.
\end{equation}
With $\LC_{\text{M}}$ from Eq.~(\ref{ld-maxwell}),
we obtain directly
\begin{equation}\label{el-maxwell}
\pfrac{f^{\mu\alpha}}{x^{\alpha}}+\frac{4\pi}{c}j^{\mu}=0.
\end{equation}
In Minkowski space,
this is the tensor form of the inhomogeneous Maxwell equation.
In order to formulate the equivalent Hamiltonian description,
we first define, according to Eq.~(\ref{p-def}),
the canonical momentum components $p^{\mu\nu}$ as the
conjugate objects of the derivatives of the $4$-vector
potential $\ba$
\begin{equation}\label{p-def1}
p^{\mu\nu}=\pfrac{\LC_{\text{M}}}{(\partial_{\nu}a_{\mu})}
\equiv\pfrac{\LC_{\text{M}}}{a_{\mu,\nu}}
\end{equation}
With the particular Lagrangian density~(\ref{ld-maxwell}),
Eq.~(\ref{p-def1}) means, in detail,
\begin{align*}
p^{\mu\nu}&=-\quarter\left(
\pfrac{f_{\alpha\beta}}{\left(\partial_{\nu}a_{\mu}\right)}f^{\alpha\beta}+
\pfrac{f^{\alpha\beta}}{\left(\partial_{\nu}a_{\mu}\right)}f_{\alpha\beta}
\right)\\
&=-\quarter\left(
\pfrac{f_{\alpha\beta}}{\left(\partial_{\nu}a_{\mu}\right)}f^{\alpha\beta}+
\pfrac{f_{\alpha\beta}}{\left(\partial_{\nu}a_{\mu}\right)}f^{\alpha\beta}
\right)\\
&=-\onehalf f^{\alpha\beta}
\pfrac{f_{\alpha\beta}}{\left(\partial_{\nu}a_{\mu}\right)}\\
&=-\onehalf f^{\alpha\beta}\pfrac{}{\left(\pfrac{a_{\mu}}
{x^{\nu}}\right)}
\left(\pfrac{a_{\beta}}{x^{\alpha}}-\pfrac{a_{\alpha}}{x^{\beta}}\right)\\
&=-\onehalf f^{\alpha\beta}\left(\delta_{\beta}^{\mu}\delta_{\alpha}^{\nu}-
\delta_{\alpha}^{\mu}\delta_{\beta}^{\nu}\right)=
-\onehalf\left(f^{\nu\mu}-f^{\mu\nu}\right)\\
&=f^{\mu\nu}.
\end{align*}
The tensor $p^{\mu\nu}$ thus matches exactly the
electromagnetic field tensor $f^{\mu\nu}$ from Eq.~(\ref{ld-maxwell})
and hence inherits the skew-symmetry of $f^{\mu\nu}$
because of the particular dependence of $\LC_{\mathrm{M}}$
on the $a_{\mu,\nu}\equiv\partial a_{\mu}/\partial x^{\nu}$.

The Legendre map is singular on the full space of sixteen derivatives $\partial_{\nu}a_{\mu}$.
Its image satisfies the primary constraints $p^{(\mu\nu)}=0$.
We therefore work with the first-order action restricted to antisymmetric momenta,
$$
S_{\mathrm M}=\int\left(p^{\alpha\beta}\partial_{\beta}a_{\alpha}-\HC_{\mathrm M}\right)\,\d^{4}x,
\qquad p^{\alpha\beta}=-p^{\beta\alpha}.
$$
Variations $\delta p^{\alpha\beta}$ are antisymmetric as well.
Whenever a derivative with respect to $p^{\mu\nu}$ is written below, it denotes its antisymmetric representative in
$\delta\HC=(\partial\HC/\partial p^{\mu\nu})\delta p^{\mu\nu}$, with the sum taken over both ordered indices.
Accordingly, the first canonical equation determines only
$a_{[\mu,\nu]}=\tfrac12(\partial_{\nu}a_{\mu}-\partial_{\mu}a_{\nu})$.
This is a constrained Legendre transformation and requires the separate treatment described here\cite{pietrzyk22}.
An equivalent formulation on an unrestricted momentum space introduces symmetric multipliers for $p^{(\mu\nu)}=0$.

As the Lagrangian density~(\ref{ld-maxwell}) now describes the dynamics
of a \emph{vector field}, $a_{\mu}$, rather than a set of scalar fields
$\phi_{I}$, the canonical momenta $p^{\mu\nu}$ now constitute
a second rank \emph{tensor} rather than a vector, $\pi_{I}^{\nu}$.
The Legendre transformation corresponding to Eq.~(\ref{H-def})
then comprises the product $p^{\alpha\beta}\partial_{\beta}a_{\alpha}$.
The skew-symmetry of the momentum tensor $p^{\mu\nu}$ picks
out the skew-symmetric part of $\partial_{\nu}a_{\mu}$ as the symmetric
part of $\partial_{\nu}a_{\mu}$ vanishes identically calculating
the product $p^{\alpha\beta}\partial_{\beta}a_{\alpha}$
$$
p^{\alpha\beta}\pfrac{a_{\alpha}}{x^{\beta}}=
\onehalf p^{\alpha\beta}\left(\pfrac{a_{\alpha}}{x^{\beta}}-
\pfrac{a_{\beta}}{x^{\alpha}}\right)+\onehalf
\underbrace{p^{\alpha\beta}\left(\pfrac{a_{\alpha}}{x^{\beta}}+
\pfrac{a_{\beta}}{x^{\alpha}}\right)}_{\equiv0}.
$$
For a skew-symmetric momentum tensor $p^{\mu\nu}$, we thus obtain
the Hamiltonian density $\HC_{\text{M}}$ as the Legendre-transformed
Lagrangian density $\LC_{\text{M}}$
$$
\HC_{\text{M}}(\ba,\bp,\bx)=\onehalf p^{\alpha\beta}\left(
\pfrac{a_{\alpha}}{x^{\beta}}-\pfrac{a_{\beta}}{x^{\alpha}}\right)-
\LC_{\text{M}}(\ba,\bpartial\ba,\bx).
$$
From this Legendre transformation prescription and the
Euler-Lagrange equations~(\ref{elgl1}), the canonical
field equations are immediately obtained as
\begin{align*}
\pfrac{\HC_{\text{M}}}{p^{\mu\nu}}&=\frac{1}{2}\left(
\pfrac{a_{\mu}}{x^{\nu}}-\pfrac{a_{\nu}}{x^{\mu}}\right)\\
\pfrac{\HC_{\text{M}}}{a_{\mu}}&=-\pfrac{\LC_{\text{M}}}{a_{\mu}}=
-\pfrac{}{x^{\alpha}}\pfrac{\LC_{\text{M}}}{(\partial_{\alpha}a_{\mu})}=
-\pfrac{p^{\mu\alpha}}{x^{\alpha}}\\
\left.\pfrac{\HC_{\text{M}}}{x^{\nu}}\right|_{\mathrm{expl}}
&=-\left.\pfrac{\LC_{\text{M}}}{x^{\nu}}\right|_{\mathrm{expl}}.
\end{align*}
The Hamiltonian density for the Lagrangian
density~(\ref{ld-maxwell}) follows as
\begin{align}
\HC_{\text{M}}(\ba,\bp,\bx)&=-\onehalf p^{\alpha\beta}p_{\alpha\beta}+
\quarter p^{\alpha\beta}p_{\alpha\beta}+\frac{4\pi}{c}j^{\alpha}(\bx)\,
a_{\alpha}\nonumber\\
&=-\quarter p^{\alpha\beta}p_{\alpha\beta}+
\frac{4\pi}{c}j^{\alpha}(\bx)\,a_{\alpha},\qquad
p^{\alpha\beta}=-p^{\beta\alpha}.
\label{hd-maxwell}
\end{align}
The first canonical field equation follows from the derivative of the
Hamiltonian density~(\ref{hd-maxwell}) with respect to $p^{\mu\nu}$
$$
\frac12\left(\pfrac{a_{\mu}}{x^{\nu}}-\pfrac{a_{\nu}}{x^{\mu}}\right)=
\pfrac{\HC_{\text{M}}}{p^{\mu\nu}}=-\onehalf p_{\mu\nu}=\onehalf p_{\nu\mu},
$$
hence
\begin{equation}\label{fg1-maxwell}
p_{\mu\nu}=\pfrac{a_{\nu}}{x^{\mu}}-\pfrac{a_{\mu}}{x^{\nu}},
\end{equation}
which reproduces the definition of $p_{\mu\nu}$ and $p^{\mu\nu}$
from Eq.~(\ref{p-def1}), as a consequence of the Legendre transformation.

The second canonical field equation
is obtained calculating the derivative of the Hamiltonian
density~(\ref{hd-maxwell}) with respect to $a_{\mu}$
$$
-\pfrac{p^{\mu\alpha}}{x^{\alpha}}=
\pfrac{\HC_{\text{M}}}{a_{\mu}}=\frac{4\pi}{c}j^{\mu}.
$$
Inserting the first canonical equation, the second order field
equation for the $a_{\mu}$ is thus obtained for the Maxwell
Hamiltonian density~(\ref{hd-maxwell}) as
\begin{equation}\label{fg2-maxwell}
\pfrac{f^{\mu\alpha}}{x^{\alpha}}+\frac{4\pi}{c}j^{\mu}=0,
\end{equation}
which agrees, as expected, with the corresponding
Euler-Lagrange equation~(\ref{el-maxwell}).
Its divergence requires $\partial_{\mu}j^{\mu}=0$, which is the consistency condition on a prescribed external current.
The identity $f_{\mu\nu}=\partial_{\mu}a_{\nu}-\partial_{\nu}a_{\mu}$ supplies the homogeneous Maxwell equations.
\subsection{The Proca Hamiltonian density}
In relativistic quantum theory, the dynamics of
particles of spin $1$ and mass $m$ is derived from the
Proca Lagrangian density $\LC_{\text{P}}$,
\begin{equation}\label{ld-proca}
\LC_{\text{P}}=-\quarter f^{\alpha\beta}f_{\alpha\beta}+
\onehalf\omega^{2}a^{\alpha}a_{\alpha},\qquad
f_{\mu\nu}=\pfrac{a_{\nu}}{x^{\mu}}-\pfrac{a_{\mu}}{x^{\nu}},\qquad
\omega=\frac{mc}{\hbar}.
\end{equation}
We observe that the kinetic term of $\LC_{\text{P}}$ agrees
with that of the Lagrangian density $\LC_{\text{M}}$ of the
electromagnetic field of Eq.~(\ref{ld-maxwell}).
Therefore, the field equations emerging from the
Euler-Lagrange equations~(\ref{elgl1}) are similar to
those of Eq.~(\ref{el-maxwell})
\begin{equation}\label{el-proca}
\pfrac{f^{\mu\alpha}}{x^{\alpha}}-\omega^{2}a^{\mu}=0.
\end{equation}
The transition to the corresponding Hamilton description is
performed by defining on the basis of the actual Lagrangian
$\LC_{\text{P}}$ the canonical momentum field tensors
$p^{\mu\nu}$ as the conjugate objects of the derivatives
of the $4$-vector $\ba$
$$
p^{\mu\nu}=\pfrac{\LC_{\text{P}}}{\left(\partial_{\nu}a_{\mu}\right)}
\equiv\pfrac{\LC_{\text{P}}}{a_{\mu,\nu}}.
$$
Similar to the preceding section, we find
$$
p^{\mu\nu}=f^{\mu\nu},\qquad p_{\mu\nu}=f_{\mu\nu}
$$
because of the particular dependence of $\LC_{\text{P}}$
on the derivatives of the $a^{\mu}$.
With $p^{\alpha\beta}$ being skew-symmetric in $\alpha,\beta$,
the product $p^{\alpha\beta}\,a_{\alpha,\beta}$ picks
out the skew-symmetric part of the partial derivative
$\partial a_{\alpha}/\partial x^{\beta}$ as the product
with the symmetric part vanishes identically.
Denoting the skew-symmetric part by $a_{[\alpha,\beta]}$,
the Legendre transformation prescription
\begin{align*}
\HC_{\text{P}}&=p^{\alpha\beta}\,a_{\alpha,\beta}-\LC_{\text{P}}\\
&=p^{\alpha\beta}\,a_{[\alpha,\beta]}-\LC_{\text{P}}\\
&=\onehalf p^{\alpha\beta}\left(\pfrac{a_{\alpha}}{x^{\beta}}-
\pfrac{a_{\beta}}{x^{\alpha}}\right)-\LC_{\text{P}},
\end{align*}
leads to the Proca Hamiltonian density by
following the path of Eq.~(\ref{hd-maxwell})
\begin{equation}\label{hd-proca}
\HC_{\text{P}}=-\quarter p^{\alpha\beta}p_{\alpha\beta}-
\onehalf\omega^{2}a^{\alpha}a_{\alpha},\qquad p^{\alpha\beta}=-p^{\beta\alpha}.
\end{equation}
The canonical field equations emerge as
\begin{align*}
\frac12\left(\pfrac{a_{\mu}}{x^{\nu}}-\pfrac{a_{\nu}}{x^{\mu}}\right)=
\pfrac{\HC_{\text{P}}}{p^{\mu\nu}}&=-\onehalf p_{\mu\nu}=\onehalf p_{\nu\mu}\\
-\pfrac{p^{\mu\alpha}}{x^{\alpha}}=\pfrac{\HC_{\text{P}}}{a_{\mu}}&=
-\omega^{2}a^{\mu}.
\end{align*}
By means of eliminating $p^{\mu\nu}$, this coupled set of
first order equations can be converted into second
order equations for the vector field $\ba(\bx)$,
$$
\pfrac{}{x_{\alpha}}\left(\pfrac{a_{\mu}}{x^{\alpha}}-
\pfrac{a_{\alpha}}{x^{\mu}}\right)+\omega^{2}a_{\mu}=0.
$$
As expected, this equation coincides with the
Euler-Lagrange equation~(\ref{el-proca}).

The antisymmetry constraint must also be respected when writing bracket equations.
The component Poisson brackets of Sect.~\ref{sec:p-k}, applied to sixteen independent $p^{\mu\nu}$, do not descend directly to this constrained space.
For example, the unreduced fundamental bracket gives
\begin{equation}\label{fpbr-proca}
\big[a_{\xi},p^{\eta\nu}+p^{\nu\eta}\big]_{\mu}
=\delta_{\mu}^{\nu}\delta_{\xi}^{\eta}
 +\delta_{\mu}^{\eta}\delta_{\xi}^{\nu},
\end{equation}
which need not vanish although $p^{\eta\nu}+p^{\nu\eta}=0$ on the constraint surface.
One must use a suitable constrained bracket construction or the restricted variational principle used above.
The latter gives directly
\begin{equation}\label{proca-eqm1-qu}
p_{\mu\nu}=\partial_{\mu}a_{\nu}-\partial_{\nu}a_{\mu}
\end{equation}
and
\begin{equation}\label{proca-eqm2-qu}
\partial_{\alpha}p^{\mu\alpha}=\omega^{2}a^{\mu}.
\end{equation}
For $\omega\ne0$, taking the divergence yields $\partial_{\mu}a^{\mu}=0$.
Thus the Proca field satisfies
$$
\left(\partial_{\alpha}\partial^{\alpha}+\omega^{2}\right)a_{\mu}=0,
\qquad \partial_{\mu}a^{\mu}=0.
$$
Here the Lorenz condition follows from the equations of motion and is not a gauge choice for the massive theory.
\subsection{\label{sec:hd-kgm}Canonical field equations
of a coupled Klein-Gordon-Maxwell system}
The Lagrangian density $\LC_{\text{KGM}}$ of a complex
Klein-Gordon field $\phi$ that couples \emph{minimally} to
an electromagnetic $4$-vector potential $\ba$ is given by
\begin{equation}\label{ld-kg-max}
\LC_{\text{KGM}}=
\left(\pfrac{\overline{\phi}}{x_{\alpha}}+\rmi q\,a^{\alpha}\overline{\phi}\right)
\left(\pfrac{\phi}{x^{\alpha}}-\rmi q\,a_{\alpha}\phi\right)-
\omega^{2}\overline{\phi}\phi-\quarter f^{\alpha\beta}f_{\alpha\beta}.
\end{equation}
We use the convention $D_{\mu}\phi=(\partial_{\mu}-\rmi q a_{\mu})\phi$ and
$\overline{D_{\mu}\phi}=\partial_{\mu}\overline{\phi}+\rmi q a_{\mu}\overline{\phi}$ throughout the gauge examples.
The components $f_{\mu\nu}$ of the electromagnetic field
tensor are defined in Eq.~(\ref{ld-maxwell}).
The conjugate fields of $\phi$ and $\ba$ are
obtained from the Lagrangian $\LC_{\text{KGM}}$ via
\begin{align*}
\overline{\pi}^{\nu}&=\pfrac{\LC_{\text{KGM}}}{\left(
\partial_{\nu}\phi\right)}\;\;=
\pfrac{\overline{\phi}}{x_{\nu}}+\rmi q\,a^{\nu}\overline{\phi}\\
\pi_{\nu}&=\pfrac{\LC_{\text{KGM}}}{\left(
\partial^{\nu}\overline{\phi}\right)}\;\,=
\pfrac{\phi}{x^{\nu}}-\rmi q\,a_{\nu}\phi\\
p^{\mu\nu}&=\pfrac{\LC_{\text{KGM}}}
{\left(\partial_{\nu}a_{\mu}\right)}=f^{\mu\nu}
\equiv\pfrac{a^{\nu}}{x_{\mu}}-\pfrac{a^{\mu}}{x_{\nu}}
\end{align*}
The corresponding Hamiltonian density $\HC_{\text{KGM}}$
now emerges from $\LC_{\text{KGM}}$ by the constrained Legendre
transformation prescription
$$
\HC_{\text{KGM}}=p^{\alpha\beta}\pfrac{a_{\alpha}}{x^{\beta}}+
\overline{\pi}^{\alpha}\pfrac{\phi}{x^{\alpha}}+
\pfrac{\overline{\phi}}{x_{\alpha}}\pi_{\alpha}-\LC_{\text{KGM}}.
$$
The Hamiltonian $\HC_{\text{KGM}}$ is found by eliminating the derivatives of $\phi$, $\overline{\phi}$ and the antisymmetric derivative of $\ba$ in favor of the respective conjugate fields,
\begin{equation}\label{hd-kg-max}
\HC_{\text{KGM}}=\overline{\pi}_{\alpha}\pi^{\alpha}+\rmi q\,a_{\alpha}\left(
\overline{\pi}^{\alpha}\phi-\overline{\phi}\pi^{\alpha}\right)+
\omega^{2}\overline{\phi}\phi-\quarter p^{\alpha\beta}p_{\alpha\beta},
\quad p^{\alpha\beta}=-p^{\beta\alpha}.
\end{equation}
Corresponding to Sect.~\ref{sec:maxwell}, the derivative
of the Hamiltonian density $\HC_{\text{KGM}}$ with respect
to the $p_{\mu\nu}$ yields the canonical equation
$$
\frac12\left(\pfrac{a_{\mu}}{x^{\nu}}-\pfrac{a_{\nu}}{x^{\mu}}\right)
=\pfrac{\HC_{\text{KGM}}}{p^{\mu\nu}}=
-\onehalf p_{\mu\nu}=\onehalf p_{\nu\mu},
$$
hence
$$
p_{\nu\mu}=\pfrac{a_{\mu}}{x^{\nu}}-\pfrac{a_{\nu}}{x^{\mu}}.
$$
From the derivatives of $\HC_{\text{KGM}}$ with respect to
the $\pi^{\mu}$ and $\overline{\pi}_{\mu}$, the following
canonical field equations emerge
\begin{align*}
\pfrac{\overline{\phi}}{x^{\mu}}=\pfrac{\HC_{\text{KGM}}}{\pi^{\mu}}&=
\overline{\pi}_{\mu}-\rmi q\,a_{\mu}\overline{\phi}\\
\pfrac{\phi}{x_{\mu}}=\pfrac{\HC_{\text{KGM}}}{\overline{\pi}_{\mu}}&=
\pi^{\mu}+\rmi q\,a^{\mu}\phi.
\end{align*}
The third group of canonical field equations results from
the derivatives of $\HC_{\text{KGM}}$ with respect to the
$a_{\mu}$, and to the $\phi$, $\overline{\phi}$ as
\begin{align*}
-\pfrac{p^{\mu\alpha}}{x^{\alpha}}=\pfrac{\HC_{\text{KGM}}}{a_{\mu}}&=
\rmi q\left(\overline{\pi}^{\mu}\phi-\overline{\phi}\pi^{\mu}\right)
\stackrel{\mathrm{Def.}}{=}j^{\mu}\\
-\pfrac{\overline{\pi}^{\alpha}}{x^{\alpha}}=\pfrac{\HC_{\text{KGM}}}{\phi}&=
\omega^{2}\overline{\phi}+\rmi q\,a_{\alpha}\overline{\pi}^{\alpha}\\
-\pfrac{\pi^{\alpha}}{x^{\alpha}}=\pfrac{\HC_{\text{KGM}}}{\overline{\phi}}&=
\omega^{2}\phi-\rmi q\,a^{\alpha}\pi_{\alpha}.
\end{align*}
Thus $\partial_{\alpha}p^{\alpha\mu}=j^{\mu}$.
The conventional factor $4\pi/c$ of the external-source example is absorbed into the normalization of the matter current in this section.
Due to the property of $p^{\mu\nu}$ being skew-symmetric
in its indices, the divergence of $j^{\mu}$ must vanish
in order for the system of canonical equations to be consistent,
$$
\pfrac{j^{\alpha}}{x^{\alpha}}=
-\ppfrac{p^{\alpha\beta}}{x^{\alpha}}{x^{\beta}}=0.
$$
This requirement is indeed fulfilled, for,
inserting the canonical equations, we find
\begin{align*}
\pfrac{j^{\alpha}}{x^{\alpha}}&=-\rmi q\pfrac{}{x^{\alpha}}\left(
\overline{\phi}\pi^{\alpha}-\overline{\pi}^{\alpha}\phi\right)\\
&=-\rmi q\left(\pfrac{\overline{\phi}}{x^{\alpha}}\pi^{\alpha}+
\overline{\phi}\,\pfrac{\pi^{\alpha}}{x^{\alpha}}-
\pfrac{\overline{\pi}^{\alpha}}{x^{\alpha}}\phi-
\overline{\pi}^{\alpha}\pfrac{\phi}{x^{\alpha}}\right)\\
&=-\rmi q\left[\left(\overline{\pi}_{\alpha}-\rmi q\,a_{\alpha}\overline{\phi}\,\right)\pi^{\alpha}-
\overline{\phi}\left(\omega^{2}\phi-\rmi q\,a^{\alpha}\pi_{\alpha}\right)\right.\\
&\quad\left.+\left(\omega^{2}\overline{\phi}+\rmi q\,a_{\alpha}\overline{\pi}^{\alpha}\right)\phi-
\overline{\pi}^{\alpha}\left(\pi_{\alpha}+\rmi q\,a_{\alpha}\phi\right)\right]\\
&=0.
\end{align*}
By eliminating the conjugate fields $p^{\mu\nu}$
and $\pi^{\mu}$, the canonical field equations can be rewritten
as second order partial differential equations,
\begin{align*}
\ppfrac{\overline{\phi}}{x^{\alpha}}{x_{\alpha}}+
\left(\omega^{2}-q^{2}a_{\alpha}a^{\alpha}\right)\overline{\phi}+
2\rmi q\,a_{\alpha}\pfrac{\overline{\phi}}{x_{\alpha}}+
\rmi q\overline{\phi}\,\pfrac{a^{\alpha}}{x^{\alpha}}&=0\\
\ppfrac{\phi}{x^{\alpha}}{x_{\alpha}}+
\left(\omega^{2}-q^{2}a_{\alpha}a^{\alpha}\right)\phi_{\phantom{*}}-
2\rmi q\,a_{\alpha}\pfrac{\phi}{x_{\alpha}}-\rmi q\phi\pfrac{a^{\alpha}}{x^{\alpha}}&=0,
\end{align*}
which coincide exactly with those that follow directly from the
Euler-Lagrange equations for the Lagrangian density $\LC_{\text{KGM}}$.
With the $4$-current density $j^{\mu}$ expressed in terms of
$\phi,\overline{\phi}$ and their derivatives,
\begin{align*}
j^{\mu}&=-\rmi q\left[\overline{\phi}\left(\pfrac{\phi}{x_{\mu}}-
\rmi q\,a^{\mu}\phi\right)-\left(\pfrac{\overline{\phi}}{x_{\mu}}+
\rmi q\,a^{\mu}\overline{\phi}\right)\phi\right]\\
&=-\rmi q\left(\overline{\phi}\pfrac{\phi}{x_{\mu}}-
\pfrac{\overline{\phi}}{x_{\mu}}\phi\right)-
2q^{2}\overline{\phi}\phi\,a^{\mu},
\end{align*}
we equally verify that $\partial j^{\alpha}/\partial x^{\alpha}=0$
by inserting the second order field equations.
\subsection{\label{sec:dirac-ham}The Dirac Hamiltonian density}
The dynamics of particles with spin $\frac{1}{2}$
and mass $m$ is described by the Dirac equation.
In units $\hbar=c=1$, with $\gamma^{\mu}$, $\mu=0,\ldots,3$ denoting the
$4\times 4$ Dirac matrices, and $\psi$ a four component Dirac
spinor, the Dirac Lagrangian density $\LC_{\text{D}}$ is given by
\begin{equation}\label{ld-dirac}
\LC_{\text{D}}=\rmi\,\overline{\psi}\gamma^{\alpha}
\pfrac{\psi}{x^{\alpha}} - m\overline{\psi}\psi,
\end{equation}
wherein
$
\overline{\psi}\equiv\psi^{\dagger}\gamma^{0}
$
denotes the adjoint spinor of $\psi$.
We use commuting classical spinor components in the ordinary derivative and Poisson-bracket calculations below.
A formulation with Grassmann-valued fields requires graded derivatives and brackets, which are not developed here.
Operating with spinors and Dirac matrices, Eq.~(\ref{ld-dirac})
is actually a shorthand for
$$
\LC_{\text{D}}=\rmi\overline{\psi}_{K}\gamma_{KJ}^{\alpha}
\pfrac{\psi_{J}}{x^{\alpha}} - m\overline{\psi}_{J}\psi_{J},
$$
where $\psi_{J}$ stands for a spinor component and $\gamma_{KJ}^{\alpha}$
for the $KJ$ matrix element of the Dirac matrix $\gamma^{\alpha}$.
In the following we summarize some fundamental relations
that apply for the Dirac matrices $\gamma^{\mu}$,
and their duals, $\gamma_{\mu}$,
\begin{align}
g^{\mu\nu}\Eins&=\frac{1}{2}\left(\gamma^{\mu}\gamma^{\nu}+
\gamma^{\nu}\gamma^{\mu}\right)\equiv\frac{1}{2}\{\gamma^{\mu},\gamma^{\nu}\}\nonumber\\
\gamma^{\alpha}\gamma_{\alpha}&=\gamma_{\alpha}\gamma^{\alpha}=4\;\Eins\nonumber\\
\gamma^{\alpha}\gamma^{\mu}\gamma_{\alpha}&=\gamma_{\alpha}\gamma^{\mu}\gamma^{\alpha}=-2\gamma^{\mu}\nonumber\\
\gamma^{\alpha}\gamma_{\mu}\gamma_{\alpha}&=\gamma_{\alpha}\gamma_{\mu}\gamma^{\alpha}=-2\gamma_{\mu}\nonumber\\
\gamma_{\alpha}\gamma^{\mu}\gamma^{\nu}\gamma^{\alpha}&=\gamma^{\alpha}\gamma^{\mu}\gamma^{\nu}\gamma_{\alpha}=
4\,g^{\mu\nu}\;\Eins\nonumber\\
\gamma_{\alpha}\gamma_{\mu}\gamma_{\nu}\gamma^{\alpha}&=\gamma^{\alpha}\gamma_{\mu}\gamma_{\nu}\gamma_{\alpha}=
4\,g_{\mu\nu}\;\Eins\nonumber\\
\sigma^{\mu\nu}&\equiv\frac{\im}{2}\left(\gamma^{\mu}\gamma^{\nu}-\gamma^{\nu}\gamma^{\mu}\right)\equiv
\frac{\im}{2}\left[\gamma^{\mu},\gamma^{\nu}\right]\nonumber\\
\det\sigma^{\mu\nu}&=1,\qquad\mu\ne\nu\nonumber\\
\tau_{\mu\nu}&=\frac{\im}{6}\left(\gamma_{\mu}\gamma_{\nu}+3\gamma_{\nu}\gamma_{\mu}\right)\nonumber\\
\tau_{\mu\alpha}\sigma^{\alpha\nu}&=\sigma^{\nu\alpha}\tau_{\alpha\mu}=
\delta_{\mu}^{\nu}\,\Eins\nonumber\\
\gamma^{\alpha}\tau_{\alpha\mu}&=\tau_{\mu\alpha}\gamma^{\alpha}=
\frac{1}{3\im}\,\gamma_{\mu}\nonumber\\
\gamma_{\alpha}\sigma^{\alpha\mu}&=\sigma^{\mu\alpha}\gamma_{\alpha}=
3\im\,\gamma^{\mu}\nonumber\\
\gamma^{\alpha}\tau_{\alpha\beta}\gamma^{\beta}&=\frac{4}{3\im}\Eins\nonumber\\
\gamma_{\alpha}\sigma^{\alpha\beta}\gamma_{\beta}&=12\im\,\Eins.
\label{dirac-algebra}
\end{align}
Herein, the symbol $\Eins$ stands for the $4\times 4$ unit matrix, and
the real numbers $g^{\mu\nu},g_{\mu\nu}\in\RB$ are the components of the
Minkowski metric $(g^{\mu\nu})=(g_{\mu\nu})=\diag(1,-1,-1,-1)$.
The matrices $(\sigma^{\mu\nu})$ and $(\tau_{\mu\nu})$ are to be understood
as $4\times 4$ block matrices, with each block $\sigma^{\mu\nu}$,
$\tau_{\mu\nu}$ representing a $4\times 4$ matrix of complex numbers.
Thus, $(\sigma^{\mu\nu})$ and its inverse $(\tau_{\mu\nu})$ are actually
$16\times 16$ matrices of complex numbers.

The Dirac Lagrangian density $\LC_{\text{D}}$ can be rendered symmetric
by combining the Lagrangian density Eq.~(\ref{ld-dirac}) with its
adjoint, which leads to
\begin{equation}\label{ld-dirac-symm}
\LC_{\text{D}}=\frac{\rmi}{2}\left(\overline{\psi}\gamma^{\alpha}
\pfrac{\psi}{x^{\alpha}}-\pfrac{\overline{\psi}}{x^{\alpha}}\gamma^{\alpha}\psi\right)
-m\overline{\psi}\psi.
\end{equation}\label{el-dirac}
The resulting Euler-Lagrange equations are identical to
those derived from Eq.~(\ref{ld-dirac}),
\begin{equation}
\begin{aligned}
\rmi\gamma^{\alpha}\pfrac{\psi}{x^{\alpha}}-m\psi&=0\\
\rmi\pfrac{\overline{\psi}}{x^{\alpha}}\gamma^{\alpha}+m\overline{\psi}&=0.
\end{aligned}
\end{equation}
As both Lagrangians~(\ref{ld-dirac}) and~(\ref{ld-dirac-symm})
are \emph{linear} in the derivatives of the fields, the
determinant of the Hessian vanishes,
\begin{equation}\label{ld-irregular}
\det\left[\ppfrac{\LC_{\text{D}}}
{\left(\partial_{\mu}\psi\right)}{\left(\partial_{\nu}\overline{\psi}\right)}\right]=0.
\end{equation}
Therefore, Legendre transformations of the Lagrangian
densities~(\ref{ld-dirac}) and~(\ref{ld-dirac-symm}) are irregular.
Nevertheless, as a Lagrangian density is determined only up to the
divergence of an arbitrary vector function $F^{\mu}$ according to
Eq.~(\ref{intbed}), one can construct an equivalent Lagrangian density
$\LC_{\text{D}}^{\prime}$ that yields identical Euler-Lagrange
equations while yielding a regular Legendre transformation.
The additional term\cite{gasiorowicz66} emerges as the divergence of a vector
function $F^{\mu}$, which may be expressed in symmetric form as
$$
F^{\mu}=\frac{\rmi}{2m}\left(\overline{\psi}\,\sigma^{\mu\alpha}\pfrac{\psi}{x^{\alpha}}+
\pfrac{\overline{\psi}}{x^{\alpha}}\sigma^{\alpha\mu}\,\psi\right).
$$
The factor $m^{-1}$ was chosen to match the dimensions, and this particular regularization assumes $m\ne0$.
For a massless field, it can be replaced by an independent nonzero inverse mass scale.
The formulas below retain $m\ne0$ and identify the regularization scale with $m$.
The vector $F^{\mu}$ depends on field derivatives.
Its divergence is therefore a total divergence on the first-jet space, not the explicit coordinate derivative of a derivative-free canonical generator.
The equivalence of the bulk equations follows for compactly supported variations, or for boundary conditions that make the variation of the integrated divergence vanish.
Explicitly, the additional term is given by
\begin{align*}
\pfrac{F^{\beta}}{x^{\beta}}&=\frac{\rmi}{2m}\left(
\pfrac{\overline{\psi}}{x^{\beta}}\sigma^{\beta\alpha}\pfrac{\psi}{x^{\alpha}}+
\overline{\psi}\sigma^{\beta\alpha}\ppfrac{\psi}{x^{\alpha}}{x^{\beta}}+
\ppfrac{\overline{\psi}}{x^{\alpha}}{x^{\beta}}\sigma^{\alpha\beta}\psi+
\pfrac{\overline{\psi}}{x^{\alpha}}\sigma^{\alpha\beta}\pfrac{\psi}{x^{\beta}}\right)\\
&=\frac{\rmi}{m}\pfrac{\overline{\psi}}{x^{\alpha}}\,\sigma^{\alpha\beta}
\pfrac{\psi}{x^{\beta}}.
\end{align*}
Note that the double sums
$\sigma^{\beta\alpha}\partial_{\beta}\partial_{\alpha}\psi$ and
$\partial_{\beta}\partial_{\alpha}\overline{\psi}\sigma^{\alpha\beta}$
vanish identically, as we sum over a symmetric
($\partial_{\mu}\partial_{\nu}\psi=\partial_{\nu}\partial_{\mu}\psi$)
and a skew-symmetric ($\sigma^{\mu\nu}=-\sigma^{\nu\mu}$) factor.
Choosing the sign of the added total divergence as written above, the equivalent Lagrangian density is given by
$\LC_{\text{D}}^{\prime}=\LC_{\text{D}}+\partial F^{\beta}/\partial x^{\beta}$,
which means, explicitly,
\begin{equation}\label{ld-dirac-regular}
\LC_{\text{D}}^{\prime}=\frac{\rmi}{2}\left(\overline{\psi}\gamma^{\alpha}
\pfrac{\psi}{x^{\alpha}}-\pfrac{\overline{\psi}}{x^{\alpha}}\gamma^{\alpha}\psi\right)
+\frac{\rmi}{m}\pfrac{\overline{\psi}}{x^{\alpha}}\,\sigma^{\alpha\beta}
\pfrac{\psi}{x^{\beta}}-m\overline{\psi}\psi.
\end{equation}
Due to the skew-symmetry of the $\sigma^{\mu\nu}$, the Euler-Lagrange
equations~(\ref{elgl}) for $\LC_{\text{D}}^{\prime}$ yield
again the Dirac equations~(\ref{el-dirac}).
To test regularity, both the spacetime and spinor indices must be included.
Writing $A,B$ for spinor indices, the mixed Hessian is the $16\times16$ matrix
$K_{(\mu A)(\nu B)}=(\rmi/m)(\sigma^{\mu\nu})_{AB}$.
Its inverse follows from Eq.~(\ref{dirac-algebra}),
\begin{equation}\label{ld-regular}
\sum_{\nu,B}\left(\frac{\rmi}{m}\sigma^{\mu\nu}\right)_{AB}
\left(\frac{m}{\rmi}\tau_{\nu\rho}\right)_{BC}
=\delta^{\mu}_{\rho}\delta_{AC}.
\end{equation}
The full Hessian in the derivatives of $\psi$ and $\overline{\psi}$ has invertible off-diagonal blocks $K$ and $K^{\mathsf T}$.
Hence the Legendre transformation of $\LC_{\text{D}}^{\prime}$ is regular.
The regularity statement concerns this full matrix rather than the determinants of its individual blocks.
It is remarkable that it is exactly a term which does \emph{not}
contribute to the Euler-Lagrange equations that makes the
Legendre transformation of $\LC_{\text{D}}^{\prime}$ \emph{regular}
and thus transfers the information on the dynamical system that
is contained in the Lagrangian to the Hamiltonian description.
The canonical momenta follow as
\begin{align}
\overline{\pi}^{\mu}&=
\pfrac{\LC_{\text{D}}^{\prime}}{\left(\partial_{\mu}
\psi\right)}=\hphantom{-}\frac{\rmi}{2}\overline{\psi}\gamma^{\mu}+
\frac{\rmi}{m}\pfrac{\overline{\psi}}{x^{\alpha}}\,\sigma^{\alpha\mu}
\nonumber\\
\pi^{\mu}&=
\pfrac{\LC_{\text{D}}^{\prime}}{\left(\partial_{\mu}
\overline{\psi}\right)}=-\frac{\rmi}{2}\gamma^{\mu}\psi+
\frac{\rmi}{m}\sigma^{\mu\alpha}\pfrac{\psi}{x^{\alpha}}.
\label{pi-dirac}
\end{align}
The Legendre transformation can now be worked out, yielding
\begin{align*}
\HC_{\text{D}}&=\overline{\pi}^{\nu}\pfrac{\psi}{x^{\nu}}+
\pfrac{\overline{\psi}}{x^{\nu}}\,\pi^{\nu}-
\LC_{\text{D}}^{\prime}\\
&=\frac{\rmi}{m}\pfrac{\overline{\psi}}{x^{\mu}}\,\sigma^{\mu\nu}
\pfrac{\psi}{x^{\nu}}+m\overline{\psi}\psi\\
&=\left(\overline{\pi}^{\nu}-\frac{\rmi}{2}\overline{\psi}\gamma^{\nu}
\right)\pfrac{\psi}{x^{\nu}}+m\overline{\psi}\psi.
\end{align*}
As the Hamiltonian density must always be expressed in terms of the
canonical momenta rather then by the velocities,
we must solve Eq.~(\ref{pi-dirac}) for $\partial_{\mu}\psi$
and $\partial_{\mu}\overline{\psi}$.
To this end, we multiply $\overline{\pi}^{\mu}$ by $\tau_{\mu\nu}$ --- hence by the inverse of $\sigma^{\alpha\mu}$ ---
from the right, and $\pi^{\mu}$ by $\tau_{\nu\mu}$ from the left,
\begin{align}\label{v-dirac}
\pfrac{\overline{\psi}}{x^{\nu}}&=\frac{m}{\rmi}\left(\overline{\pi}^{\mu}-
\frac{\rmi}{2}\overline{\psi}\gamma^{\mu}\right)\tau_{\mu\nu}\nonumber\\
\pfrac{\psi}{x^{\nu}}&=\frac{m}{\rmi}\,\tau_{\nu\mu}\left(
\pi^{\mu}+\frac{\rmi}{2}\gamma^{\mu}\psi\right).
\end{align}
The Dirac Hamiltonian density is then finally obtained as
\begin{equation}\label{hd-dirac}
\HC_{\text{D}}=\frac{m}{\rmi}
\left(\overline{\pi}^{\nu}-\frac{\rmi}{2}\overline{\psi}\gamma^{\nu}
\right)\tau_{\nu\mu}\left(\pi^{\mu}+\frac{\rmi}{2}\gamma^{\mu}
\psi\right)+m\overline{\psi}\psi.
\end{equation}
We may expand the products in Eq.~(\ref{hd-dirac}) using
Eqs.~(\ref{dirac-algebra}) to find
\begin{equation}\label{hd-dirac2}
\HC_{\text{D}}=\frac{\rmi m}{3}\left(\frac{1}{2}\overline{\psi}\gamma_{\nu}\pi^{\nu}
-3\overline{\pi}^{\mu}\tau_{\mu\nu}\pi^{\nu}
-\frac{1}{2}\overline{\pi}^{\nu}\gamma_{\nu}\psi\right)+\frac{2m}{3}\,\overline{\psi}\psi.
\end{equation}
In order to show that the Hamiltonian density $\HC_{\text{D}}$
describes the same dynamics as $\LC_{\text{D}}$ from
Eq.~(\ref{ld-dirac}), we set up the canonical equations
\begin{align*}
\pfrac{\overline{\psi}}{x^{\nu}}&=
\pfrac{\HC_{\text{D}}}{\pi^{\nu}}=\hphantom{-}
\rmi m\left(\frac{1}{6}\overline{\psi}\,\gamma_{\nu}-\overline{\pi}^{\mu}\tau_{\mu\nu}\right)\\
\pfrac{\psi}{x^{\mu}}&=
\pfrac{\HC_{\text{D}}}{\overline{\pi}^{\mu}}
=-\rmi m\left(\frac{1}{6}\gamma_{\mu}\psi+\tau_{\mu\nu}\pi^{\nu}\right).
\end{align*}
These equations reproduce the definition of the
canonical momenta from Eqs.~(\ref{pi-dirac}) in their inverted
form given by Eqs.~(\ref{v-dirac}).
The second set of canonical equations follows from the $\psi$ and
$\overline{\psi}$ dependence of the Hamiltonian $\HC_{\text{D}}$,
\begin{align*}
\pfrac{\overline{\pi}^{\alpha}}{x^{\alpha}}=
-\pfrac{\HC_{\text{D}}}{\psi}&=
\frac{\rmi m}{6}\overline{\pi}^{\mu}\gamma_{\mu}-\frac{2m}{3}\,\overline{\psi}\\
&=\frac{\rmi m}{6}\left(
\frac{\rmi}{2}\overline{\psi}\gamma^{\mu}+
\frac{\rmi}{m}\pfrac{\overline{\psi}}{x^{\nu}}\,\sigma^{\nu\mu}\right)\gamma_{\mu}-
\frac{2m}{3}\,\overline{\psi}\\
&=-m\overline{\psi}-\frac{\rmi}{2}\pfrac{\overline{\psi}}{x^{\nu}}\gamma^{\nu}\\
\pfrac{\pi^{\alpha}}{x^{\alpha}}=-\pfrac{\HC_{\text{D}}}{\overline{\psi}}&=
-\frac{\rmi m}{6}\gamma_{\mu}\pi^{\mu}-\frac{2m}{3}\,\psi\\
&=-\frac{\rmi m}{6}\gamma_{\mu}\left(
-\frac{\rmi}{2}\gamma^{\mu}\psi+\frac{\rmi}{m}
\sigma^{\mu\nu}\pfrac{\psi}{x^{\nu}}\right)-\frac{2m}{3}\,\psi\\
&=-m\psi+\frac{\rmi}{2}\gamma^{\nu}\pfrac{\psi}{x^{\nu}}.
\end{align*}
The divergences of the canonical momenta follow equally
from the derivatives of the first canonical equations, or,
equivalently, from the derivatives of Eqs.~(\ref{pi-dirac}),
\begin{align*}
\pfrac{\overline{\pi}^{\alpha}}{x^{\alpha}}&=
\hphantom{-}\frac{\rmi}{2}\pfrac{\overline{\psi}}{x^{\alpha}}\gamma^{\alpha}+
\frac{\rmi}{m}\ppfrac{\overline{\psi}}{x^{\alpha}}{x^{\beta}}\,\sigma^{\alpha\beta}\\
&=\hphantom{-}\frac{\rmi}{2}\pfrac{\overline{\psi}}{x^{\alpha}}\gamma^{\alpha}\\
\pfrac{\pi^{\alpha}}{x^{\alpha}}&=
-\frac{\rmi}{2}\gamma^{\alpha}\pfrac{\psi}{x^{\alpha}}+
\frac{\rmi}{m}\,\sigma^{\alpha\beta}\,\ppfrac{\psi}{x^{\alpha}}{x^{\beta}}\\
&=-\frac{\rmi}{2}\gamma^{\alpha}\pfrac{\psi}{x^{\alpha}}.
\end{align*}
The terms containing the second derivatives of $\psi$ and
$\overline{\psi}$ vanish due to the skew-symmetry of $\sigma^{\mu\nu}$.
Equating finally the expressions for the divergences of the
canonical momenta, we encounter, as expected, the Dirac
equations~(\ref{el-dirac})
\begin{align*}
\frac{\rmi}{2}\pfrac{\overline{\psi}}{x^{\alpha}}\gamma^{\alpha}&=-m\overline{\psi}-
\frac{\rmi}{2}\pfrac{\overline{\psi}}{x^{\nu}}\gamma^{\nu}\\
-\frac{\rmi}{2}\gamma^{\alpha}\pfrac{\psi}{x^{\alpha}}&=
-m\psi+\frac{\rmi}{2}\gamma^{\nu}\pfrac{\psi}{x^{\nu}}.
\end{align*}
It should be mentioned that this section is similar to the
derivation of the Dirac Hamiltonian density in Ref.~\cite{vonRieth84}.
Section~\ref{sec:Dirac-shift} discusses the momentum shifts induced by the added total divergence.
They have identically vanishing divergences, but the change in Legendre rank prevents their interpretation as an invertible canonical map between the original constrained phase space and the regularized phase space.
Section~\ref{sec:Dirac-interchange} examines why a proposed interchange of spinor fields and vector-spinor momenta is likewise not an invertible transformation on the full phase space.

We finally note that the additional term in the Dirac Lagrangian
density $\LC_{\text{D}}^{\prime}$ from Eq.~(\ref{ld-dirac-regular})
--- as compared to the Lagrangian $\LC_{\text{D}}$ from
Eq.~(\ref{ld-dirac-symm}) --- entails additional terms in the
energy-momentum tensor, defined in Eq.~(\ref{e-i-def}), namely,
$$
\Theta\indices{_{\mu}^{\nu}}-\theta\indices{_{\mu}^{\nu}}\equiv
j_{\mu}^{\nu}(\bx)=\frac{\rmi}{m}\left(
\partial_{\alpha}\overline{\psi}\sigma^{\alpha\nu}\partial_{\mu}\psi+
\partial_{\mu}\overline{\psi}\sigma^{\nu\alpha}\partial_{\alpha}\psi-
\delta_{\mu}^{\nu}\partial_{\alpha}\overline{\psi}\sigma^{\alpha\beta}
\partial_{\beta}\psi\right).
$$
We easily convince ourselves by direct calculation that the
divergences of $\Theta\indices{_{\mu}^{\nu}}$ and $\theta\indices{_{\mu}^{\nu}}$ coincide,
\begin{align*}
\pfrac{j_{\mu}^{\lambda}}{x^{\lambda}}&=\frac{\rmi}{m}\Big(
\partial_{\lambda}\partial_{\alpha}\overline{\psi}\sigma^{\alpha\lambda}\partial_{\mu}\psi+
\partial_{\alpha}\overline{\psi}\sigma^{\alpha\lambda}\partial_{\lambda}\partial_{\mu}\psi+
\partial_{\lambda}\partial_{\mu}\overline{\psi}\sigma^{\lambda\alpha}\partial_{\alpha}\psi\\ &\qquad\quad\mbox{}
+\partial_{\mu}\overline{\psi}\sigma^{\lambda\alpha}\partial_{\lambda}\partial_{\alpha}\psi-
\delta_{\mu}^{\lambda}\partial_{\lambda}\partial_{\alpha}\overline{\psi}\sigma^{\alpha\beta}\partial_{\beta}\psi-
\delta_{\mu}^{\lambda}\partial_{\alpha}\overline{\psi}\sigma^{\alpha\beta}\partial_{\lambda}\partial_{\beta}\psi\Big)\\
&=\frac{\rmi}{m}\Big(
\partial_{\alpha}\overline{\psi}\sigma^{\alpha\lambda}\partial_{\lambda}\partial_{\mu}\psi+
\partial_{\lambda}\partial_{\mu}\overline{\psi}\sigma^{\lambda\alpha}\partial_{\alpha}\psi
-\partial_{\mu}\partial_{\alpha}\overline{\psi}\sigma^{\alpha\beta}\partial_{\beta}\psi-
\partial_{\alpha}\overline{\psi}\sigma^{\alpha\beta}\partial_{\mu}\partial_{\beta}\psi\Big)\\
&=0,
\end{align*}
Thus the two energy-momentum tensors differ by an identically conserved improvement term.
Their integrated four-momenta agree when the corresponding boundary contribution vanishes.
This conservation identity holds without imposing the Dirac equations and does not provide an additional independent dynamical conservation law.

We now derive the equations of motion for the sets of fields
${\psi}_{I},{\bpi}_{I}$ using the covariant Heisenberg
equations~(\ref{Heisenberg-eqs})
$$
\pfrac{{\psi}_{I}}{x^{\mu}}=\big[{\psi}_{I},{\HC}\big]_{\mu},\quad
\delta_{\mu}^{\nu}\,\pfrac{{\pi}_{I}^{\alpha}}{x^{\alpha}}=
\big[{\pi}_{I}^{\nu},{\HC}\big]_{\mu},\quad
\pfrac{{\overline{\psi}}_{I}}{x^{\mu}}=\big[\,{\overline{\psi}}_{I},{\HC}\big]_{\mu},\quad
\delta_{\mu}^{\nu}\,\pfrac{{\overline{\pi}}_{I}^{\alpha}}{x^{\alpha}}=
\big[{\overline{\pi}}_{I}^{\nu},{\HC}\big]_{\mu},
$$
in conjunction with the \emph{fundamental Poisson bracket rules}
from Eqs.~(\ref{fundpk1}).
According to the definition~(\ref{pi-dirac}) of the canonical momenta
of the $\psi_{I}$ and the $\overline{\psi}_{I}$, the fields
$({\psi}_{I},{\overline{\bpi}}_{I})$ and, correspondingly,
$(\,{\overline{\psi}}_{I},{\bpi}_{I})$ constitute
the pairs of canonical conjugate quantities.
Again, ${\psi}_{I}$ and ${\overline{\psi}}_{I}$
as well for the conjugate fields, ${\bpi}_{I}$ and
${\overline{\bpi}}_{I}$ are to be regarded as \emph{independent fields}.

\begingroup
\allowdisplaybreaks[2]
The equations for ${\psi}$ and ${\overline{\psi}}$ follow as
\begin{align*}
\pfrac{{\psi}}{x^{\mu}}&=\left[{\psi},{\HC}_{\text{D}}
\right]_{\mu}\\
&=\left[{\psi},\rmi m\left(\frac{1}{6}{\overline{\psi}}\gamma_{\alpha}{\pi}^{\alpha}
-\overline{\pi}^{\alpha}\tau_{\alpha\beta}{\pi}^{\beta}-
\frac{1}{6}{\overline{\pi}}^{\alpha}\gamma_{\alpha}{\psi}\right)+
\frac{2}{3}m\,{\overline{\psi}}{\psi}\right]_{\mu}\\
&=\frac{\rmi m}{6}\left[{\psi},{\overline{\psi}}
\gamma_{\alpha}{\pi}^{\alpha}\right]_{\mu}
-\rmi m\left[{\psi},{\overline{\pi}}^{\alpha}
\tau_{\alpha\beta}{\pi}^{\beta}\right]_{\mu}\\
&\quad -
\frac{\rmi m}{6}\left[{\psi},{\overline{\pi}}^{\alpha}
\gamma_{\alpha}{\psi}\right]_{\mu}+
\frac{2m}{3}\left[{\psi},{\overline{\psi}}{\psi}\right]_{\mu}\\
&=-\rmi m{\overline{\pi}}^{\alpha}
\tau_{\alpha\beta}\underbrace{\left[{\psi},{\pi}^{\beta}\right]_{\mu}}_{=0}
-\rmi m\underbrace{\left[{\psi},{\overline{\pi}}^{\alpha}\right]_{\mu}}_{=\delta_{\mu}^{\alpha}}
\tau_{\alpha\beta}{\pi}^{\beta}\\
&\quad -\frac{\rmi m}{6}{\overline{\pi}}^{\alpha}
\gamma_{\alpha}\underbrace{\left[{\psi},{\psi}\right]_{\mu}}_{=0}-
\frac{\rmi m}{6}\underbrace{\left[{\psi},{\overline{\pi}}^{\alpha}
\right]_{\mu}}_{=\delta_{\mu}^{\alpha}}\gamma_{\alpha}{\psi}\\
&\quad+\frac{\rmi m}{6}\,{\overline{\psi}}
\gamma_{\alpha}\underbrace{\left[{\psi},{\pi}^{\alpha}\right]_{\mu}}_{=0}+
\frac{\rmi m}{6}\underbrace{\left[{\psi},{\overline{\psi}}\,\right]_{\mu}}_{=0}
\gamma_{\alpha}{\pi}^{\alpha}\\
&\quad +
\frac{2m}{3}{\overline{\psi}}\underbrace{\left[{\psi},{\psi}\right]_{\mu}}_{=0}+
\frac{2m}{3}\underbrace{\left[{\psi},{\overline{\psi}}\,\right]_{\mu}}_{=0}{\psi}\\
&=-\rmi m\left(\tau_{\mu\alpha}{\pi}^{\alpha}+\frac{1}{6}\,\gamma_{\mu}{\psi}\right),
\end{align*}
\begin{align*}
\pfrac{{\overline{\psi}}}{x^{\mu}}&=\left[\,\overline{\psi},{\HC}_{\text{D}}
\right]_{\mu}\\
&=\left[\overline{\psi},\rmi m\left(-\overline{\pi}^{\alpha}\tau_{\alpha\beta}{\pi}^{\beta}
-\frac{1}{6}{\overline{\pi}}^{\alpha}\gamma_{\alpha}{\psi}
+\frac{1}{6}{\overline{\psi}}\gamma_{\alpha}{\pi}^{\alpha}\right)
+\frac{2}{3}m\,{\overline{\psi}}{\psi}\right]_{\mu}\\
&=-\rmi m\left[\,\overline{\psi},{\overline{\pi}}^{\alpha}\tau_{\alpha\beta}{\pi}^{\beta}\right]_{\mu}-
\frac{\rmi m}{6}\left[\,\overline{\psi},{\overline{\pi}}^{\alpha}\gamma_{\alpha}{\psi}\right]_{\mu}\\
&\quad +
\frac{\rmi m}{6}\left[\,\overline{\psi},{\overline{\psi}}\gamma_{\alpha}{\pi}^{\alpha}\right]_{\mu}+
\frac{2m}{3}\left[\,\overline{\psi},{\overline{\psi}}{\psi}\right]_{\mu}\\
&=-\rmi m{\overline{\pi}}^{\alpha}\tau_{\alpha\beta}\underbrace{\left[\,{\overline{\psi}},{\pi}^{\beta}\right]_{\mu}}_{=\delta_{\mu}^{\beta}}-
\rmi m\underbrace{\left[\,{\overline{\psi}},{\overline{\pi}}^{\alpha}\right]_{\mu}}_{=0}\tau_{\alpha\beta}{\pi}^{\beta}\\
&\quad -
\frac{\rmi m}{6}{\overline{\pi}}^{\alpha}\gamma_{\alpha}\underbrace{\left[\,{\overline{\psi}},{\psi}\right]_{\mu}}_{=0}-
\frac{\rmi m}{6}\underbrace{\left[\,{\overline{\psi}},{\overline{\pi}}^{\alpha}\right]_{\mu}}_{=0}\gamma_{\alpha}{\psi}\\
&\quad+\frac{\rmi m}{6}{\overline{\psi}}\gamma_{\alpha}\underbrace{\left[\,{\overline{\psi}},{\pi}^{\alpha}
\right]_{\mu}}_{=\delta_{\mu}^{\alpha}}+\frac{\rmi m}{6}\underbrace{\left[\,{\overline{\psi}},{\overline{\psi}}\,\right]_{\mu}}_{=0}\gamma_{\alpha}{\pi}^{\alpha}\\
&\quad +
\frac{2m}{3}{\overline{\psi}}\underbrace{\left[\,{\overline{\psi}},{\psi}\right]_{\mu}}_{=0}+
\frac{2m}{3}\underbrace{\left[\,{\overline{\psi}},{\overline{\psi}}\right]_{\mu}}_{=0}{\psi}\\
&=\rmi m\left(\frac{1}{6}{\overline{\psi}}\,\gamma_{\mu}-{\overline{\pi}}^{\alpha}\tau_{\alpha\mu}\right).
\end{align*}
Similarly, the equations for the divergences of
${\bpi}_{I}$ and ${\overline{\bpi}}_{I}$ are
\begin{align*}
\delta_{\mu}^{\nu}\pfrac{{\pi}^{\alpha}}{x^{\alpha}}&=
\left[{\pi}^{\nu},{\HC}_{\text{D}}\right]_{\mu}\\
&=\left[{\pi}^{\nu},\rmi m\left(-{\overline{\pi}}^{\alpha}
\tau_{\alpha\beta}{\pi}^{\beta}-\frac{1}{6}{\overline{\pi}}^{\alpha}\gamma_{\alpha}{\psi}+
\frac{1}{6}{\overline{\psi}}\gamma_{\alpha}{\pi}^{\alpha}\right)+\frac{2}{3}m\,{\overline{\psi}}{\psi}\right]_{\mu}\\
&=-\rmi m\left[{\pi}^{\nu},{\overline{\pi}}^{\alpha}\tau_{\alpha\beta}{\pi}^{\beta}\right]_{\mu}-
\frac{\rmi m}{6}\left[{\pi}^{\nu},{\overline{\pi}}^{\alpha}\gamma_{\alpha}{\psi}\right]_{\mu}\\
&\quad +
\frac{\rmi m}{6}\left[{\pi}^{\nu},{\overline{\psi}}\gamma_{\alpha}{\pi}^{\alpha}\right]_{\mu}+
\frac{2m}{3}\left[{\pi}^{\nu},{\overline{\psi}}{\psi}\right]_{\mu}\\
&=-\rmi m{\overline{\pi}}^{\alpha}\tau_{\alpha\beta}\underbrace{\left[{\pi}^{\nu},{\pi}^{\beta}\right]_{\mu}}_{=0}-
\rmi m\underbrace{\left[{\pi}^{\nu},{\overline{\pi}}^{\alpha}\right]_{\mu}}_{=0}
\tau_{\alpha\beta}{\pi}^{\beta}\\
&\quad -\frac{\rmi m}{6}{\overline{\pi}}^{\alpha}
\gamma_{\alpha}\underbrace{\left[{\pi}^{\nu},{\psi}\right]_{\mu}}_{=0}-
\frac{\rmi m}{6}\underbrace{\left[{\pi}^{\nu},{\overline{\pi}}^{\alpha}
\right]_{\mu}}_{=0}\gamma_{\alpha}{\psi}\\
&\quad+\frac{\rmi m}{6}\,{\overline{\psi}}\gamma_{\alpha}\underbrace{\left[{\pi}^{\nu},{\pi}^{\alpha}\right]_{\mu}}_{=0}+
\frac{\rmi m}{6}\underbrace{\left[{\pi}^{\nu},{\overline{\psi}}\,\right]_{\mu}}_{=-\delta_{\mu}^{\nu}}\gamma_{\alpha}{\pi}^{\alpha}\\
&\quad +
\frac{2m}{3}{\overline{\psi}}\underbrace{\left[{\pi}^{\nu},{\psi}\right]_{\mu}}_{=0}+\frac{2m}{3}\underbrace{\left[{\pi}^{\nu},{\overline{\psi}}
\,\right]_{\mu}}_{=-\delta_{\mu}^{\nu}}{\psi}\\
&=-\delta_{\mu}^{\nu}m\left(\frac{\rmi}{6}\gamma_{\alpha}{\pi}^{\alpha}+\frac{2}{3}\,{\psi}\right)
\end{align*}
\begin{align*}
\delta_{\mu}^{\nu}\pfrac{{\overline{\pi}}^{\alpha}}{x^{\alpha}}
&=\left[{\overline{\pi}}^{\nu},{\HC}_{\text{D}}\right]_{\mu}\\
&=\left[{\overline{\pi}}^{\nu},\rmi m\left(-{\overline{\pi}}^{\alpha}\tau_{\alpha\beta}{\pi}^{\beta}-
\frac{1}{6}{\overline{\pi}}^{\alpha}\gamma_{\alpha}{\psi}+
\frac{1}{6}{\overline{\psi}}\gamma_{\alpha}{\pi}^{\alpha}\right)+
\frac{2}{3}m\,{\overline{\psi}}{\psi}\right]_{\mu}\\
&=-\rmi m\left[{\overline{\pi}}^{\nu},{\overline{\pi}}^{\alpha}
\tau_{\alpha\beta}{\pi}^{\beta}\right]_{\mu}-
\frac{\rmi m}{6}\left[{\overline{\pi}}^{\nu},{\overline{\pi}}^{\alpha}\gamma_{\alpha}{\psi}\right]_{\mu}\\
&\quad +
\frac{\rmi m}{6}\left[{\overline{\pi}}^{\nu},{\overline{\psi}}
\gamma_{\alpha}{\pi}^{\alpha}\right]_{\mu}+
\frac{2m}{3}\left[{\overline{\pi}}^{\nu},{\overline{\psi}}{\psi}\right]_{\mu}\\
&=-\rmi m{\overline{\pi}}^{\alpha}\tau_{\alpha\beta}\underbrace{\left[{\overline{\pi}}^{\nu},{\pi}^{\beta}\right]_{\mu}}_{=0}
-\rmi m\underbrace{\left[{\overline{\pi}}^{\nu},{\overline{\pi}}^{\alpha}\right]_{\mu}}_{=0}
\tau_{\alpha\beta}{\pi}^{\beta}\\
&\quad -\frac{\rmi m}{6}{\overline{\pi}}^{\alpha}
\gamma_{\alpha}\underbrace{\left[{\overline{\pi}}^{\nu},{\psi}\right]_{\mu}}_{=-\delta_{\mu}^{\nu}}-
\frac{\rmi m}{6}\underbrace{\left[{\overline{\pi}}^{\nu},{\overline{\pi}}^{\alpha}\right]_{\mu}}_{=0}\gamma_{\alpha}{\psi}\\
&\quad+\frac{\rmi m}{6}\,{\overline{\psi}}\gamma_{\alpha}\underbrace{\left[{\overline{\pi}}^{\nu},{\pi}^{\alpha}\right]_{\mu}}_{=0}
+\frac{\rmi m}{6}\underbrace{\left[{\overline{\pi}}^{\nu},{\overline{\psi}}\,\right]_{\mu}}_{=0}
\gamma_{\alpha}{\pi}^{\alpha}\\
&\quad +
\frac{2m}{3}{\overline{\psi}}\underbrace{\left[{\overline{\pi}}^{\nu},{\psi}\right]_{\mu}}_{=-\delta_{\mu}^{\nu}}+
\frac{2m}{3}\underbrace{\left[{\overline{\pi}}^{\nu},{\overline{\psi}}\,\right]_{\mu}}_{=0}{\psi}\\
&=\delta_{\mu}^{\nu}m\left(\frac{\rmi}{6}{\overline{\pi}}^{\alpha}\gamma_{\alpha}-\frac{2}{3}\,{\overline{\psi}}\right).
\end{align*}
\endgroup
The complete set of equations is now
\begin{align}
\pfrac{{\psi}}{x^{\mu}}&=-\rmi m\left(\tau_{\mu\alpha}{\pi}^{\alpha}+
\frac{1}{6}\,\gamma_{\mu}{\psi}\right)\label{dirac-eqm1-qu}\\
\pfrac{{\overline{\psi}}}{x^{\mu}}&=-\rmi m\left({\overline{\pi}}^{\alpha}\tau_{\alpha\mu}-
\frac{1}{6}{\overline{\psi}}\,\gamma_{\mu}\right)\label{dirac-eqm2-qu}\\
\pfrac{{\pi}^{\alpha}}{x^{\alpha}}&=-m\left(\frac{\rmi}{6}\gamma_{\alpha}{\pi}^{\alpha}+
\frac{2}{3}\,{\psi}\right)\label{dirac-eqm3-qu}\\
\pfrac{{\overline{\pi}}^{\alpha}}{x^{\alpha}}&=\hphantom{-}m\left(\frac{\rmi}{6}
{\overline{\pi}}^{\alpha}\gamma_{\alpha}-\frac{2}{3}\,{\overline{\psi}}\right).
\label{dirac-eqm4-qu}
\end{align}
In order to derive the equations for ${\psi}$ and ${\overline{\psi}}$,
we must get rid of their actual dependence on the ${\bpi}$ and ${\overline{\bpi}}$.
To this end, we solve Eqs.~(\ref{dirac-eqm1-qu}) and (\ref{dirac-eqm2-qu}) for
${\overline{\pi}}^{\mu}$ and ${\pi}^{\mu}$, respectively, then calculate
their divergences, and finally equate the divergences with the corresponding
divergence of Eqs.~(\ref{dirac-eqm3-qu}) and (\ref{dirac-eqm4-qu}).
Thus, in the first step, we find
\begin{align*}
\frac{1}{\rmi m}\,\sigma^{\nu\mu}\pfrac{{\psi}}{x^{\mu}}+
\frac{1}{6}\,\sigma^{\nu\mu}\gamma_{\mu}{\psi}=
-\sigma^{\nu\mu}\tau_{\mu\alpha}{\pi}^{\alpha}&=-{\pi}^{\nu}\\
\frac{1}{\rmi m}\pfrac{{\overline{\psi}}}{x^{\mu}}\sigma^{\mu\nu}-
\frac{1}{6}{\overline{\psi}}\,\gamma_{\mu}\sigma^{\mu\nu}=
-\overline{\pi}^{\alpha}\tau_{\alpha\mu}\sigma^{\mu\nu}&=-\overline{\pi}^{\nu},
\end{align*}
hence
\begin{align}
{\pi}^{\nu}&=\frac{\rmi}{m}\,\sigma^{\nu\mu}\pfrac{{\psi}}{x^{\mu}}-
\frac{\rmi}{2}\,\gamma^{\nu}{\psi}
\label{dirac-eqm5-qu}\\
{\overline{\pi}}^{\nu}&=
\frac{\rmi}{m}\,\pfrac{{\overline{\psi}}}{x^{\mu}}\sigma^{\mu\nu}+
\frac{\rmi}{2}{\overline{\psi}}\,\gamma^{\nu}.
\label{dirac-eqm6-qu}
\end{align}
The divergences are then
\begin{align*}
\pfrac{{\pi}^{\alpha}}{x^{\alpha}}&=\frac{\rmi}{m}\cancel{\sigma^{\alpha\mu}
\ppfrac{\psi}{x^{\mu}}{x^{\alpha}}}-\frac{\rmi}{2}\gamma^{\alpha}\pfrac{{\psi}}{x^{\alpha}}\\
\pfrac{{\overline{\pi}}^{\alpha}}{x^{\alpha}}&=
\frac{\rmi}{m}\cancel{\ppfrac{{\overline{\psi}}}{x^{\mu}}{x^{\alpha}}\sigma^{\mu\alpha}}+
\frac{\rmi}{2}\,\pfrac{{\overline{\psi}}}{x^{\alpha}}\gamma^{\alpha}.
\end{align*}
The second derivative terms vanish due to the skew-symmetry of $\sigma^{\nu\mu}$.
In Eqs.~(\ref{dirac-eqm3-qu}) and (\ref{dirac-eqm4-qu}), we eliminate their
dependence on the canonical momentum fields by inserting Eqs.~(\ref{dirac-eqm5-qu})
and (\ref{dirac-eqm6-qu}), respectively.
This gives
\begin{align*}
\pfrac{{\pi}^{\alpha}}{x^{\alpha}}&=-\frac{\rmi m}{6}\gamma_{\alpha}\left(
\frac{\rmi}{m}\sigma^{\alpha\mu}\pfrac{{\psi}}{x^{\mu}}-
\frac{\rmi}{2}\,\gamma^{\alpha}{\psi}\right)-\frac{2m}{3}\,{\psi}\\
&=\frac{\rmi}{2}\gamma^{\alpha}\pfrac{{\psi}}{x^{\alpha}}-m\psi\\
\pfrac{{\overline{\pi}}^{\alpha}}{x^{\alpha}}&=\frac{\rmi m}{6}
\left(\frac{\rmi}{m}\pfrac{{\overline{\psi}}}{x^{\mu}}\sigma^{\mu\alpha}+
\frac{\rmi}{2}{\overline{\psi}}\,\gamma^{\alpha}\right)
\gamma_{\alpha}-\frac{2m}{3}\,{\overline{\psi}}\\
&=-\frac{\rmi}{2}\pfrac{{\overline{\psi}}}{x^{\alpha}}\gamma^{\alpha}-m\overline{\psi}.
\end{align*}
Equating these expressions with the divergences obtained directly from the momentum definitions gives
\begin{align*}
\rmi\gamma^{\alpha}\pfrac{{\psi}}{x^{\alpha}}-m\,{\psi}&=0\\
\rmi\pfrac{{\overline{\psi}}}{x^{\alpha}}\gamma^{\alpha}+m\,{\overline{\psi}}&=0,
\end{align*}
which are the Dirac equations.
\section{\label{examples-ct}Examples of canonical
transformations in covariant Hamiltonian field theory}
\subsection{Point transformation}
Canonical transformations for which the transformed fields
$\Phi_{I}$ only depend on the original fields
$\phi_{I}$, and possibly on the independent variables $x^{\mu}$,
but \emph{not} on the original conjugate fields
$\bpi_{I}$ are referred to as point transformations.
A useful family of $4$-vector generating functions
$\bF_{2}$ that defines such transformations has the components
\begin{equation}\label{genPt}
F_{2}^{\mu}(\phi_{I},\bPi_{I},\bx)=
f_{J}(\phi_{I},\bx)\,\Pi_{J}^{\mu}.
\end{equation}
Herein, $f_{J}=f_{J}(\phi_{I},\bx)$ denotes a set of
differentiable functions with a nonsingular field Jacobian
$\det(\partial f_{J}/\partial\phi_{I})\ne0$.
The more general admissible generator may also contain an additive term $b^{\mu}(\bphi,x)$, as discussed in Sect.~\ref{sec:genf2}.
According to the general rules~(\ref{genF2}) for generating
functions of type $\bF_{2}$, the transformed field
$\Phi_{K}$ follows as
$$
\Phi_{K}\,\delta^{\mu}_{\nu}=
\pfrac{F_{2}^{\mu}}{\Pi_{K}^{\nu}}=
f_{J}(\phi_{I},\bx)\,\pfrac{\Pi_{J}^{\mu}}{\Pi_{K}^{\nu}}=
f_{J}(\phi_{I},\bx)\,\delta_{JK}\delta_{\nu}^{\mu}.
$$
The complete set of transformation rules is then
$$
\pi_{I}^{\mu}=\Pi_{J}^{\mu}\pfrac{f_{J}}{\phi_{I}},\qquad
\Phi_{K}=f_{K}(\phi_{I},\bx),\qquad
\HC^{\prime}=\HC+\Pi_{J}^{\alpha}{\left.\pfrac{f_{J}}
{x^{\alpha}}\right\vert}_{\text{expl}}.
$$
As a trivial example of a point transformation, we consider
the generating function of the \emph{identical} transformation.
Defining $f_{J}(\phi_{I})\equiv\phi_{J}$ in the generating function~(\ref{genPt})
\begin{equation}\label{genId}
F_{2}^{\mu}(\phi_{I},\bPi_{I})=\phi_{J}\,\Pi_{J}^{\mu},
\end{equation}
the pertaining transformation rules for this particular case are
$$
\pi_{I}^{\mu}=\Pi_{I}^{\mu},\qquad\Phi_{I}=\phi_{I},\qquad\HC^{\prime}=\HC.
$$
The existence of a neutral element is a necessary condition
for the set of canonical transformations to form a group.
\subsection{\label{sec:beispiel3}
Canonical shift of the conjugate momentum vector
field $\bpi_{I}$}
The generator of a canonical transformation that
shifts the conjugate $4$-vector field $\bpi_{I}(\bx)$
into a new conjugate $4$-vector field $\bPi_{I}(\bx)$
can be defined in terms of a function of type
$\bF_{2}(\phi_{I},\bPi_{I},\bx)$ as
\begin{equation}\label{genEich}
F_{2}^{\mu}=\phi_{J}\Pi_{J}^{\mu}+\phi_{J}\pfrac{}{x_{\alpha}}
\left(\pfrac{h_{J}^{\mu}}{x^{\alpha}}-\delta_{\alpha}^{\mu}
\pfrac{h_{J}^{\beta}}{x^{\beta}}\right),
\end{equation}
with arbitrary differentiable $x^{\mu}$-dependent parameter
functions $h_{I}^{\mu}(\bx)$.
From the general transformation rules~(\ref{genF2}),
the particular rules for this generating function are
\begin{align*}
\pi_{I}^{\mu}&=\pfrac{F_{2}^{\mu}}{\phi_{I}}=
\Pi_{I}^{\mu}+\pfrac{}{x_{\alpha}}
\left(\pfrac{h_{I}^{\mu}}{x^{\alpha}}-\delta_{\alpha}^{\mu}
\pfrac{h_{I}^{\beta}}{x^{\beta}}\right)\\
\Phi_{I}\,\delta_{\nu}^{\mu}&=\pfrac{F_{2}^{\mu}}{\Pi_{I}^{\nu}}=
\phi_{I}\,\delta_{\nu}^{\mu}\\
\HC^{\prime}-\HC&={\left.\pfrac{F_{2}^{\gamma}}
{x^{\gamma}}\right\vert}_{\text{expl}}=\phi_{I}
\left(\pppfrac{h_{I}^{\gamma}}{x^{\alpha}}{x_{\alpha}}
{x^{\gamma}}-\delta_{\alpha}^{\gamma}
\pppfrac{h_{I}^{\beta}}{x^{\beta}}{x_{\alpha}}{x^{\gamma}}\right)\equiv0.
\end{align*}
hence
\begin{equation}\label{transEich}
\Pi_{I}^{\mu}=\pi_{I}^{\mu}+
\ppfrac{h_{I}^{\alpha}}{x_{\mu}}{x^{\alpha}}-
\ppfrac{h_{I}^{\mu}}{x_{\alpha}}{x^{\alpha}},
\qquad\Phi_{I}=\phi_{I},\qquad \HC^{\prime}=\HC.
\end{equation}
The divergences of the fields $\bpi_{I}$ and $\bPi_{I}$ coincide since
$$
\pfrac{\Pi_{I}^{\beta}}{x^{\beta}}=\pfrac{\pi_{I}^{\beta}}{x^{\beta}}+
\pppfrac{h_{I}^{\alpha}}{x^{\beta}}{x_{\beta}}{x^{\alpha}}-
\pppfrac{h_{I}^{\beta}}{x^{\beta}}{x_{\alpha}}{x^{\alpha}}=
\pfrac{\pi_{I}^{\beta}}{x^{\beta}}.
$$
This equality of divergences does not imply that the two momentum fields solve the same prescribed Hamiltonian equations with the same functional Hamiltonian.
Writing
\[
\eta_{I}^{\mu}(x)=\pfrac{}{x_{\alpha}}
\left(\pfrac{h_{I}^{\mu}}{x^{\alpha}}-\delta_{\alpha}^{\mu}
\pfrac{h_{I}^{\beta}}{x^{\beta}}\right),
\]
the equality $\HC^{\prime}=\HC$ in Eq.~(\ref{transEich}) is evaluated at related phase-space points and means
\[
\HC^{\prime}(\bPhi,\bPi,x)=\HC(\bPhi,\bPi+\boldsymbol{\eta}(x),x).
\]
The transformation preserves the canonical form of the equations with this transformed Hamiltonian.
It is a symmetry of a fixed Hamiltonian function only if that function is invariant under the momentum shift.
\subsection{\label{sec:beispiel4}
Local and global gauge transformation of the fields $\phi_{I}$}
A common phase transformation of the fields $\phi_{I}(\bx)$ of the form
\begin{equation}\label{lokale-et}
\phi_{I}(\bx)\mapsto\Phi_{I}(\bx)=
\phi_{I}(\bx)\,e^{i\theta(\bx)}
\end{equation}
is commonly called a ``local gauge transformation''.
We can conceive this as a point transformation
that is generated by a $4$-vector function of type $\bF_{2}$
\begin{equation}\label{gen-lokale-et}
F_{2}^{\mu}(\phi_{I},\bPi_{I},\bx)=
\Pi_{I}^{\mu}\,e^{i\theta(\bx)}\,\phi_{I}.
\end{equation}
For complex fields in a real action, the complex-conjugate canonical sector is included as well.
The rules displayed here refer to one of these sectors.
The pertaining transformation rules follow directly
from the general rules of Eqs.~(\ref{genF2})
$$
\Phi_{I}=\phi_{I}\,e^{i\theta(\bx)},\qquad
\Pi_{I}^{\mu}=\pi_{I}^{\mu}e^{-i\theta(\bx)},\qquad
\HC^{\prime}=\HC+i\,\pi_{I}^{\alpha}
\pfrac{\theta(\bx)}{x^{\alpha}}\,\phi_{I}.
$$
In the particular case that $\theta$
does \emph{not} depend on the $x^{\mu}$, hence if
\mbox{$\theta=\text{const.}$}, then the gauge
transformation is referred to as ``global''.
In that case, the generating function~(\ref{gen-lokale-et})
itself does no longer explicitly depend on the $x^{\mu}$.
For a global transformation the Hamiltonian has the same value at corresponding phase-space points,
$\HC^{\prime}(\bPhi,\bPi,x)=\HC(\bphi,\bpi,x)$.
Preservation of its specified functional form is an additional symmetry condition and does not follow merely from the absence of explicit $x$ dependence in the generator.
The transformation rules are
$$
\Phi_{I}=\phi_{I}\,e^{i\theta},\qquad
\Pi_{I}^{\mu}=\pi_{I}^{\mu}e^{-i\theta},\qquad
\HC^{\prime}=\HC.
$$
A generalization of the gauge transformation~(\ref{lokale-et})
of the fields $\phi_{I}(\bx)$ is straightforward.
With $S_{IJ}(\bx)$ an invertible matrix, we may define
the generating function
\begin{equation}\label{gen-lokale-et-gen}
F_{2}^{\mu}(\phi_{I},\bPi_{I},\bx)=\Pi_{I}^{\mu}\,S_{IJ}(\bx)\,\phi_{J}.
\end{equation}
With $T_{IJ}(\bx)$ the inverse matrix of $S_{IJ}(\bx)$, hence
$S_{IK}T_{KJ}=\delta_{IJ}$, the transformation rules follow as
$$
\Phi_{I}=S_{IJ}(\bx)\,\phi_{J},\qquad
\Pi_{I}^{\mu}=\pi_{K}^{\mu}\,T_{KI}(\bx),\qquad
\HC^{\prime}=\HC+\pi_{K}^{\alpha}\,T_{KI}\,
\pfrac{S_{IJ}}{x^{\alpha}}\,\phi_{J}.
$$
\subsection{\label{sec:Noether}Infinitesimal canonical symmetries and Noether's theorem}
Canonical transformations preserve the form of the Hamiltonian variational principle while transforming the Hamiltonian according to the generator.
A \emph{symmetry} additionally preserves the specified functional form of the Hamiltonian.
Only this additional condition leads to the conservation law considered here.
We first work on an unconstrained phase-space domain, where all polymomenta can be varied independently.

The generating function of an infinitesimal transformation differs from that of the identity by a term of first order in $\epsilon$,
\begin{equation}\label{gen-infini}
F_2^\mu(\bphi,\bPi,x)=\phi_J\Pi_J^\mu+\epsilon g^\mu(\bphi,\bPi,x).
\end{equation}
The admissibility condition~(\ref{admissibleF2}) requires
$$
g^\mu(\bphi,\bPi,x)=h_I(\bphi,x)\Pi_I^\mu+b^\mu(\bphi,x),
\qquad
\pfrac{g^\mu}{\Pi_I^\nu}=h_I\delta_\nu^\mu.
$$
In particular, an arbitrary divergence-free function of all polymomenta is not automatically an admissible generator.
To first order, the transformation rules are
$$
\Pi_I^\mu=\pi_I^\mu-\epsilon\pfrac{g^\mu}{\phi_I},\qquad
\Phi_I=\phi_I+\epsilon h_I,\qquad
\HC'=\HC+\epsilon\left.\pfrac{g^\alpha}{x^\alpha}\right|_{\mathrm{expl}}.
$$
As in the original variation convention, put
$\delta\pi_I^\mu=\pi_I^\mu-\Pi_I^\mu$,
$\delta\phi_I=\phi_I-\Phi_I$, and
${\delta\HC|}_{\mathrm{CT}}=\HC-\HC'$.
Then
\begin{equation}\label{gl1}
\delta\pi_I^\mu=\epsilon\pfrac{g^\mu}{\phi_I},\qquad
\delta\phi_I=-\epsilon h_I,\qquad
{\delta\HC|}_{\mathrm{CT}}
=-\epsilon\left.\pfrac{g^\alpha}{x^\alpha}\right|_{\mathrm{expl}}.
\end{equation}
Within first-order terms, the momenta appearing in $g^\mu$ may be replaced by $\bpi$.
All quantities refer to the same space-time point.
The chain rule and the canonical field equations yield, on a solution,
\begin{align*}
\epsilon\pfrac{g^\alpha}{x^\alpha}
&=\delta\pi_I^\alpha\pfrac{\phi_I}{x^\alpha}
-\delta\phi_I\pfrac{\pi_I^\alpha}{x^\alpha}
-{\delta\HC|}_{\mathrm{CT}}+\CO(\epsilon^2)\\
&=\pfrac{\HC}{\pi_I^\alpha}\delta\pi_I^\alpha
+\pfrac{\HC}{\phi_I}\delta\phi_I
-{\delta\HC|}_{\mathrm{CT}}+\CO(\epsilon^2).
\end{align*}
For a symmetry, $\HC'(\bPhi,\bPi,x)=\HC(\bPhi,\bPi,x)$ as a functional identity to the order under consideration.
Consequently, the change prescribed by the canonical transformation agrees with the change obtained by inserting the transformed variables into the same Hamiltonian function,
\begin{equation}\label{div-g}
{\delta\HC|}_{\mathrm{CT}}
=\pfrac{\HC}{\phi_I}\delta\phi_I
+\pfrac{\HC}{\pi_I^\alpha}\delta\pi_I^\alpha
\quad\Longrightarrow\quad
\pfrac{g^\alpha}{x^\alpha}=0
\quad\hbox{on solutions}.
\end{equation}
This conservation condition characterizes a canonical symmetry within the admissible class, rather than canonicity by itself.
For clarity, the corresponding identity away from solutions is
\begin{align*}
\pfrac{g^\alpha}{x^\alpha}
={}&\pfrac{g^\alpha}{\phi_I}
\left(\pfrac{\phi_I}{x^\alpha}-\pfrac{\HC}{\pi_I^\alpha}\right)
+h_I\left(\pfrac{\pi_I^\alpha}{x^\alpha}+\pfrac{\HC}{\phi_I}\right)\\
&+\pfrac{g^\alpha}{\phi_I}\pfrac{\HC}{\pi_I^\alpha}
-h_I\pfrac{\HC}{\phi_I}
+\left.\pfrac{g^\alpha}{x^\alpha}\right|_{\mathrm{expl}}.
\end{align*}
The last line vanishes identically when the Hamiltonian symmetry condition holds.
The first line then vanishes by the field equations, establishing the Noether current
\begin{equation}\label{div-g1}
j^\mu(x)=g^\mu(\bphi(x),\bpi(x),x),\qquad
\pfrac{j^\alpha}{x^\alpha}=0.
\end{equation}
Conversely, conservation on a sufficiently rich family of solutions implies the symmetry condition only when their values cover the phase-space domain on which that identity is asserted.
The result is therefore a Noether theorem for the stated class of local canonical symmetries, rather than the most general form of Noether's theorem.
For the constrained gauge systems considered below, the transformation must preserve the primary constraint surface.
The symmetry condition is then imposed on that surface, or on the total Hamiltonian with the corresponding transformation of its constraint multipliers.
Only the independent momentum variations enter the proof.
The gauge-current conservation laws below are also checked directly from the constrained field equations.

\subsection{\label{infini-rz}Space-time translations and Hamiltonian evolution}
The energy-momentum tensor is
\begin{equation}\label{e-i-def-h-alt}
\theta\indices{_\alpha^\mu}
=\delta_\alpha^\mu\HC+\pi_I^\mu\pfrac{\phi_I}{x^\alpha}
-\delta_\alpha^\mu\pi_I^\beta\pfrac{\phi_I}{x^\beta}.
\end{equation}
For a constant infinitesimal displacement $\delta x^\alpha$, the corresponding current obeys
$$
\pfrac{}{x^\mu}\left(\theta\indices{_\alpha^\mu}\delta x^\alpha\right)
=\delta x^\alpha\left.\pfrac{\HC}{x^\alpha}\right|_{\mathrm{expl}}.
$$
Thus translation invariance of the specified Hamiltonian yields the expected conserved energy-momentum current.

It is tempting to insert this current into the type-$F_2$ expression
\begin{equation}\label{infinigenH}
F_2^\mu=\phi_I\Pi_I^\mu+\theta\indices{_\alpha^\mu}\delta x^\alpha.
\end{equation}
In more than one space-time dimension, this expression is generally \emph{not} an admissible derivative-free generator of the form~(\ref{admissibleF2}).
The derivatives $\partial_\alpha\phi_I$ in $\theta$ are not prescribed external functions when one varies arbitrary field configurations.
If they are instead eliminated with the regular Legendre map, the resulting momentum dependence generally violates the off-diagonal conditions in Eq.~(\ref{genF2}).
For example, for a real Klein-Gordon field with $\HC=\tfrac12\pi^\alpha\pi_\alpha+V(\phi)$, the translation current is
$$
g^\mu=\pi^\mu\pi_\alpha\delta x^\alpha
-\delta x^\mu\left(\tfrac12\pi^\alpha\pi_\alpha-V\right).
$$
For $\nu\ne\mu$, one obtains
$$
\pfrac{g^\mu}{\pi^\nu}=\pi^\mu\delta x_\nu-\delta x^\mu\pi_\nu,
$$
which does not generally vanish.
Treating the field derivatives as external coefficients and imposing the field equations only afterwards does not repair this failure of the generating identity away from solutions.

The actual displacement of a solution gives
\begin{equation}\label{infinipi}
\delta\pi_I^\mu=\pfrac{\pi_I^\mu}{x^\alpha}\delta x^\alpha,
\end{equation}
\begin{equation}\label{infiniphi}
\delta\phi_I=\pfrac{\phi_I}{x^\alpha}\delta x^\alpha.
\end{equation}
In particular, the first of these is generally different from
$(\partial_\alpha\pi_I^\alpha)\delta x^\mu$.
The total change of the Hamiltonian evaluated along that solution is
\begin{equation}\label{infinih}
\delta\HC=\pfrac{\HC}{x^\alpha}\delta x^\alpha,
\end{equation}
where the derivative includes the variation of its field and momentum arguments.
It cannot in general be replaced by its explicit coordinate derivative.
In point mechanics, there is only one independent variable, these distinctions disappear, and $F_2=\phi_I\Pi_I+H\delta t$ is the familiar infinitesimal evolution generator.

The covariant preservation law for field evolution is expressed by the symplectic current\cite{forgerromero05}.
Let $\delta_1$ and $\delta_2$ be two commuting variations satisfying the linearized canonical field equations about a regular solution.
Define
$$
\omega^\mu(\delta_1,\delta_2)
=\delta_1\phi_I\,\delta_2\pi_I^\mu
-\delta_2\phi_I\,\delta_1\pi_I^\mu.
$$
Taking its divergence and using the linearized equations gives
$$
\pfrac{\omega^\mu}{x^\mu}=0.
$$
The cancellation follows from the symmetry of the Hessian of $\HC$ in its complete set of canonical arguments.
Consequently,
$$
\Omega_\Sigma(\delta_1,\delta_2)
=\int_\Sigma\omega^\mu(\delta_1,\delta_2)\,\d\Sigma_\mu
$$
is independent of the chosen Cauchy hypersurface when the flux through the remaining boundary vanishes.
For a well-posed regular field theory this expresses the symplectic character of evolution on Cauchy data.
Gauge theories additionally require the treatment of their constraints and the degeneracy associated with gauge transformations.
This conservation law does not assert that evaluating a field at a neighboring space-time point is an invertible canonical transformation of the finite-dimensional variables $(\bphi,\bpi)$ at a single point.
\subsection{Gauge transformations of the Maxwell Hamiltonian density}
The Hamiltonian density $\HC_{\mathrm{M}}$ of the electromagnetic field was derived in Sec.~\ref{sec:maxwell}.
Its first constrained canonical equation determines the curl of $a_{\mu}$ and leaves the usual electromagnetic gauge freedom.
For a twice continuously differentiable function $\Lambda(x)$, define
\begin{equation}\label{lor-eich}
A_{\mu}=a_{\mu}+\pfrac{\Lambda}{x^{\mu}}.
\end{equation}
The corresponding field strength is unchanged,
\begin{equation}\label{lor-eich1}
P_{\mu\nu}=\pfrac{A_{\nu}}{x^{\mu}}-\pfrac{A_{\mu}}{x^{\nu}}
=\pfrac{a_{\nu}}{x^{\mu}}-\pfrac{a_{\mu}}{x^{\nu}}=p_{\mu\nu}.
\end{equation}
This is a gauge transformation.
The Lorenz gauge is the additional condition $\partial_{\mu}a^{\mu}=0$.
A transformation between two potentials satisfying that condition must also obey $\partial_{\mu}\partial^{\mu}\Lambda=0$.

For a prescribed conserved source, $\partial_{\mu}j^{\mu}=0$, a generating function that preserves the specified Maxwell Hamiltonian is
\begin{equation}\label{pt1}
F_{2}^{\mu}=P^{\alpha\mu}\left(a_{\alpha}+\pfrac{\Lambda}{x^{\alpha}}\right)
+\frac{4\pi}{c}j^{\mu}(x)\Lambda(x).
\end{equation}
The last term is absent in the source-free case.
The generator depends on the phase-space variables and on $x$, without derivatives of the dynamical momenta.
We may first differentiate its unrestricted extension and then restrict the resulting identity to $P^{\alpha\mu}=-P^{\mu\alpha}$.
The transformation preserves this primary constraint.
The rules
\begin{equation}\label{genF2b}
p^{\nu\mu}=\pfrac{F_{2}^{\mu}}{a_{\nu}},\qquad
A_{\nu}\delta^{\mu}_{\beta}=\pfrac{F_{2}^{\mu}}{P^{\nu\beta}},\qquad
\HC^{\prime}-\HC=\left.\pfrac{F_{2}^{\mu}}{x^{\mu}}\right|_{\mathrm{expl}}
\end{equation}
then give
\begin{align}
p^{\nu\mu}&=P^{\nu\mu},& A_{\nu}&=a_{\nu}+\pfrac{\Lambda}{x^{\nu}},\nonumber\\
\HC^{\prime}-\HC
&=P^{\alpha\mu}\ppfrac{\Lambda}{x^{\alpha}}{x^{\mu}}
+\frac{4\pi}{c}\pfrac{j^{\mu}}{x^{\mu}}\Lambda
+\frac{4\pi}{c}j^{\mu}\pfrac{\Lambda}{x^{\mu}}
=\frac{4\pi}{c}j^{\mu}\pfrac{\Lambda}{x^{\mu}}.\label{maxwell-gt}
\end{align}
The first term vanishes by antisymmetry and the second by conservation of the prescribed source.
Consequently,
\[
\HC_{\mathrm{M}}^{\prime}(A,P,x)
=-\quarter P_{\alpha\beta}P^{\alpha\beta}+\frac{4\pi}{c}j^{\alpha}A_{\alpha}
=\HC_{\mathrm{M}}(a,p,x)+\frac{4\pi}{c}j^{\alpha}\partial_{\alpha}\Lambda.
\]
Thus the canonical transformation preserves the functional form of the Hamiltonian, while its numerical value changes by the displayed explicit-divergence term.
The source coupling in the Lagrangian changes by a total divergence.

For $\Lambda\mapsto\epsilon\Lambda$, the infinitesimal generator is
\begin{equation}\label{pt2}
F_{2}^{\mu}=a_{\alpha}P^{\alpha\mu}+\epsilon g^{\mu},\qquad
 g^{\mu}=p^{\alpha\mu}\partial_{\alpha}\Lambda+\frac{4\pi}{c}j^{\mu}\Lambda.
\end{equation}
Its divergence satisfies
\begin{equation}\label{nt2}
\partial_{\mu}g^{\mu}
=\left(\partial_{\mu}p^{\alpha\mu}+\frac{4\pi}{c}j^{\alpha}\right)
\partial_{\alpha}\Lambda+\frac{4\pi}{c}(\partial_{\mu}j^{\mu})\Lambda=0
\end{equation}
on solutions of Maxwell's equation.
On those solutions,
\[
g^{\mu}=\partial_{\alpha}\big(p^{\alpha\mu}\Lambda\big).
\]
The current is therefore an antisymmetric superpotential divergence on shell.
Its identically conserved superpotential representative must be distinguished from a derivative-free phase-space generator.
\subsection{Gauge transformation of the coupled
Klein-Gordon-Maxwell field, local gauge invariance}
The Hamiltonian density $\HC_{\text{KGM}}$ of a complex
Klein-Gordon field that couples to an electromagnetic
$4$-vector potential $\bA$ was introduced in
Sec.~\ref{sec:hd-kgm} by Eq.~(\ref{hd-kg-max}).
We now define for this Hamiltonian density a \emph{local} gauge
transformation by means of the generating function
\begin{equation}\label{gen-erwet}
F_{2}^{\mu}=\overline{\Pi}^{\mu}\phi\,e^{\rmi\Lambda(\bx)}+
\overline{\phi}\,\Pi^{\mu}e^{-\rmi\Lambda(\bx)}+
P^{\alpha\mu}\left(a_{\alpha}+\frac{1}{q}\pfrac{\Lambda(\bx)}{x^{\alpha}}\right).
\end{equation}
In this context, the notation ``local'' refers to the fact that
the generator~(\ref{gen-erwet}) depends \emph{explicitly} on $\bx$.
The general transformation rules~(\ref{genF2}), (\ref{genF2b})
applied to the actual generating function yield for the fields
\begin{align}
P^{\mu\nu}&=p^{\mu\nu},&
A_{\mu}&=a_{\mu}+\frac{1}{q}\pfrac{\Lambda}{x^{\mu}}\nonumber\\
\Pi^{\mu}&=\pi^{\mu}e^{\rmi\Lambda(\bx)},&
\Phi&=\phi\,e^{\rmi\Lambda(\bx)}\label{eq:u1-fini}\\
\overline{\Pi}^{\mu}&=\overline{\pi}^{\mu}\,e^{-\rmi\Lambda(\bx)},&
\overline{\Phi}&=\overline{\phi}\,e^{-\rmi\Lambda(\bx)}\nonumber
\end{align}
and for the Hamiltonian
\begin{align*}
\HC^{\prime}&=\HC+\rmi\left(\overline{\pi}^{\alpha}\,\phi-
\overline{\phi}\,\pi^{\alpha}\right)\pfrac{\Lambda(\bx)}{x^{\alpha}}\\
&=\HC+\rmi q\left(\overline{\pi}^{\alpha}\,\phi-
\overline{\phi}\,\pi^{\alpha}\right)\left(A_{\alpha}-a_{\alpha}\right)\\
&=\HC+\rmi q\,\big(\,\overline{\Pi}^{\alpha}\,\Phi-
\overline{\Phi}\,\Pi^{\alpha}\big)A_{\alpha}-
\rmi q\left(\overline{\pi}^{\alpha}\,\phi-
\overline{\phi}\,\pi^{\alpha}\right)a_{\alpha}.
\end{align*}
In the transformation rule for the Hamiltonian density, the term
$P^{\alpha\beta}\partial^{2}\Lambda/\partial x^{\alpha}\partial x^{\beta}$
vanishes as the momentum tensor $P^{\alpha\beta}$ is skew-symmetric.
The transformed Hamiltonian density $\HC_{\text{KGM}}^{\prime}$
is now obtained by inserting the transformation rules into the Hamiltonian
density $\HC_{\text{KGM}}$ of Eq.~(\ref{hd-kg-max}),
$$
\HC_{\text{KGM}}^{\prime}=\overline{\Pi}_{\alpha}\,\Pi^{\alpha}+
\rmi qA_{\alpha}\left(\overline{\Pi}^{\alpha}\Phi-
\overline{\Phi}\,\Pi^{\alpha}\right)+\omega^{2}\overline{\Phi}\,\Phi-
\quarter P^{\alpha\beta}\,P_{\alpha\beta}.
$$
We observe that the Hamiltonian density (\ref{hd-kg-max}) is
\emph{form-invariant} under the local canonical transformation
generated by $\bF_{2}$ from Eq.~(\ref{gen-erwet}).

In order to derive the conserved Noether current that is
associated with this symmetry transformation, we first
set up the generating function of the \emph{infinitesimal}
canonical transformation corresponding to~(\ref{eq:u1-fini})
by letting $\Lambda\to\epsilon\Lambda$ and expanding
the exponential function up to the linear term in $\epsilon$
\begin{align*}
P^{\mu\nu}&=p^{\mu\nu},&
A_{\mu}&=a_{\mu}+\frac{\epsilon}{q}\pfrac{\Lambda}{x^{\mu}}\\
\Pi^{\mu}&=\pi^{\mu}\big(1+\rmi\epsilon\Lambda(x)\big),&
\Phi&=\phi\,\big(1+\rmi\epsilon\Lambda(x)\big)\\
\overline{\Pi}^{\mu}&=\overline{\pi}^{\mu}\,\big(1-\rmi\epsilon\Lambda(x)\big),&
\overline{\Phi}&=\overline{\phi}\,\big(1-\rmi\epsilon\Lambda(x)\big).
\end{align*}
These transformation rules are deduced from the generating function
\begin{align*}
F_{2}^{\mu}&=\overline{\Pi}^{\mu}\phi(1+\rmi\epsilon\Lambda)+
\overline{\phi}\,\Pi^{\mu}(1-\rmi\epsilon\Lambda)+
P^{\alpha\mu}\left(a_{\alpha}+\frac{\epsilon}{q}\pfrac{\Lambda}{x^{\alpha}}\right)\\
&=\overline{\Pi}^{\mu}\phi+\overline{\phi}\,\Pi^{\mu}+P^{\alpha\mu}a_{\alpha}+
\frac{\epsilon}{q}\left[\rmi q\left(\overline{\pi}^{\mu}\phi-\overline{\phi}\,\pi^{\mu}\right)\Lambda+
p^{\alpha\mu}\pfrac{\Lambda}{x^{\alpha}}\right].
\end{align*}
According to Noether's theorem from Eq.~(\ref{div-g1}), the expression in
brackets represents the conserved Noether current $g^{\mu}(x)$
$$
g^{\mu}(x)=\rmi q\left(\overline{\pi}^{\mu}\phi-
\overline{\phi}\,\pi^{\mu}\right)\Lambda+p^{\beta\mu}\,\pfrac{\Lambda}{x^{\beta}}.
$$
Its divergence is given by
\begin{align}
\pfrac{g^{\alpha}}{x^{\alpha}}&=\Lambda\left[\pfrac{}{x^{\alpha}}\rmi q\left(
\overline{\pi}^{\alpha}\phi-\overline{\phi}\pi^{\alpha}\right)\right]+
\pfrac{\Lambda}{x^{\beta}}\left[\rmi q\left(\overline{\pi}^{\beta}\phi-
\overline{\phi}\,\pi^{\beta}\right)+\pfrac{p^{\beta\alpha}}{x^{\alpha}}\right]+
\ppfrac{\Lambda}{x^{\beta}}{x^{\alpha}}\,p^{\beta\alpha}.
\label{eq:noether-current}
\end{align}
With $\Lambda(x)$ an \emph{arbitrary} function of space-time, the divergence
of $g^{\mu}(x)$ vanishes if and only if the three terms associated with
$\Lambda(x)$ and its derivatives in Eq.~(\ref{eq:noether-current}) separately vanish.
Regarding the canonical field equations from Sec.~\ref{sec:hd-kgm}
that emerge from the Hamiltonian $\HC_{\text{KGM}}$, we verify that indeed
$\partial j^{\alpha}/\partial x^{\alpha}=0$.

This means in particular that the matter contribution $j^{\mu}_{1}$ to the Noether current
$$
j^{\mu}_{1}=\rmi q\left(\overline{\pi}^{\mu}\phi-\overline{\phi}\,\pi^{\mu}\right),\qquad
\pfrac{j^{\alpha}_{1}}{x^{\alpha}}=0
$$
is separately conserved, whereas the second in conjunction with the third term,
$$
\pfrac{p^{\alpha\mu}}{x^{\alpha}}=j^{\mu}_{1},\qquad p^{\alpha\mu}=-p^{\mu\alpha},
$$
depicts the inhomogeneous Maxwell equation which satisfies the consistency requirement
$$
\ppfrac{p^{\alpha\beta}}{x^{\alpha}}{x^{\beta}}=-\ppfrac{p^{\beta\alpha}}{x^{\alpha}}{x^{\beta}}=
\pfrac{j^{\alpha}_{1}}{x^{\alpha}}=0.
$$
The matter current $j_{1}^{\mu}$ coincides with $j^{\mu}$ defined in Sect.~\ref{sec:hd-kgm}.
On solutions of the coupled equations, the arbitrary-parameter current has the superpotential form
$$
g^{\mu}=\Lambda j_{1}^{\mu}+p^{\alpha\mu}\partial_{\alpha}\Lambda
=\partial_{\alpha}\big(p^{\alpha\mu}\Lambda\big).
$$
Thus the family of local gauge currents is constrained by the field equations and the associated differential identity.
\subsection{\label{sec:gen-gauge}General local U$(N)$ gauge transformation}
As an interesting example of a canonical transformation in the
covariant Hamiltonian description of classical fields, the general
local U$(N)$ gauge transformation is treated in this section.
The main feature of the approach is that the terms to be
added to a given Hamiltonian $\HC$ in order to render it
\emph{locally} gauge invariant only depends on the
\emph{type of fields} contained in the Hamiltonian $\HC$
and not on the particular form of the original Hamiltonian itself.
The only precondition is that $\HC$ must be invariant under the
corresponding \emph{global} gauge transformation, hence a
transformation \emph{not} depending explicitly on $\bx$.
\subsubsection{External gauge field}
We consider a system consisting of a vector of $N$ complex fields
$\phi_{I},\;I=1,\ldots,N$, and the adjoint field vector,
$\overline{\bphi}$,
$$
\bphi=\begin{pmatrix}\phi_{1}\\\vdots\\\phi_{N}
\end{pmatrix},\qquad
\overline{\bphi}=\left(\overline{\phi}_{1}\cdots\overline{\phi}_{N}\right).
$$
A general local linear transformation may be expressed in terms
of a dimensionless complex matrix $U(\bx)=(u_{IJ}(\bx))$ and its adjoint,
$U^{\dagger}$ that may depend explicitly on the independent
variables, $x^{\mu}$, as
\begin{align}
\bPhi&=U\,\bphi,&
\overline{\bPhi}&=\overline{\bphi}\,U^{\dagger}\nonumber\\
\Phi_{I}&=u_{IJ}\:\phi_{J},&
\overline{\Phi}_{I}&=\overline{\phi}_{J}\,u_{IJ}^{*}.
\label{general-pointtra}
\end{align}
Here a star denotes ordinary complex conjugation, so that $(U^{\dagger})_{IJ}=u_{JI}^{*}$.

With this notation, $\phi_{I}$ may stand for a set of
$I=1,\ldots,N$ complex scalar fields $\phi_{I}$ or Dirac spinors.
In other words, $U$ is supposed to define an isomorphism
within the space of the $\phi_{I}$, hence to linearly map the
$\phi_{I}$ into objects of the same type.
The uppercase Latin letter indexes label the field or spinor number.
Their transformation in iso-space are not associated with any metric.
We, therefore, do not use superscripts for these indexes as there
is no distinction between covariant and contravariant components.
In contrast, Greek indexes are used for those components that
\emph{are} associated with a metric --- such as the derivatives
with respect to a space-time variable, $x^{\mu}$.
As usual, summation is understood for indexes occurring in pairs.

We restrict ourselves to transformations that preserve the
internal bilinear $\overline{\bphi}\bphi$
\begin{align*}
\overline{\bPhi}\bPhi&=
\overline{\bphi}\,U^{\dagger}U\,\bphi=\overline{\bphi}\bphi
&\Longrightarrow\qquad U^{\dagger}U&=\Eins=UU^{\dagger}\\
\overline{\Phi}_{I}\Phi_{I}&=
\overline{\phi}_{J}u_{IJ}^{*}\,u_{IK}\phi_{K}=
\overline{\phi}_{K}\phi_{K}&\Longrightarrow\qquad
u_{IJ}^{*}\,u_{IK}&=\delta_{JK}=u_{JI}\,u_{KI}^{*}.
\end{align*}
This means that $U^{\dagger}=U^{-1}$, hence that the
matrix $U$ is supposed to be \emph{unitary}.
The transformation~(\ref{general-pointtra}) follows from a generating
function that --- corresponding to $\HC$ --- must be a real-valued
function of the generally complex fields $\bphi$ and their canonical
conjugates, $\bpi^{\mu}$,
\begin{align}
\label{gen-pointtra}
F_{2}^{\mu}(\bphi,\overline{\bphi},\bPi^{\mu},\overline{\bPi}^{\mu},\bx)
&=\overline{\bPi}^{\mu}U\,\bphi+
\overline{\bphi}\,U^{\dagger}\,\bPi^{\mu}\nonumber\\
&=\overline{\Pi}_{K}^{\mu}\,u_{KJ}\,\phi_{J}+
\overline{\phi}_{K}\,u_{JK}^{*}\,\Pi_{J}^{\mu}.
\end{align}
According to Eqs.~(\ref{genF2}) the set of transformation
rules follows as
\begin{align*}
\overline{\pi}_{I}^{\mu}=\pfrac{F_{2}^{\mu}}{\phi_{I}}&=
\overline{\Pi}_{K}^{\mu}u_{KJ}\delta_{IJ},&
\overline{\Phi}_{I}\delta_{\nu}^{\mu}&=
\pfrac{F_{2}^{\mu}}{\Pi_{I}^{\nu}}=\overline{\phi}_{K}
u_{JK}^{*}\delta_{\nu}^{\mu}\delta_{IJ}\\
\pi_{I}^{\mu}=\pfrac{F_{2}^{\mu}}{\overline{\phi}_{I}}&=
\delta_{IK}u_{JK}^{*}\Pi_{J}^{\mu},&
\Phi_{I}\delta_{\nu}^{\mu}&=
\pfrac{F_{2}^{\mu}}{\overline{\Pi}_{I}^{\nu}}=
\delta_{\nu}^{\mu}\delta_{IK}u_{KJ}\phi_{J}.
\end{align*}
The complete set of transformation rules and their
inverses then read in component notation
\begin{align}
\Phi_{I}&=u_{IJ}\,\phi_{J},\qquad
\overline{\Phi}_{I}=\overline{\phi}_{J}\,u_{IJ}^{*},\qquad
\Pi_{I}^{\mu}=u_{IJ}\,\pi_{J}^{\mu},\qquad
\overline{\Pi}_{I}^{\mu}=\overline{\pi}_{J}^{\mu}\,u_{IJ}^{*}\nonumber\\
\phi_{I}&=u_{JI}^{*}\,\Phi_{J},\qquad
\overline{\phi}_{I}=\overline{\Phi}_{J}u_{JI},\qquad\,
\pi_{I}^{\mu}=u_{JI}^{*}\,\Pi_{J}^{\mu},\qquad
\overline{\pi}_{I}^{\mu}=\overline{\Pi}_{J}^{\mu}u_{JI}.
\label{pointtra-rules}
\end{align}
We assume the Hamiltonian $\HC$ to be \emph{form-invariant}
under the \emph{global} gauge transformation~(\ref{general-pointtra}),
which is given for $U=\mathrm{const}$, hence for all $u_{IJ}$
\emph{not} depending on the independent variables, $x^{\mu}$.
In contrast, if $U=U(\bx)$, the transformation~(\ref{pointtra-rules})
is referred to as a \emph{local} gauge transformation.
The transformation rule for the Hamiltonian is then determined by
the explicitly $x^{\mu}$-dependent terms of the generating
function $F_{2}^{\mu}$ according to
\begin{align}
\HC^{\prime}-\HC=\left.\pfrac{F_{2}^{\alpha}}{x^{\alpha}}\right\vert
_{\text{expl}}&=
\overline{\Pi}_{I}^{\alpha}\pfrac{u_{IJ}}{x^{\alpha}}\,\phi_{J}+
\overline{\phi}_{I}\pfrac{u_{JI}^{*}}{x^{\alpha}}\,\Pi_{J}^{\alpha}\nonumber\\
&=\overline{\pi}_{K}^{\alpha}\,u_{IK}^{*}\pfrac{u_{IJ}}{x^{\alpha}}
\phi_{J}+\overline{\phi}_{I}
\pfrac{u_{JI}^{*}}{x^{\alpha}}\,u_{JK}\pi_{K}^{\alpha}\nonumber\\
&=\left(\overline{\pi}_{K}^{\alpha}\,\phi_{J}-
\overline{\phi}_{K}\pi_{J}^{\alpha}\right)
u_{IK}^{*}\pfrac{u_{IJ}}{x^{\alpha}}.\label{pointtra-ham}
\end{align}
In the last step, the identity
$$
\pfrac{u_{IJ}^{*}}{x^{\mu}}\,u_{IK}+
u_{IJ}^{*}\,\pfrac{u_{IK}}{x^{\mu}}=\pfrac{}{x^{\mu}}\left(u_{IJ}^{*}u_{IK}\right)=
\pfrac{}{x^{\mu}}\delta_{JK}=0
$$
was inserted.
If we want to set up a Hamiltonian $\bar{\HC}$ that is
\emph{form-invariant} under the \emph{local}, hence
$x^{\mu}$-dependent transformation generated by~(\ref{gen-pointtra}),
then we must compensate the additional terms~(\ref{pointtra-ham})
that emerge from the explicit $x^{\mu}$-dependence of the generating
function~(\ref{gen-pointtra}).
A sufficient way to achieve this is to \emph{adjoin} the Hamiltonian $\HC$
of our system with terms that correspond to~(\ref{pointtra-ham})
with regard to their dependence on the canonical variables,
$\bphi,\overline{\bphi},\bpi^{\mu},\overline{\bpi}^{\mu}$.
With a \emph{unitary} matrix $U$, the $u_{IJ}$-dependent terms
in Eq.~(\ref{pointtra-ham}) are \emph{skew-hermitian},
\begin{align*}
{\left(u_{IK}^{*}\,\pfrac{u_{IJ}}{x^{\mu}}\right)}^{*}&=
\pfrac{u_{IJ}^{*}}{x^{\mu}}\,u_{IK}=
-u_{IJ}^{*}\,\pfrac{u_{IK}}{x^{\mu}},\\
{\left(\pfrac{u_{KI}}{x^{\mu}}\,u_{JI}^{*}\right)}^{*}&=
u_{JI}\pfrac{u_{KI}^{*}}{x^{\mu}}=
-\pfrac{u_{JI}}{x^{\mu}}u_{KI}^{*},
\end{align*}
or in matrix notation
$$
{\left(U^{\dagger}\pfrac{U}{x^{\mu}}\right)}^{\dagger}=
\pfrac{U^{\dagger}}{x^{\mu}}U=-U^{\dagger}\pfrac{U}{x^{\mu}},\qquad
{\left(\pfrac{U}{x^{\mu}}U^{\dagger}\right)}^{\dagger}=
U\pfrac{U^{\dagger}}{x^{\mu}}=-\pfrac{U}{x^{\mu}}U^{\dagger}.
$$
The $u$-dependent terms in Eq.~(\ref{pointtra-ham}) can thus be compensated
by a \emph{Hermitian} matrix $(\ba_{KJ})$ of ``$4$-vector gauge fields'',
with each off-diagonal matrix element, $\ba_{KJ},\;K\neq J$,
a complex $4$-vector field with components $a_{KJ\mu},\;\mu=0,\ldots,3$
$$
a_{KJ\mu}=a_{JK\mu}^{*}.
$$
The number of independent gauge fields thus amounts
to $N^{2}$ real $4$-vectors.
The amended Hamiltonian $\bar{\HC}$ thus reads
\begin{equation}\label{tildeHC}
\bar{\HC}=\HC+\HC_{\mathrm{a}},\qquad \HC_{\mathrm{a}}=
\rmi q\left(\overline{\pi}_{K}^{\alpha}\phi_{J}-
\overline{\phi}_{K}\pi_{J}^{\alpha}\right)a_{KJ\alpha}.
\end{equation}
With the real coupling constant $q$, the interaction
Hamiltonian $\HC_{\mathrm{a}}$ is thus real.
Usually, $q$ is defined to be dimensionless.
We then infer the dimension of the gauge fields $\ba_{KJ}$ to be
$$
[q]=1,\qquad [\ba_{KJ}]=[L]^{-1}=[m]=[\partial_{\mu}].
$$
Here $\HC(\bPhi,\overline{\bPhi},\bPi,\overline{\bPi},x)$ means the same globally invariant matter function evaluated on transformed arguments.
It is distinguished from the canonically transformed bare Hamiltonian $\HC^{\prime}$ of Eq.~(\ref{pointtra-ham}), which also contains the explicit-divergence term.
In contrast to the given system Hamiltonian $\HC$, the \emph{amended}
Hamiltonian $\bar{\HC}$ is supposed to be \emph{invariant in its form}
under the canonical transformation, hence
\begin{gather}\label{tildeHCp}
\bar{\HC}^{\prime}=\HC(\bPhi,\overline{\bPhi},\bPi,\overline{\bPi},x)+\HC_{\mathrm{a}}^{\prime},\\
\HC_{\mathrm{a}}^{\prime}=
\rmi q\left(\overline{\Pi}_{K}^{\alpha}\Phi_{J}-
\overline{\Phi}_{K}\Pi_{J}^{\alpha}\right)A_{KJ\alpha}.
\end{gather}
Submitting the amended Hamiltonian $\bar{\HC}$ from
Eq.~(\ref{tildeHC}) to the canonical
transformation generated by Eq.~(\ref{gen-pointtra}), the new
Hamiltonian $\bar{\HC}^{\prime}$ emerges with
Eqs.~(\ref{pointtra-ham}) and (\ref{tildeHCp}) as
\begin{align*}
\bar{\HC}^{\prime}&=\bar{\HC}+\left.\pfrac{F_{2}^{\alpha}}{x^{\alpha}}
\right\vert_{\text{expl}}=
\HC+\HC_{\mathrm{a}}+\left.\pfrac{F_{2}^{\alpha}}{x^{\alpha}}
\right\vert_{\text{expl}}\\
&=\HC+\left(\overline{\pi}_{K}^{\alpha}\phi_{J}-
\overline{\phi}_{K}\pi_{J}^{\alpha}\right)\left(\rmi q\,a_{KJ\alpha}+
u_{IK}^{*}\pfrac{u_{IJ}}{x^{\alpha}}\right)\\
&\stackrel{!}{=}\HC(\bPhi,\overline{\bPhi},\bPi,\overline{\bPi},x)+\left(\overline{\Pi}_{K}^{\alpha}\Phi_{J}-
\overline{\Phi}_{K}\Pi_{J}^{\alpha}\right)\rmi q\,A_{KJ\alpha}.
\end{align*}
The original base fields, $\phi_{J},\overline{\phi}_{K}$ and their conjugates
can now be expressed in terms of the transformed ones according to
the rules~(\ref{pointtra-rules}), which yields, after index relabeling, the conditions
\begin{gather*}
\HC(\bPhi,\overline{\bPhi},\bPi^{\mu},\overline{\bPi}^{\mu},x^{\mu})
\stackrel{\text{global GT}}{=}\HC(\bphi,\overline{\bphi},\bpi^{\mu},\overline{\bpi}^{\mu},x^{\mu})\\
\left(\overline{\Pi}_{K}^{\alpha}\Phi_{J}-
\overline{\Phi}_{K}\Pi_{J}^{\alpha}\right)\rmi q\,A_{KJ\alpha}=
\left(\overline{\Pi}_{K}^{\alpha}\Phi_{J}-
\overline{\Phi}_{K}\Pi_{J}^{\alpha}\right)\left(\rmi q\,u_{KL}\,a_{LI\alpha}\,u_{JI}^{*}+
\pfrac{u_{KI}}{x^{\alpha}}u_{JI}^{*}\right).
\end{gather*}
This means that the system Hamiltonian must be invariant under
the \emph{global} gauge transformation defined by Eq.~(\ref{pointtra-rules}),
whereas the gauge fields $A_{IJ\mu}$ must satisfy the transformation rule
\begin{equation}\label{gauge-tra1}
A_{KJ\mu}=u_{KL}\,a_{LI\mu}\,u_{JI}^{*}+
\frac{1}{\rmi q}\,\pfrac{u_{KI}}{x^{\mu}}\,u_{JI}^{*}.
\end{equation}
We observe that for any type of canonical field variables
$\phi_{I}$ and for any globally invariant matter Hamiltonian $\HC$ in this representation, the
transformation of the $4$-vector gauge fields $\ba_{IJ}(\bx)$
is determined by Eq.~(\ref{gauge-tra1}) for the chosen minimal coupling and
the transformation matrix $U(\bx)$ for the $N$ fields $\phi_{I}$.
In the notation of the $4$-vector gauge fields
$\ba_{KJ}(\bx),\;K,J=1,\ldots,N$,
the transformation rule is equivalently expressed as
$$
\bA_{KJ}=u_{KL}\,\ba_{LI}\,u_{JI}^{*}+
\frac{1}{\rmi q}\,\pfrac{u_{KI}}{\bx}\,u_{JI}^{*},
$$
or, in matrix notation
\begin{equation}\label{gauge-tra2}
\hat{A}_{\mu}=U\hat{a}_{\mu}U^{\dagger}+
\frac{1}{\rmi q}\,\pfrac{U}{x^{\mu}}U^{\dagger},\qquad
\hat{\bA}=U\,\hat{\ba}\,U^{\dagger}+
\frac{1}{\rmi q}\,\pfrac{U}{\bx}\,U^{\dagger},
\end{equation}
with $\hat{a}_{\mu}$ denoting the $N\times N$ matrices
of the $\mu$-components of the $4$-vectors $\ba_{IK}(\bx)$,
and, finally, $\hat{\ba}$ the $N\times N$ matrix of
gauge $4$-vectors $\ba_{IK}(\bx)$.
The matrix $U(\bx)$ is \emph{unitary},
hence belongs to the group U$(N)$
$$
U^{\dagger}(\bx)=U^{-1}(\bx),\qquad |\det{U(\bx)}|=1.
$$
For $\det{U(\bx)}=+1$, the matrix $U(\bx)$ is
an element of the group SU$(N)$.

Equation~(\ref{gauge-tra2}) is the general transformation
law for gauge bosons.
$U$ and $\hat{a}_{\mu}$ need not commute when $N>1$.
The full group U$(N)$ is non-Abelian in that case, although particular field configurations may belong to commuting subalgebras.
As the matrices $\hat{a}_{\mu}$ are Hermitian, the number
of independent gauge $4$-vectors $\ba_{IK}$ amounts to
$N$ real vectors on the main diagonal, and $(N^{2}-N)/2$
independent complex off-diagonal vectors, which corresponds
to a total number of $N^{2}$ independent real gauge $4$-vectors
for a U$(N)$ symmetry transformation, and hence $N^{2}-1$
real gauge $4$-vectors for a SU$(N)$ symmetry transformation.

To describe the Lie algebra, introduce a fixed Hermitian basis $T_{a}$, with $a=1,\ldots,N^{2}$ for U$(N)$.
The gauge potential is expanded as $\hat a_{\mu}=a_{\mu}^{a}T_{a}$.
The structure constants carry Lie-algebra indices,
\begin{equation}\label{eq:a-algebra}
[T_{a},T_{b}]=\rmi f_{ab}{}^{c}T_{c},\qquad
[\hat a_{\mu},\hat a_{\nu}]
=\rmi a_{\mu}^{a}a_{\nu}^{b}f_{ab}{}^{c}T_{c}.
\end{equation}
They are independent of the gauge-field configuration and satisfy
\[
f_{ab}{}^{d}f_{dc}{}^{e}+f_{bc}{}^{d}f_{da}{}^{e}
+f_{ca}{}^{d}f_{db}{}^{e}=0.
\]
This follows from the Jacobi identity for matrix commutators.
The four spacetime components $\hat a_{\mu}$ are not in general a basis of the internal Lie algebra and need not close among themselves.
For SU$(N)$ one restricts both the basis and the connection to traceless Hermitian matrices.
\subsubsection{Including the gauge field dynamics}
The transformation of the gauge fields from Eq.~(\ref{gauge-tra1}) is included by extending the generating function to
\begin{equation}\label{gen-gaugetra}
F_{2}^{\mu}=\overline{\Pi}_{K}^{\mu}u_{KJ}\phi_{J}
+\overline{\phi}_{K}u_{JK}^{*}\Pi_{J}^{\mu}
+P_{JK}^{\alpha\mu}\left(u_{KL}a_{LI\alpha}u_{JI}^{*}
+\frac{1}{\rmi q}\pfrac{u_{KI}}{x^{\alpha}}u_{JI}^{*}\right).
\end{equation}
The matter-field transformations remain those of Eqs.~(\ref{pointtra-rules}).
Differentiation with respect to $P_{JK}^{\alpha\nu}$ gives Eq.~(\ref{gauge-tra1}), while
\begin{equation}\label{general-pointtra-gf-deri}
p_{IL}^{\alpha\mu}=\pfrac{F_{2}^{\mu}}{a_{LI\alpha}}
=u_{JI}^{*}P_{JK}^{\alpha\mu}u_{KL},\qquad
\hat P^{\alpha\mu}=U\hat p^{\alpha\mu}U^{\dagger}.
\end{equation}
The generating identity can first be formed on unrestricted momentum components and then restricted to the primary constraint surface
\[
p_{JK}^{(\alpha\beta)}=0,\qquad P_{JK}^{(\alpha\beta)}=0.
\]
Equation~(\ref{general-pointtra-gf-deri}) preserves these constraints.
For the Yang--Mills kinetic term they express the absence of symmetric derivatives of $a_{KJ\alpha}$ in the Lagrangian.
Gauge symmetry alone does not select this kinetic term or imply regularity of its Legendre transformation.

The explicit divergence of the generating function is
\begin{align}\label{H-deri-expl}
\left.\pfrac{F_{2}^{\beta}}{x^{\beta}}\right|_{\mathrm{expl}}
&=\overline{\Pi}_{I}^{\beta}\partial_{\beta}u_{IJ}\,\phi_{J}
+\overline{\phi}_{I}\partial_{\beta}u_{JI}^{*}\,\Pi_{J}^{\beta}\nonumber\\
&\quad+P_{JK}^{\alpha\beta}\left[
(\partial_{\beta}u_{KL})a_{LI\alpha}u_{JI}^{*}
+u_{KL}a_{LI\alpha}\partial_{\beta}u_{JI}^{*}\right]\nonumber\\
&\quad+\frac{P_{JK}^{\alpha\beta}}{\rmi q}\left[
(\partial_{\alpha}u_{KI})\partial_{\beta}u_{JI}^{*}
+(\partial_{\beta}\partial_{\alpha}u_{KI})u_{JI}^{*}\right].
\end{align}
The last term vanishes on the antisymmetric constraint surface.
Using the transformation rules and $U^{\dagger}U=\Eins$ in the remaining terms gives
\begin{align*}
\left.\pfrac{F_{2}^{\beta}}{x^{\beta}}\right|_{\mathrm{expl}}
&=\rmi q\left[
(\overline{\Pi}_{K}^{\alpha}\Phi_{J}-\overline{\Phi}_{K}\Pi_{J}^{\alpha})A_{KJ\alpha}
-(\overline{\pi}_{K}^{\alpha}\phi_{J}-\overline{\phi}_{K}\pi_{J}^{\alpha})a_{KJ\alpha}\right]\\
&\quad-\frac{\rmi q}{2}P_{JK}^{\alpha\beta}
(A_{KI\alpha}A_{IJ\beta}-A_{KI\beta}A_{IJ\alpha})\\
&\quad+\frac{\rmi q}{2}p_{JK}^{\alpha\beta}
(a_{KI\alpha}a_{IJ\beta}-a_{KI\beta}a_{IJ\alpha}).
\end{align*}
No derivatives of the dynamical gauge fields have been introduced as independent Hamiltonian coefficients.

A possible gauge-field kinetic Hamiltonian is
$-\quarter p_{JK}^{\alpha\beta}p_{KJ\alpha\beta}$.
Its sign agrees with the Maxwell convention of Sec.~\ref{sec:maxwell}.
Its invariance follows from
\begin{equation}\label{ptimesp-trans}
p_{JK}^{\alpha\beta}p_{KJ\alpha\beta}
=\tr(\hat p^{\alpha\beta}\hat p_{\alpha\beta})
=\tr(\hat P^{\alpha\beta}\hat P_{\alpha\beta})
=P_{JK}^{\alpha\beta}P_{KJ\alpha\beta}.
\end{equation}
The resulting locally gauge-invariant Hamiltonian is
\begin{align}\label{H-tilde}
\bar{\HC}&=\HC+\HC_{\mathrm g},\nonumber\\
\HC_{\mathrm g}
&=\rmi q(\overline{\pi}_{K}^{\alpha}\phi_{J}-\overline{\phi}_{K}\pi_{J}^{\alpha})a_{KJ\alpha}
-\quarter p_{JK}^{\alpha\beta}p_{KJ\alpha\beta}\nonumber\\
&\quad-\frac{\rmi q}{2}p_{JK}^{\alpha\beta}
(a_{KI\alpha}a_{IJ\beta}-a_{KI\beta}a_{IJ\alpha}),\qquad
p_{JK}^{\alpha\beta}=-p_{JK}^{\beta\alpha}.
\end{align}
Here the matter Hamiltonian $\HC$ is invariant under global U$(N)$ transformations.
The kinetic term is a physical choice compatible with this symmetry.
Additional gauge-invariant interactions are not excluded by the canonical construction.

Varying the first-order action with respect to antisymmetric $p_{LM}^{\mu\nu}$ gives only the antisymmetric part of the first canonical equation,
\[
\onehalf\left(\partial_{\nu}a_{ML\mu}-\partial_{\mu}a_{ML\nu}\right)
=-\onehalf p_{ML\mu\nu}
-\frac{\rmi q}{2}(a_{MI\mu}a_{IL\nu}-a_{MI\nu}a_{IL\mu}).
\]
Thus
\begin{equation}\label{can-momentum-gf}
p_{KJ\mu\nu}=\partial_{\mu}a_{KJ\nu}-\partial_{\nu}a_{KJ\mu}
+\rmi q(a_{KI\nu}a_{IJ\mu}-a_{KI\mu}a_{IJ\nu}).
\end{equation}
The symmetric derivatives remain undetermined by this equation.
Equivalently, one may use unrestricted momenta and add symmetric Lagrange multipliers enforcing $p_{JK}^{(\mu\nu)}=0$.
They must be varied independently rather than identified in advance with derivatives of the gauge field.

Using antisymmetry, Eq.~(\ref{H-tilde}) can also be written as
\begin{align}\label{H-g2}
\HC_{\mathrm g}&=-\quarter p_{JK}^{\alpha\beta}p_{KJ\alpha\beta}
+\rmi q\left(\overline{\pi}_{K}^{\alpha}a_{KJ\alpha}\phi_{J}
-\overline{\phi}_{K}a_{KJ\alpha}\pi_{J}^{\alpha}
-p_{JK}^{\alpha\beta}a_{KI\alpha}a_{IJ\beta}\right),\nonumber\\
p_{JK}^{\mu\nu}&=-p_{JK}^{\nu\mu}.
\end{align}
The matter-field equations give
\[
\partial_{\mu}\phi_{I}=\pfrac{\bar\HC}{\overline\pi_{I}^{\mu}}
=\pfrac{\HC}{\overline\pi_{I}^{\mu}}+\rmi q a_{IJ\mu}\phi_{J}.
\]
Accordingly, this minimal canonical construction replaces ordinary derivatives by
$D_{\mu}\phi_{I}=\partial_{\mu}\phi_{I}-\rmi q a_{IJ\mu}\phi_{J}$.
This is the usual minimal coupling rule for the chosen representation.
\subsection{Locally gauge-invariant Lagrangian}
\subsubsection{Legendre transformation for a general system Hamiltonian}
The equivalent gauge-invariant Lagrangian $\bar{\LC}$
is derived by Legendre-transforming the gauge-invariant
Hamiltonian $\bar{\HC}$, defined in Eqs.~(\ref{H-tilde}) and (\ref{H-g2})
$$
\bar{\LC}=\overline{\pi}_{K}^{\alpha}\pfrac{\phi_{K}}{x^{\alpha}}+
\pfrac{\overline{\phi}_{K}}{x^{\alpha}}\pi_{K}^{\alpha}+
p_{JK}^{\alpha\beta}\pfrac{a_{KJ\alpha}}{x^{\beta}}-\bar{\HC},\qquad
\bar{\HC}=\HC+\HC_{\mathrm{g}}.
$$
With $p_{JK}^{\mu\nu}$ from Eq.~(\ref{can-momentum-gf}) and $\HC_{\mathrm{g}}$
from Eq.~({\ref{H-tilde}}), we thus have
\begin{align*}
p_{JK}^{\alpha\beta}\pfrac{a_{KJ\alpha}}{x^{\beta}}-\HC_{\mathrm{g}}&=
\onehalf p_{JK}^{\alpha\beta}\left(\pfrac{a_{KJ\alpha}}{x^{\beta}}-
\pfrac{a_{KJ\beta}}{x^{\alpha}}\right)+
\onehalf p_{JK}^{\alpha\beta}\left(\pfrac{a_{KJ\alpha}}{x^{\beta}}+
\pfrac{a_{KJ\beta}}{x^{\alpha}}\right)-\HC_{\mathrm{g}}\\
&=-\onehalf p_{JK}^{\alpha\beta}\,p_{KJ\alpha\beta}-
\onehalf\rmi q\,p_{JK}^{\alpha\beta}\left(a_{KI\alpha}\,a_{IJ\beta}-
a_{KI\beta}\,a_{IJ\alpha}\right)\\
&\quad\mbox{}+
\onehalf p_{JK}^{\alpha\beta}\left(\pfrac{a_{KJ\alpha}}{x^{\beta}}+
\pfrac{a_{KJ\beta}}{x^{\alpha}}\right)-\HC_{\mathrm{g}}\\
&=-\rmi q\left(\overline{\pi}_{K}^{\alpha}\phi_{J}-
\overline{\phi}_{K}\pi_{J}^{\alpha}\right)\,a_{KJ\alpha}-
\quarter p_{JK}^{\alpha\beta}\,p_{KJ\alpha\beta}.
\end{align*}
The locally gauge-invariant Lagrangian $\bar{\LC}$
for a given system Hamiltonian $\HC$ is then
\begin{equation}\label{general-invariant-lagrangian}
\bar{\LC}=-\quarter p_{JK}^{\alpha\beta}\,p_{KJ\alpha\beta}-
\rmi q\left(\overline{\pi}_{K}^{\alpha}\phi_{J}-
\overline{\phi}_{K}\pi_{J}^{\alpha}\right)a_{KJ\alpha}+
\overline{\pi}_{K}^{\alpha}\pfrac{\phi_{K}}{x^{\alpha}}+
\pfrac{\overline{\phi}_{K}}{x^{\alpha}}\pi_{K}^{\alpha}-\HC.
\end{equation}
As implied by the Lagrangian formalism, the dynamical variables
are given by both the fields, $\overline{\phi}_{K}$, $\phi_{J}$,
and $a_{KJ\alpha}$, in conjunction with their respective partial
derivatives with respect to the independent variables, $x^{\mu}$.
Therefore, the $\bp_{KJ}$ in $\bar{\LC}$ from
Eq.~(\ref{general-invariant-lagrangian}) are now merely abbreviations
for a combination of the Lagrangian dynamical variables.
Independently of the given system Hamiltonian $\HC$, the correlation
of the $\bp_{KJ}$ with the gauge fields $\ba_{KJ}$ and their derivatives
is given by the first canonical equation~(\ref{can-momentum-gf}).

The correlation of the momenta $\bpi_{I},\overline{\bpi}_{I}$
to the base fields $\phi_{I},\overline{\phi}_{I}$ and their derivatives
are derived from Eq.~(\ref{general-invariant-lagrangian})
for the given system Hamiltonian $\HC$ via
\begin{equation}\label{pi-phip}
\pfrac{\HC}{\overline{\pi}_{I}^{\mu}}=\pfrac{\phi_{I}}{x^{\mu}}-
\rmi q\,a_{IJ\mu}\phi_{J},\qquad
\pfrac{\HC}{\pi_{I}^{\mu}}=\pfrac{\overline{\phi}_{I}}{x^{\mu}}+
\rmi q\,\overline{\phi}_{J}\,a_{JI\mu}.
\end{equation}
Eliminating the matter momenta from these relations requires local invertibility of the matter Legendre map.
If that condition fails, the first-order action and its matter constraints must be retained.
Global gauge invariance alone does not supply this regularity.
Thus, for any \emph{globally} gauge-invariant system Hamiltonian
$\HC(\overline{\phi}_{I},\phi_{I},\overline{\bpi}_{I},\bpi_{I},x)$, the amen\-ded
Lagrangian $\bar{\LC}$ from Eq.~(\ref{general-invariant-lagrangian})
with the $\overline{\bpi}_{I},\bpi_{I}$ to be determined from Eqs.~(\ref{pi-phip})
describes in the Lagrangian formalism the associated physical system
that is invariant under \emph{local} gauge transformations.
\subsubsection{Klein-Gordon system Hamiltonian}
The generalized Klein-Gordon Hamiltonian $\HC_{\text{KG}}$
describing $N$ complex scalar fields $\phi_{I}$ that are associated
with equal masses $m$ is
$$
\HC_{\text{KG}}(\bpi_{\mu},\bpi^{*\,\mu},\bphi,\bphi^{*})=
\pi_{I\alpha}^{*}\pi_{I}^{\alpha}+m^{2}\,\phi_{I}^{*}\phi_{I}.
$$
This Hamiltonian is clearly form-invariant under the global
gauge-transformation defined by Eqs.~(\ref{pointtra-rules}).
Following Eqs.~(\ref{H-tilde}) and (\ref{H-g2}), the corresponding locally
gauge-invariant Hamiltonian $\bar{\HC}_{\text{KG}}$ is then
\begin{align*}
\bar{\HC}_{\mathrm{KG}}&=\pi_{I\alpha}^{*}\pi_{I}^{\alpha}+
m^{2}\,\phi_{I}^{*}\phi_{I}-\quarter p_{JK}^{\alpha\beta}\,p_{KJ\alpha\beta}\\
&\quad\mbox{}+\rmi q\left(\pi_{K}^{*\,\alpha}\,a_{KJ\alpha}\,\phi_{J}-
\phi_{K}^{*}\,a_{KJ\alpha}\,\pi_{J}^{\alpha}-
p_{JK}^{\alpha\beta}\,a_{KI\alpha}\,a_{IJ\beta}\right),\qquad
p_{JK}^{\mu\nu}&\stackrel{!}{=}-p_{JK}^{\nu\mu}.
\end{align*}
To derive the equivalent locally gauge-invariant Lagrangian
$\bar{\LC}_{\mathrm{KG}}$, we set up the first canonical equation for the
gauge-invariant Hamiltonian $\bar{\HC}_{\text{KG}}$ of our actual example
$$
\pfrac{\phi_{I}}{x^{\mu}}=\pfrac{\bar{\HC}_{\text{KG}}}{\pi_{I}^{*\,\mu}}=
\pi_{I\mu}+\rmi q\,a_{IJ\mu}\phi_{J},\qquad
\pfrac{\phi_{I}^{*}}{x^{\mu}}=\pfrac{\bar{\HC}_{\text{KG}}}{\pi_{I}^{\mu}}=
\pi_{I\mu}^{*}-\rmi q\,\phi_{J}^{*}\,a_{JI\mu}.
$$
Inserting $\partial\phi_{I}/\partial x^{\mu}$ and
$\partial\phi_{I}^{*}/\partial x^{\mu}$ into Eq.~(\ref{general-invariant-lagrangian}),
we directly encounter the \emph{locally} gauge-invariant Lagrangian $\bar{\LC}_{\text{KG}}$ as
$$
\bar{\LC}_{\text{KG}}=\pi_{I\alpha}^{*}\pi_{I}^{\alpha}-
m^{2}\,\phi_{I}^{*}\phi_{I}-\quarter p_{JK}^{\alpha\beta}\,p_{KJ\alpha\beta},
$$
with the abbreviations
\begin{align*}
\pi_{I\mu}&=\pfrac{\phi_{I}}{x^{\mu}}-\rmi q\,a_{IJ\mu}\phi_{J},\qquad
\pi_{I\mu}^{*}=\pfrac{\phi_{I}^{*}}{x^{\mu}}+\rmi q\,\phi_{J}^{*}\,a_{JI\mu}\\
p_{KJ\mu\nu}&=\pfrac{a_{KJ\nu}}{x^{\mu}}-\pfrac{a_{KJ\mu}}{x^{\nu}}+
\rmi q\left(a_{KI\nu}\,a_{IJ\mu}-a_{KI\mu}\,a_{IJ\nu}\right).
\end{align*}
In explicit form, $\bar{\LC}_{\text{KG}}$ is thus given by
\begin{align*}
\bar{\LC}_{\text{KG}}&=\left(\pfrac{\phi_{I}^{*}}{x^{\alpha}}+
\rmi q\,\phi_{J}^{*}\,a_{JI\alpha}\right)
\left(\pfrac{\phi_{I}}{x_{\alpha}}-\rmi q\,a_{IK}^{\alpha}\phi_{K}\right)-
m^{2}\,\phi_{I}^{*}\phi_{I}-\quarter p_{JK}^{\alpha\beta}\,p_{KJ\alpha\beta}
\end{align*}
The expressions in parentheses are the gauge-covariant derivatives of the scalar fields.
In this Lagrangian they coincide with the corresponding polymomenta.
By inserting $\bar{\LC}_{\text{KG}}$ into the Euler-Lagrange equations,
and $\bar{\HC}_{\text{KG}}$ into the canonical equations, we may convince
ourselves that the emerging field equations for $\phi_{I}^{*}$,
$\phi_{I}$, and $\ba_{JK}$ agree.
This means that $\bar{\HC}_{\text{KG}}$ and $\bar{\LC}_{\text{KG}}$
describe the \emph{same physical system}.
\subsubsection{Dirac system Hamiltonian}
For $m\ne0$, the regularized Dirac Hamiltonian of Eq.~(\ref{hd-dirac}), generalized to $N$ spinors of equal mass, is
\[
\HC_{\mathrm D}=-\rmi m\left(\overline\pi_{I}^{\alpha}-\frac{\rmi}{2}\overline\psi_{I}\gamma^{\alpha}\right)
\tau_{\alpha\beta}\left(\pi_{I}^{\beta}+\frac{\rmi}{2}\gamma^{\beta}\psi_{I}\right)
+m\overline\psi_{I}\psi_{I}.
\]
The unitary matrices act on the internal index and commute with the Dirac matrices.
Consequently, $\HC_{\mathrm D}$ is invariant under the global transformations of Eqs.~(\ref{pointtra-rules}).
The minimal canonical extension of this particular regularized Hamiltonian is
$\bar\HC_{\mathrm D}=\HC_{\mathrm D}+\HC_{\mathrm g}$, with $\HC_{\mathrm g}$ from Eq.~(\ref{H-g2}).

Introduce
\[
D_{\mu}\psi_{I}=\partial_{\mu}\psi_{I}-\rmi q a_{IJ\mu}\psi_{J},\qquad
D_{\mu}\overline\psi_{I}=\partial_{\mu}\overline\psi_{I}+\rmi q\overline\psi_{J}a_{JI\mu}.
\]
The first canonical equations become
\begin{align}\label{pi-phip-dirac}
D_{\mu}\psi_{I}
&=-\rmi m\tau_{\mu\beta}\left(\pi_{I}^{\beta}+\frac{\rmi}{2}\gamma^{\beta}\psi_{I}\right),\nonumber\\
D_{\mu}\overline\psi_{I}
&=-\rmi m\left(\overline\pi_{I}^{\alpha}-\frac{\rmi}{2}\overline\psi_{I}\gamma^{\alpha}\right)\tau_{\alpha\mu}.
\end{align}
Their inverses are
\begin{align*}
\pi_{I}^{\mu}&=-\frac{\rmi}{2}\gamma^{\mu}\psi_{I}
+\frac{\rmi}{m}\sigma^{\mu\alpha}D_{\alpha}\psi_{I},\\
\overline\pi_{I}^{\mu}&=\frac{\rmi}{2}\overline\psi_{I}\gamma^{\mu}
+\frac{\rmi}{m}(D_{\alpha}\overline\psi_{I})\sigma^{\alpha\mu}.
\end{align*}
The Legendre transformation first yields
\begin{equation}\label{ld-intermediate}
\bar\LC_{\mathrm D}=-\quarter p_{JK}^{\alpha\beta}p_{KJ\alpha\beta}
-\rmi m\overline\pi_{I}^{\alpha}\tau_{\alpha\beta}\pi_{I}^{\beta}
-\frac{2m}{3}\overline\psi_{I}\psi_{I}.
\end{equation}
Substituting the momenta gives
\begin{align*}
\bar\LC_{\mathrm D}
&=\frac{\rmi}{2}\left[\overline\psi_{I}\gamma^{\alpha}D_{\alpha}\psi_{I}
-(D_{\alpha}\overline\psi_{I})\gamma^{\alpha}\psi_{I}\right]
-m\overline\psi_{I}\psi_{I}\\
&\quad+\frac{\rmi}{m}(D_{\alpha}\overline\psi_{I})\sigma^{\alpha\beta}D_{\beta}\psi_{I}
-\quarter p_{JK}^{\alpha\beta}p_{KJ\alpha\beta}.
\end{align*}
The sign of the term quadratic in covariant derivatives agrees with the positive $\rmi/m$ coefficient in Eq.~(\ref{ld-dirac-regular}).
Here $p_{KJ\alpha\beta}$ is the field strength in Eq.~(\ref{can-momentum-gf}).

The free regularizing term is a total divergence, but its covariantized counterpart also contains a field-strength interaction.
Indeed,
\[
[D_{\alpha},D_{\beta}]\psi_{I}=-\rmi q p_{IJ\alpha\beta}\psi_{J},
\]
so that antisymmetry of $\sigma^{\alpha\beta}$ gives the exact identity
\begin{align*}
\frac{\rmi}{m}(D_{\alpha}\overline\psi_{I})\sigma^{\alpha\beta}D_{\beta}\psi_{I}
&=\frac{\rmi}{m}\partial_{\alpha}
\left(\overline\psi_{I}\sigma^{\alpha\beta}D_{\beta}\psi_{I}\right)
-\frac{q}{2m}\overline\psi_{I}\sigma^{\alpha\beta}p_{IJ\alpha\beta}\psi_{J}.
\end{align*}
Modulo the displayed total divergence, the resulting theory is therefore the minimally coupled Dirac theory with the additional Pauli interaction
$-q\overline\psi_{I}\sigma^{\alpha\beta}p_{IJ\alpha\beta}\psi_{J}/(2m)$.
This interaction depends on the chosen regularization of the free theory.
It is not fixed by local gauge invariance alone, and this model must not be identified with ordinary minimal Dirac coupling without that qualification.
In particular, two free Lagrangians that differ by a total divergence need not remain dynamically equivalent after ordinary derivatives in that divergence are replaced by covariant derivatives.

The spinor equations are
\begin{align*}
\rmi\gamma^{\alpha}D_{\alpha}\psi_{I}-m\psi_{I}
-\frac{q}{2m}p_{IK\alpha\beta}\sigma^{\alpha\beta}\psi_{K}&=0,\\
\rmi(D_{\alpha}\overline\psi_{I})\gamma^{\alpha}+m\overline\psi_{I}
+\frac{q}{2m}\overline\psi_{K}\sigma^{\alpha\beta}p_{KI\alpha\beta}&=0.
\end{align*}
For a single spinor, the first equation reduces to
\[
\left[\rmi\gamma^{\alpha}\partial_{\alpha}
+q\gamma^{\alpha}a_{\alpha}-m
-\frac{q}{2m}(\partial_{\alpha}a_{\beta}-\partial_{\beta}a_{\alpha})\sigma^{\alpha\beta}\right]\psi=0.
\]
The field-strength term is invariant under U$(1)$ gauge transformations, while the full equation transforms covariantly with the spinor.
\subsubsection{Canonical equations of the coupled system}
For the fundamental matter fields, the component brackets of Sec.~\ref{sec:p-k} reproduce the canonical equations for $\bar\HC=\HC+\HC_{\mathrm g}$.
The formulas below are classical canonical identities and do not prescribe a quantum commutator.
The gauge-field equations must separately respect the antisymmetric momentum constraints.
Not elaborating on terms that finally vanish by virtue of the Poisson bracket
relations, the equations for ${\psi}$ and ${\overline{\psi}}$ follow as
\begin{align*}
\pfrac{{\psi}_{I}}{x^{\mu}}&=\left[{\psi}_{I},{\bar{\HC}}\right]_{\mu}\\
&=\left[{\psi}_{I},{\HC}\right]_{\mu}+\rmi q\underbrace{\left[{\psi}_{I},
{\overline{\pi}}_{K}^{\alpha}\right]_{\mu}}_{=\delta_{\mu}^{\alpha}\delta_{IK}}
{a}_{KJ\alpha}\,{\psi}_{J}\\
&=\left[{\psi}_{I},{\HC}\right]_{\mu}+\rmi q\,{a}_{IJ\mu}\,{\psi}_{J}.
\end{align*}
Similarly,
\begin{align*}
\pfrac{{\overline{\psi}}_{I}}{x^{\mu}}&=\left[{\overline{\psi}}_{I},{\bar{\HC}}\right]_{\mu}\\
&=\left[{\overline{\psi}}_{I},{\HC}\right]_{\mu}-
\rmi q\,{\overline{\psi}}_{K}{a}_{KJ\alpha}\,\underbrace{\left[{\overline{\psi}}_{I},
{\pi}_{J}^{\alpha}\right]_{\mu}}_{=\delta_{\mu}^{\alpha}\delta_{IJ}}\\
&=\left[{\overline{\psi}}_{I},{\HC}\right]_{\mu}-\rmi q\,{\overline{\psi}}_{K}\,{a}_{KI\mu}.
\end{align*}
The equations for the canonical momenta follow as
\begin{align*}
\delta_{\mu}^{\nu}\pfrac{{\pi}_{I}^{\alpha}}{x^{\alpha}}&=
\left[{\pi}_{I}^{\nu},{\bar{\HC}}\right]_{\mu}\\
&=\left[{\pi}_{I}^{\nu},{\HC}\right]_{\mu}-\rmi q\underbrace{\left[{\pi}_{I}^{\nu},
{\overline{\psi}}_{K}\right]_{\mu}}_{=-\delta_{\mu}^{\nu}\delta_{IK}}
{a}_{KJ\alpha}\,{\pi}_{J}^{\alpha}\\
&=\left[{\pi}_{I}^{\nu},{\HC}\right]_{\mu}+\delta_{\mu}^{\nu}\,
\rmi q\,{a}_{IJ\alpha}\,{\pi}_{J}^{\alpha}
\end{align*}
and
\begin{align*}
\delta_{\mu}^{\nu}\pfrac{{\overline{\pi}}_{I}^{\alpha}}{x^{\alpha}}&=
\left[{\overline{\pi}}_{I}^{\nu},{\bar{\HC}}\right]_{\mu}\\
&=\left[{\overline{\pi}}_{I}^{\nu},{\HC}\right]_{\mu}+
\rmi q\,{\overline{\pi}}_{K}^{\alpha}{a}_{KJ\alpha}\underbrace{\left[{\overline{\pi}}_{I}^{\nu},
{\psi}_{J}\right]_{\mu}}_{=-\delta_{\mu}^{\nu}\delta_{IJ}}\\
&=\left[{\overline{\pi}}_{I}^{\nu},{\HC}\right]_{\mu}-\delta_{\mu}^{\nu}\,
\rmi q\,{\overline{\pi}}_{K}^{\alpha}\,{a}_{KI\alpha}.
\end{align*}
These equations implement the ``minimum coupling rule''.
Since the globally gauge-invariant matter Hamiltonian $\HC$ does not depend on the gauge fields, their equations have the same form for all such matter systems.
The first follows by antisymmetric momentum variation and is Eq.~(\ref{can-momentum-gf}).
An unrestricted bracket equation for the full gradient $\partial_{\mu}a_{MN\nu}$ would impose extra conditions and is not equivalent to this constrained equation.
Variation with respect to $a_{MN\xi}$ gives
\[
\partial_{\alpha}p_{NM}^{\xi\alpha}
=\rmi q\left(\overline\psi_{M}\pi_{N}^{\xi}-\overline\pi_{M}^{\xi}\psi_{N}
+a_{NJ\alpha}p_{JM}^{\xi\alpha}-p_{NJ}^{\xi\alpha}a_{JM\alpha}\right).
\]
Defining the matter current matrix by
$\mathsf J_{NM}^{\mu}=\rmi q(\overline\pi_{M}^{\mu}\psi_{N}-\overline\psi_{M}\pi_{N}^{\mu})$,
this equation is equivalently
\[
D_{\alpha}\hat p^{\alpha\mu}=\hat{\mathsf J}^{\mu},\qquad
D_{\alpha}X=\partial_{\alpha}X-\rmi q[\hat a_{\alpha},X].
\]
\subsection{\label{sec:noether-gauge}Noether current of the U$(N)$ gauge transformation}
Write an infinitesimal unitary transformation as
\[
U(x)=\Eins+\rmi\epsilon u(x)+\CO(\epsilon^{2}),\qquad u^{\dagger}=u.
\]
The entries of $u$ are generally complex and obey $u_{IJ}=\overline{u_{JI}}$.
For SU$(N)$ one also requires $\tr u=0$.
Then
\[
\delta\phi_{I}=\rmi\epsilon u_{IJ}\phi_{J},\qquad
\delta\overline\phi_{I}=-\rmi\epsilon\overline\phi_{J}u_{JI},\qquad
\delta\hat a_{\alpha}=\epsilon\left(\rmi[u,\hat a_{\alpha}]+\frac{1}{q}\partial_{\alpha}u\right).
\]
Expanding Eq.~(\ref{gen-gaugetra}) to first order gives
\begin{equation}\label{gen-gaugetra-infini}
F_{2}^{\mu}=\overline\Pi_{J}^{\mu}\phi_{J}+\overline\phi_{J}\Pi_{J}^{\mu}
+P_{JK}^{\alpha\mu}a_{KJ\alpha}+\frac{\epsilon}{q}j^{\mu},
\end{equation}
where
\begin{align}\label{noether-gen-gaugetra}
j^{\mu}&=\rmi q\left(\overline\pi_{K}^{\mu}u_{KJ}\phi_{J}
-\overline\phi_{K}u_{KJ}\pi_{J}^{\mu}\right)\nonumber\\
&\quad+\rmi q p_{JK}^{\alpha\mu}(u_{KI}a_{IJ\alpha}-a_{KI\alpha}u_{IJ})
+p_{JK}^{\alpha\mu}\partial_{\alpha}u_{KJ}.
\end{align}
The corresponding total matrix current is
\[
j_{JK}^{\mu}=\rmi q\left(\overline\pi_{K}^{\mu}\phi_{J}-\overline\phi_{K}\pi_{J}^{\mu}
+a_{JI\alpha}p_{IK}^{\alpha\mu}-p_{JI}^{\alpha\mu}a_{IK\alpha}\right).
\]
The gauge-field equation and antisymmetry imply
\[
\partial_{\alpha}p_{JK}^{\alpha\mu}=j_{JK}^{\mu},\qquad
\partial_{\mu}j_{JK}^{\mu}=0.
\]
Consequently, the parameter-dependent Noether current has the on-shell superpotential representation
\[
j^{\mu}=u_{KJ}j_{JK}^{\mu}+p_{JK}^{\alpha\mu}\partial_{\alpha}u_{KJ}
=\partial_{\alpha}(p_{JK}^{\alpha\mu}u_{KJ}),\qquad
\partial_{\mu}j^{\mu}=0.
\]
For SU$(N)$ these relations are understood after projection onto the traceless Lie algebra.
The matter current alone obeys covariant conservation, $D_{\mu}\hat{\mathsf J}^{\mu}=0$.
The ordinary conserved current $j_{JK}^{\mu}$ also contains the gauge-field contribution and is gauge dependent.
It is therefore not a set of gauge-invariant local color observables.
For SU$(3)$ the internal currents are color currents, whereas the general U$(N)$ construction is not specifically quantum chromodynamics.
The appearance of the gauge fields in the source expresses their non-Abelian self-interaction.
\subsection{\label{sec:higgs}The Abelian Higgs mechanism and canonical field shifts}
The gauge Hamiltonian~(\ref{H-g2}) contains no explicit Proca mass term.
Adding such a term alone would destroy its local gauge invariance.
The Higgs mechanism instead starts from a gauge-invariant matter system whose scalar potential has a minimum at nonzero field modulus.
A canonical change of variables can describe fluctuations about that vacuum, but a change of variables alone cannot create a new mass spectrum.

For $N=1$, consider the complete Hamiltonian
\[
\bar\HC=\overline\pi_{\alpha}\pi^{\alpha}
+\lambda\left(\overline\phi\phi-\frac{v^{2}}{2}\right)^{2}
+\rmi q(\overline\pi^{\alpha}\phi-\overline\phi\pi^{\alpha})a_{\alpha}
-\quarter p^{\alpha\beta}p_{\alpha\beta},\qquad
\lambda>0,\quad v>0.
\]
The scalar potential has minima at $\overline\phi\phi=v^{2}/2$.
Choose a representative vacuum $\varphi=v/\sqrt{2}$ and introduce
\begin{equation}\label{higgs-condition}
\Phi=\phi-\varphi,\qquad
\partial_{\mu}\Phi=\partial_{\mu}\phi.
\end{equation}
A canonical generator for this shift is
\begin{equation}\label{gen-higgs}
F_{2}^{\mu}=\overline\Pi^{\mu}(\phi-\varphi)
+(\overline\phi-\overline\varphi)\Pi^{\mu}+P^{\alpha\mu}a_{\alpha}.
\end{equation}
It gives
\[
\begin{gathered}
\phi=\Phi+\varphi,\qquad \overline\phi=\overline\Phi+\overline\varphi,\\
\pi^{\mu}=\Pi^{\mu},\qquad \overline\pi^{\mu}=\overline\Pi^{\mu},\\
a_{\mu}=A_{\mu},\qquad p^{\mu\nu}=P^{\mu\nu}.
\end{gathered}
\]
In particular, the antisymmetric gauge-momentum constraint is preserved.
No field-dependent symmetric shift of $p^{\mu\nu}$ is admissible on this constraint surface.
As the generator has no explicit $x$ dependence, the transformed Hamiltonian has the same value,
\begin{align}\label{hd-higgs-p}
\bar\HC^{\prime}
&=\overline\Pi_{\alpha}\Pi^{\alpha}
+\lambda\left[(\overline\Phi+\overline\varphi)(\Phi+\varphi)-\frac{v^{2}}{2}\right]^{2}
-\quarter P^{\alpha\beta}P_{\alpha\beta}\nonumber\\
&\quad+\rmi q\left[\overline\Pi^{\alpha}(\Phi+\varphi)
-(\overline\Phi+\overline\varphi)\Pi^{\alpha}\right]A_{\alpha}.
\end{align}
The shifted field still transforms under the original gauge symmetry, now nonlinearly in $\Phi$.

To display the physical fluctuations, use the equivalent Lagrangian
\[
\bar\LC=(D_{\alpha}\phi)^{*}D^{\alpha}\phi
-\lambda\left(\overline\phi\phi-\frac{v^{2}}{2}\right)^{2}
-\quarter p^{\alpha\beta}p_{\alpha\beta}.
\]
Locally where $\phi\ne0$, write
\[
\phi=\frac{v+h}{\sqrt{2}}e^{\rmi\theta/v},\qquad
A_{\mu}=a_{\mu}-\frac{1}{qv}\partial_{\mu}\theta.
\]
Choosing unitary gauge removes the phase from the scalar field and gives
\[
\bar\LC_{\mathrm u}=-\quarter P^{\alpha\beta}P_{\alpha\beta}
+\onehalf\partial_{\alpha}h\partial^{\alpha}h
+\onehalf q^{2}(v+h)^{2}A_{\alpha}A^{\alpha}
-\frac{\lambda}{4}\left[(v+h)^{2}-v^{2}\right]^{2}.
\]
With $\pi_{h}^{\mu}=\partial^{\mu}h$, its constrained covariant Hamiltonian is
\begin{equation}\label{hd-higgs-p2}
\bar\HC_{\mathrm u}=\onehalf\pi_{h\alpha}\pi_{h}^{\alpha}
-\quarter P^{\alpha\beta}P_{\alpha\beta}
-\onehalf q^{2}(v+h)^{2}A_{\alpha}A^{\alpha}
+\frac{\lambda}{4}\left[(v+h)^{2}-v^{2}\right]^{2}.
\end{equation}
The vector mass term has the same Hamiltonian sign as the Proca term.
Expanding about $h=0$ yields
\[
m_{A}^{2}=q^{2}v^{2},\qquad m_{h}^{2}=2\lambda v^{2}.
\]
The two physical polarizations of the massless vector and the two real scalar components reorganize into three polarizations of a massive vector and one radial scalar mode.
The local gauge redundancy remains present in the gauge-invariant formulation.
The mass terms arise from the covariant scalar kinetic term evaluated about the chosen nonzero vacuum, rather than from the canonical shift by itself.
\section{Transformations of the Dirac Hamiltonian}
\subsection{\label{sec:Dirac-shift}Change of momenta under the Dirac boundary term}
The canonical momenta of the original Dirac Lagrangian
$\LC_{\mathrm D}$ in Eq.~(\ref{ld-dirac-symm}) satisfy the primary constraints
\[
\overline{\pi}^{\mu}=\frac{\rmi}{2}\overline{\psi}\gamma^{\mu},
\qquad
\pi^{\mu}=-\frac{\rmi}{2}\gamma^{\mu}\psi.
\]
The momenta of the regularized Lagrangian
$\LC_{\mathrm D}^{\prime}$ in Eq.~(\ref{ld-dirac-regular}) are instead
\begin{align*}
\overline{\Pi}^{\mu}&=\frac{\rmi}{2}\overline{\psi}\gamma^{\mu}
+\frac{\rmi}{m}\pfrac{\overline{\psi}}{x^{\alpha}}\sigma^{\alpha\mu},\\
\Pi^{\mu}&=-\frac{\rmi}{2}\gamma^{\mu}\psi
+\frac{\rmi}{m}\sigma^{\mu\alpha}\pfrac{\psi}{x^{\alpha}}.
\end{align*}
Their relation to the original constrained momenta is
\begin{equation}\label{dirac-canmo}
\overline{\Pi}^{\mu}=\overline{\pi}^{\mu}
+\frac{\rmi}{m}\pfrac{\overline{\psi}}{x^{\alpha}}\sigma^{\alpha\mu},
\qquad
\Pi^{\mu}=\pi^{\mu}
+\frac{\rmi}{m}\sigma^{\mu\alpha}\pfrac{\psi}{x^{\alpha}}.
\end{equation}
The signs agree with the positive derivative term in Eq.~(\ref{ld-dirac-regular}).
For smooth fields in Cartesian coordinates, the divergences of the additional terms vanish because $\sigma^{\mu\nu}$ is constant and antisymmetric.
Consequently,
\[
\partial_{\mu}\overline{\Pi}^{\mu}=\partial_{\mu}\overline{\pi}^{\mu},
\qquad
\partial_{\mu}\Pi^{\mu}=\partial_{\mu}\pi^{\mu}.
\]
The equality of divergences follows from the derivative-dependent boundary term
\[
F^{\mu}=\frac{\rmi}{2m}\left(
\overline{\psi}\sigma^{\mu\alpha}\partial_{\alpha}\psi
+(\partial_{\alpha}\overline{\psi})\sigma^{\alpha\mu}\psi\right),
\qquad
\LC_{\mathrm D}^{\prime}=\LC_{\mathrm D}+\partial_{\mu}F^{\mu}.
\]
For compactly supported variations, the two Lagrangians therefore yield the same free Dirac equations.

This change must be distinguished from the canonical momentum shift of Sect.~\ref{sec:beispiel3}.
The derivatives of the dynamical fields are not prescribed functions of $x$ that can be held fixed as external coefficients in a phase-space generator.
Equation~(\ref{dirac-canmo}) relates two Legendre maps of different ranks.
It is not an invertible canonical map between their unrestricted momentum spaces.
In particular, the Hamiltonian of the original first-order Dirac Lagrangian is constrained, whereas the regularized Hamiltonian is the function in Eq.~(\ref{hd-dirac}) with $\pi$ replaced by $\Pi$.
Equality of the momentum divergences does not identify these Hamiltonian functions or eliminate the constraints of the original formulation.
\subsection{\label{sec:Dirac-interchange}Limits of an interchange of the dynamical variables}
Consider the formal type-$1$ expression
\begin{equation}\label{gen-dirac-swap}
F_{1}^{\mu}(\psi,\Psi,\overline{\psi},\overline{\Psi})
=-\sqrt{\onehalf}\;\overline{\psi}\gamma^{\mu}\Psi
+\sqrt{\onehalf}\;\overline{\Psi}\gamma^{\mu}\psi.
\end{equation}
The formal coefficient rules~(\ref{genF1}) give
\begin{align}\label{rule-dirac-swap}
\pi^{\mu}&=-\sqrt{\onehalf}\gamma^{\mu}\Psi,&
\Pi^{\mu}&=-\sqrt{\onehalf}\gamma^{\mu}\psi,\nonumber\\
\overline{\pi}^{\mu}&=\sqrt{\onehalf}\;\overline{\Psi}\gamma^{\mu},&
\overline{\Pi}^{\mu}&=\sqrt{\onehalf}\;\overline{\psi}\gamma^{\mu}.
\end{align}
However, the four components of each vector-spinor momentum in the regularized theory are independent phase-space variables.
The first relation in Eq.~(\ref{rule-dirac-swap}) implies
\[
\Psi=-\frac{1}{2\sqrt{2}}\gamma_{\alpha}\pi^{\alpha},
\qquad
\pi^{\mu}=\frac14\gamma^{\mu}\gamma_{\alpha}\pi^{\alpha},
\]
and its adjoint implies
\[
\overline{\pi}^{\mu}
=\frac14\overline{\pi}^{\alpha}\gamma_{\alpha}\gamma^{\mu}.
\]
These are additional algebraic restrictions on the momenta.
Each unrestricted $16$-component vector-spinor is confined to the $4$-dimensional image of the map $\Psi\mapsto\gamma^{\mu}\Psi$.
The formal relations therefore do not define a locally invertible map on the full phase-space fiber.

The regularized Dirac Hamiltonian has the form
\begin{equation}\label{hd-dirac3}
\HC_{\mathrm D}=\rmi m\left(
\frac16\overline{\psi}\gamma_{\alpha}\pi^{\alpha}
-\overline{\pi}^{\alpha}\tau_{\alpha\beta}\pi^{\beta}
-\frac16\overline{\pi}^{\alpha}\gamma_{\alpha}\psi\right)
+\frac{2m}{3}\overline{\psi}\psi.
\end{equation}
Substitution of Eq.~(\ref{rule-dirac-swap}) can rearrange this expression on the restricted momentum submanifold.
It does not establish a canonical symmetry of the unrestricted regularized Dirac Hamiltonian.
An interpretation on a reduced space would require the additional momentum constraints and their induced canonical structure to be specified separately.
\subsection{\label{sec:Dirac-gauge}Local U$(1)$ gauge transformation of the Dirac Hamiltonian}
For a single Dirac spinor, the U$(N)$ transformation reduces to
\begin{align*}
\Psi&=e^{\rmi\theta(x)}\psi,&
\overline{\Psi}&=e^{-\rmi\theta(x)}\overline{\psi},\\
\Pi^{\mu}&=e^{\rmi\theta(x)}\pi^{\mu},&
\overline{\Pi}^{\mu}&=e^{-\rmi\theta(x)}\overline{\pi}^{\mu}.
\end{align*}
The real gauge potential transforms as
\[
A_{\mu}=a_{\mu}+\frac1q\partial_{\mu}\theta.
\]
With
$D_{\mu}\psi=\partial_{\mu}\psi-\rmi q a_{\mu}\psi$ and
$D_{\mu}\overline{\psi}=\partial_{\mu}\overline{\psi}+\rmi q a_{\mu}\overline{\psi}$,
the gauge-invariant Hamiltonian associated with the regularized Dirac theory is
\begin{align}\label{ham-qed}
\bar{\HC}_{\mathrm D}
&=-\rmi m\left(\overline{\pi}^{\alpha}\tau_{\alpha\beta}\pi^{\beta}
+\frac16\overline{\pi}^{\alpha}\gamma_{\alpha}\psi
-\frac16\overline{\psi}\gamma_{\alpha}\pi^{\alpha}\right)
+\frac{2m}{3}\overline{\psi}\psi\nonumber\\
&\quad+\rmi q\left(\overline{\pi}^{\alpha}\psi
-\overline{\psi}\pi^{\alpha}\right)a_{\alpha}
-\frac14p^{\alpha\beta}p_{\alpha\beta},
\qquad p^{\mu\nu}=-p^{\nu\mu}.
\end{align}
The antisymmetry of the gauge momentum is a primary constraint inherited from the Maxwell Legendre map.
The variation of the canonical action with respect to the independent antisymmetric momentum components gives
\begin{equation}\label{ham2-qed}
\frac12\left(\partial_{\nu}a_{\mu}-\partial_{\mu}a_{\nu}\right)
=-\frac12p_{\mu\nu},
\qquad
p_{\mu\nu}=F_{\mu\nu}
=\partial_{\mu}a_{\nu}-\partial_{\nu}a_{\mu}.
\end{equation}
Thus the Hamiltonian determines the antisymmetric derivative of the gauge potential.
A term containing its symmetric derivatives is not needed in the Hamiltonian function.
It would introduce field derivatives as purported independent Hamiltonian coefficients and would obscure the primary constraints.

Eliminating the momenta yields
\begin{align*}
\bar\LC_{\mathrm D}
&=\frac{\rmi}{2}\left(
\overline{\psi}\gamma^{\mu}D_{\mu}\psi
-(D_{\mu}\overline{\psi})\gamma^{\mu}\psi\right)
-m\overline{\psi}\psi\\
&\quad+\frac{\rmi}{m}(D_{\mu}\overline{\psi})\sigma^{\mu\nu}D_{\nu}\psi
-\frac14F^{\mu\nu}F_{\mu\nu}.
\end{align*}
The derivative term was a total divergence in the free theory, but it is not a total divergence after covariantization.
Using
$[D_{\mu},D_{\nu}]\psi=-\rmi qF_{\mu\nu}\psi$ gives the identity
\[
\frac{\rmi}{m}(D_{\mu}\overline{\psi})\sigma^{\mu\nu}D_{\nu}\psi
=\partial_{\mu}\left(\frac{\rmi}{m}\overline{\psi}\sigma^{\mu\nu}D_{\nu}\psi\right)
-\frac{q}{2m}\overline{\psi}\sigma^{\mu\nu}F_{\mu\nu}\psi.
\]
The resulting interacting model therefore contains a Pauli coupling in addition to the minimally coupled Dirac and Maxwell terms.
Its coefficient follows from the chosen free regularization coefficient $1/m$.
The interaction is gauge invariant, but it is not equivalent to the minimally coupled Maxwell--Dirac theory unless a compensating interaction removes the Pauli term.

\end{document}